\documentclass{elsart}
\usepackage{epsfig}
\usepackage{amssymb}
\usepackage{graphicx}
\usepackage{bm}\usepackage{citesort}
\usepackage{amssymb,amsmath}
\usepackage{makeidx}
\usepackage{multicol}

\newcommand{\nn}{\nonumber\\ }
\newcommand{\beq}{\begin{eqnarray}}
\newcommand{\eeq}{\end{eqnarray}}
\newcommand{\be}{\begin{eqnarray}}
\newcommand{\ee}{\end{eqnarray}}
\newcommand{\BQ}{\begin{equation}}
\newcommand{\EQ}{\end{equation}}
\newcommand{\BQA}{\begin{eqnarray}}
\newcommand{\EQA}{\end{eqnarray}}
\newcommand{\NN}{\nonumber \\}

\def\simge{\mathrel{%
   \rlap{\raise 0.511ex \hbox{$>$}}{\lower 0.511ex \hbox{$\sim$}}}}
\def\simle{\mathrel{
   \rlap{\raise 0.511ex \hbox{$<$}}{\lower 0.511ex \hbox{$\sim$}}}}
\def\bigs{\mathrel{
   \rlap{\raise 0.531ex \hbox{$>$}}{\lower 0.531ex \hbox{$<$}}}}

\def\grad{\nabla}                               
\def\del{\partial}                              

\newcommand{\kk}{k_\perp}
\newcommand{\rmd}{{\rm d}}

\begin{document}

\begin{flushright}
{\small\sf SPhT-T04/028}
\end{flushright}
\vspace{0.8cm}
\begin{frontmatter}

\title{Cronin effect and high--$p_\perp$ suppression
in the nuclear gluon distribution at small $x$}

\author{E. Iancu},
\author{K. Itakura},
\author{and D. N. Triantafyllopoulos}
\address{Service de Physique Th\'eorique, CEA/DSM/SPhT,
91191 Gif-sur-Yvette Cedex, France}

\date{\today}
\vspace{0.8cm}
\begin{abstract}
We present a systematic, and fully analytic, study of the ratio
${\mathcal R}_{pA}$ between the gluon distribution in a nucleus
and that in a proton scaled up by the atomic number $A$. We
consider initial conditions of the McLerran--Venugopalan type, and
quantum evolution in the Color Glass Condensate, with both fixed
and running coupling. We perform an analytic study of the Cronin
effect in the initial conditions and point out an interesting
difference between saturating effects and twist effects in the
nuclear gluon distribution. We show that the distribution of the
gluons which make up the condensate in the initial conditions is
localized at low momenta, but this particular feature does not
survive after the quantum evolution. We demonstrate that the rapid
suppression of the ratio ${\mathcal R}_{pA}$ in the early stages
of the evolution is due to the DGLAP--like evolution of the
proton, whose gluon distribution grows much faster than that in
the nucleus because of the large separation between the respective
saturation momenta. The flattening of the Cronin peak, on the
other hand, is due to the evolution of the nucleus. We show that
the running coupling effects slow down the evolution, but
eventually lead to a stronger suppression in ${\mathcal R}_{pA}$
at sufficiently large energies.

\end{abstract}
\end{frontmatter}
\newpage\tableofcontents

\newpage
\section{Introduction}
\setcounter{equation}{0}

Motivated especially by the recent experimental results for
gold--gold (Au--Au) \cite{QM,d'Enterria} and deuteron--gold
(d--Au) collisions at RHIC
\cite{RHIC-dAu-mid,Brahms-data,RHIC-dAu-forward,RHIC-dAu-forward-other},
there is currently a large interest in the physics of collective
phenomena in the wavefunction of an energetic nucleus, which could
explain, for instance, why the gluon distribution of a large
nucleus (with atomic number $A\gg 1$) is not simply the incoherent
sum of the gluon distributions of the $A$ constituent nucleons.
For sufficiently high energies (and, marginally at least, also for
the energies at RHIC), one expects these collective phenomena to
be associated with non--linear effects favored by the high gluon
density, which rises rapidly with the energy, and also with $A$
\cite{GLR,MQ85,BM87,MV,AMCargese,CGCreviews}. Thus, these effects
are expected to be more pronounced in a (large) nucleus as
compared to a proton, and this difference should be particularly
important at not so high energies, where the proton is still in a
linear regime.

This difference could explain some of the remarkable regularities
observed in the results for d--Au collisions at RHIC, like the
Cronin enhancement in particle production at central rapidity
($y=0$)\cite{RHIC-dAu-mid}, and the `high--$\kk$ suppression' in
the particle yield at `forward rapidities' ($y=2-3$ in the
deuteron fragmentation region)
\cite{Brahms-data,RHIC-dAu-forward,RHIC-dAu-forward-other}.
Specifically, the Cronin peak \cite{CroninExp} can be understood
as the result of Glauber--like multiple scattering off the gluon
distribution produced by uncorrelated ``valence quarks''
\cite{KM98} (as described, e.g., by the McLerran--Venugopalan
model \cite{MV}, see also \cite{AG03}), whereas the high--$\kk$
suppression can result from non--linear effects in the quantum
evolution with increasing energy \cite{KLM02}. (See also
\cite{Vitev03} for alternative interpretations of these results.)
On the other hand it is most likely that a similar suppression
seen in Au--Au collisions at central rapidity \cite{d'Enterria} is
due to jet quenching through final state interactions
\cite{JetQuenching}.

More precisely, the experimental results are used to construct the
ratio $R_{dAu}(\kk,y)$ between the number of particles produced in
a d--Au collision (with a given transverse momentum $\kk$ and at a
given rapidity $y$) and the corresponding number for a
proton--proton collision scaled up by the number of collisions.
The deviation of this ratio from one --- like a Cronin peak
($R_{dAu}> 1$ at intermediate momenta), or the high--$\kk$
suppression ($R_{dAu}< 1$ at generic, hard, momenta) --- can be
attributed to the difference between the gluon spectrum (or
``unintegrated gluon distribution") in the nucleus, and that in
the proton scaled up by $A$. In turn, this difference is measured
by the ratio:\beq\label{Rdef} {\mathcal
R}_{pA}(k_\perp,y)\,\equiv\,\frac{\varphi_A(k_\perp,y)}{A^{1/3} \,
\varphi_p(k_\perp,y)}\,,\eeq between the {\sf gluon occupation
factors} (see Eq.~(\ref{phidef})) in the nucleus and the
proton\footnote{Our conventions are such that the more standard
``unintegrated gluon distribution" is equal to $S_A
\varphi_A(k_\perp,y)$, with $S_A$ the area of the nuclear disk.}.
Previous studies in the literature show that it is indeed possible
to compute the cross--section for particle production in terms of
the gluon distributions in the target and the projectile
\cite{GLR,KM98,KT02,DM01,Braun,BGV04}, and, moreover, any
qualitative trend seen in the ratio (\ref{Rdef}) of the gluon
distributions gets transmitted to the corresponding ratio for the
cross--sections \cite{KLM02,Baier03,KKT,Nestor03}, which is the
quantity relevant for RHIC.

Motivated by this observation, and also by the conceptual
importance of the ratio (\ref{Rdef}) as a direct measure of
collective effects in the nuclear wavefunction, we shall devote
this work to a systematic analysis of this quantity within the
effective theory for the Color Glass Condensate (CGC) \cite{RGE},
which is the appropriate framework to describe non--linear effects
in the gluon distribution within QCD.

Although inspired by the RHIC phenomenology, our analysis does not
aim to the description of the data (for such attempts, see e.g.
\cite{KLN02,J04}), but is rather oriented towards the fundamental
understanding of phenomena like the emergence of the Cronin peak
in the ratio ${\mathcal R}_{pA}$ and the suppression of this peak
with increasing energy. Our main objective is to present a
calculation for such phenomena from {\sf first principles}, and
which is {\sf fully analytic}. These requirements entail strong
constraints on the kinematical range which is accessible to our
analysis, and also a significant loss of accuracy, due on the one
hand to the limited accuracy of the perturbative framework that we
shall use, and on the other hand to the additional approximations
that we shall perform within this framework, to allow for analytic
calculations. In particular, our focus on the gluon distribution,
rather than on the cross-section for multiparticle production, is
also motivated by the requirement of (analytic) calculability,
together with our concern to avoid theoretical uncertainties as
much as possible.

As already mentioned, our theoretical framework will be the CGC
effective theory\cite{MV,RGE,K96,JKMW97,JKLW97,B,K,W,SAT,GAUSS}
(see also \cite{AMCargese,CGCreviews} for recent review papers),
which is a lowest--order (in $\alpha_s$) formalism in QCD,
although obtained after elaborate resummations: These resummations
include only those diagrams of perturbative QCD in which the
powers of $\alpha_s$ are enhanced by either large logarithms of
the energy (i.e., by powers of $y\sim \ln s$), or by high density
effects (e.g., by the gluon occupation factor,  which at
saturation is of order $1/\alpha_s$). Still, it should be
emphasized that this formalism becomes more and more accurate with
increasing energy because, first, it does include the leading
effects at large $y$, and, second, the {\sf saturation momentum}
$Q_s(y)$ --- the typical momentum of the gluons which make the
condensate --- increases very fast with $y$
\cite{GLR,AM99,SCALING,MT02,DT02,MP03,AM03}, so the relevant
coupling constant decreases  at high energy:
$\alpha_s(Q_s(y))\propto 1/\sqrt{y}$ at large $y$. Thus, the
results that we shall obtain here represent the actual prediction
of QCD for sufficiently large $y$.

This discussion also shows that the inclusion of running coupling
effects in the formalism is crucial in order for the asymptotic
freedom of QCD to become operative at high energies. The running
of the coupling is a higher order effect in $\alpha_s$, so its
inclusion in a leading order formalism is by definition ambiguous.
But physical intuition about the typical scales involved in the
interactions, and also the experience with BFKL equation
\cite{BFKL} --- which in the context of the CGC formalism is the
linear equation describing the approach towards saturation, and is
presently known to next--to--leading order in $\alpha_s$
\cite{NLBFKL,Salam99} ---, will permit us to effectively use a
one--loop running coupling, with the scale for running set either
by the gluon transverse momentum $k_\perp$, or by the saturation
momentum.

As mentioned above, another source of accuracy loss are the
further approximations needed to make the CGC formalism tractable
via analytic calculations. Whereas the initial condition that we
shall use at low energy --- namely, the McLerran--Venugopalan
model \cite{MV} --- is sufficiently simple to allow for exact
(analytic) calculations, the subsequent evolution with increasing
$y$ is described by complicated non--linear equations which couple
$n$--point correlations with any number of points $n$
\cite{B,W,RGE}. This is similar to the infinite hierarchy of
Schwinger--Dyson equations, and within the CGC theory it can be
summarized into a closed {\sf functional} equation for the
generating functional of the correlations \cite{JKLW97,W,RGE},
also known as the JIMWLK equation. Exact numerical solutions to
this functional equation have recently become available
\cite{RW03}, but these are still difficult to handle, so most
numerical analyses (including that of the ${\mathcal
R}_{pA}$--ratio in Ref. \cite{Nestor03}) have instead focused on a
simpler non--linear equation, the Kovchegov equation \cite{K},
which is a kind of mean field approximation to the general JIMWLK
equation \cite{IM03,MS04}, and is expected to work reasonably well
for a large nucleus at not so high energies (for numerical
analyses of the Kovchegov equation, see
\cite{AB01,Motyka,GBS03,LL01}).

But whereas the numerical calculations are undoubtedly important,
they cannot fully supplant the analytic studies as far as our
physical insight and fundamental understanding are concerned. The
numerical results in Ref. \cite{Nestor03} have revealed very
interesting features --- notably, an extremely fast suppression of
the ratio ${\mathcal R}_{pA}(\kk,y)$ at generic momenta with
increasing $y$, and also the complete disappearance of the Cronin
peak after only a short evolution ---, but these observations have
also raised new questions, since the precise mechanism behind such
a rapid evolution has remained unclear. In particular, it was not
clear whether the two effects alluded to above (the high--$\kk$
suppression and the flattening of the Cronin peak) were caused by
the same mechanism, or not. Also, the specific roles played by the
proton and by the nucleus during evolution have not been
elucidated.

The previous analytic investigations in the literature appeared
before the numerical results in Ref. \cite{Nestor03}, and cannot
fully explain the latter. For instance, while the arguments in
Refs. \cite{KLM02,KKT} do explain the suppression at relatively
large $y$ and for high $\kk$, they cannot explain why this
suppression is so rapid in the early stages of the evolution, or
why the Cronin peak flattens out so fast. Indeed, the
approximations used in these previous studies do not apply for
transverse momenta in, or near, the nuclear saturation region, nor
for very small values of $y$. Besides, running coupling effects
have never been considered in the previous studies, either
analytic or numerical. As we shall see, these effects lead to
important changes, especially for the evolution at late stages.
More generally, what seems to be still missing is a coherent,
qualitative and quantitative, picture explaining what are the
specific features of the MV model which imply the emergence of a
well pronounced peak at low energies, and also what are the
generic features of the quantum evolution which lead to the rapid
flattening of this peak and to a general suppression in the ratio
${\mathcal R}_{pA}$ at all non--asymptotic momenta.

Our subsequent analysis is intended to fill this gap, through a
systematic study of the ratio ${\mathcal R}_{pA}$ in the MV model
and of its evolution with $y$, for both fixed and running
coupling. Being systematic, our analysis has necessarily some
overlap with previous studies in the literature, especially with
Ref. \cite{KKT}, with which we share some results and conclusions,
notably about the high--$\kk$ suppression. In our opinion, what
distinguishes the present analysis from such previous
investigations, in addition to its aim to completeness and the
treatment of the running coupling case, is the constant effort
towards elucidating the fundamental reasons for the observed
behavior, and also the use of a coherent scheme of approximations,
which will be carefully justified, and whose limitations will be
discussed.

In what follows, we shall briefly describe the picture which
emerges from these calculations when increasing $y$ from $y=0$,
and at the same time present the structure of the paper :

{\sf i) Initial conditions at $y=0$ : Cronin peak in the MV model
(cf. Sect. \ref{CRONINMV})}

Although a Cronin peak in the ratio ${\mathcal R}_{pA}(\kk)$ has
been clearly seen in numerical simulations of the MV model
\cite{GJ02,Baier03,KKT,JNV,Nestor03,BGV04}, its analytic study has
been hindered so far by the non--linear aspects of the problem,
which must be treated exactly. In Sect. \ref{CRONINMV}, we shall
present a complete, analytic, study of the nuclear gluon
distribution in this model, which will enable us to compute the
magnitude and the location of the peak, and clarify the conditions
for its existence: {\sf The  emergence of a well pronounced peak
in the MV model reflects the peculiar redistribution of gluons in
transverse phase space, under the influence of the non--linear
effects responsible for saturation.} In turn, this issue has two
aspects:

The {\sf very existence} of a peak can be anticipated on the basis
of very general arguments (like the sum rule introduced in Ref.
\cite{KKT}; see also Refs. \cite{Baier03,Nestor03} and Sect.
\ref{SUM-RULE} below), which reflect two basic properties of the
MV model: (a) the fact that the gluon distribution saturates, at a
value of order $1/\alpha_s$, for low momenta $\kk \le Q_s(A)$, and
(b) the fact that, at large momenta $\kk \gg Q_s(A)$, the nuclear
gluon distribution is simply the incoherent sum of the individual
distributions of the $A$ nucleons (so that ${\mathcal
R}_{pA}(\kk)\to 1$ when $\kk \to \infty$). As we shall see in
Sect. \ref{CroninMV}, these two properties alone imply the
existence of a peak at intermediate momenta. The height of the
peak is of order $\rho_A \equiv \ln Q_s^2(A)/\Lambda^2_{\rm QCD}
\sim \ln A^{1/3}$, and thus is {\sf parametrically enhanced at
large $A$}.

Moreover, {\sf the fact that the peak is so pronounced, and
located in the vicinity of the saturation momentum $Q_s(A)$,} is
the consequence of the fact that the distribution of the saturated
gluons which make up the condensate is {\sf compact}, in the sense
that {\sf it vanishes exponentially at momenta above the
saturation scale}. This feature has not been recognized in
previous studies of the MV model. Of course, on top of this
compact distribution there is also a {\sf power--law tail}, which
represents the sum of the `twist' contributions usually mentioned
when discussing the high--$\kk$ behavior of the distribution. But
as we shall see, for momenta $\kk \simle Q_s(A)$, the sum of the
twist terms is {\sf parametrically suppressed at large $A$} as
compared to the compact distribution which represents the
condensate. Because of that, the overall distribution has a rapid,
exponential, fall--off at momenta just above $Q_s(A)$, leading to
the pronounced peak seen in numerical simulations
\cite{GJ02,Baier03,JNV,Nestor03,BGV04}.

{\sf ii) $y > 0$ : Non--linear quantum evolution in the CGC
 (cf. Sect. \ref{EVOLUTION})}

In Sect. \ref{EVOLUTION} we shall present our approximations to
the solution to the non--linear equations which describe the
evolution of the gluon occupation factor with $y$.

In the saturation region at low momenta, physics is fully
non--linear, but for $\kk\ll Q_s(A,y)$ the solution can be
constructed in a mean field approximation \cite{SAT,GAUSS}. The
result is an universal, and slowly varying, function of $z\equiv
\kk^2/Q_s^2(A,y)$ : $\varphi_A \propto (1/\alpha_s) \ln 1/z$ for
$z\ll 1$ \cite{AM99,LT99,SAT,GAUSS}. The solution in the
transition region around $Q_s(A,y)$ is not known in general (see
however Sect. \ref{FLAT}), but we do know that, along the {\sf
saturation line} $\kk= Q_s(A,y)$, $\varphi_A$ is constant and of
order $1/\alpha_s$.

At high momenta $\kk\gg Q_s(A,y)$, the gluon density is low, and
the general evolution equations reduce to the BFKL equation
\cite{BFKL}, which, although linear, is still sensitive to
saturation effects, through its boundary conditions
\cite{GLR,AM99,SCALING,MT02,DT02,MP03}. The solution to this
equation in the presence of saturation has attracted much
attention in the recent literature
\cite{SCALING,MT02,DT02,MP03,MS04}, where analytic approximations
have been constructed for both fixed and running coupling. Here,
we shall simply summarize the relevant results, and discuss their
range of applicability. Still, a few steps in the construction of
these approximations will be outlined, to clarify some subtle
points and reveal the generic features of the evolution which will
be needed in the subsequent discussion of the ratio ${\mathcal
R}_{pA}$.

We shall quite generally distinguish between two physical regimes,
which are described by different approximations: (I) a ``DLA"
(``double--log accuracy") regime at very large momenta $\kk\gg
Q_g(A,y)$, in which the evolution is dominated by the transverse
phase--space $\rho(A,\kk) \equiv \ln \kk^2/Q_s^2(A)$, and (II) a
``BFKL regime" within the range $Q_s(A,y)\ll \kk\ll Q_g(A,y)$,
where physics is linear but influenced by saturation, and the
gluon distribution preserves the {\sf geometric scaling}
\cite{geometric} property characteristic of saturation: it depends
upon the kinematical variables $\kk$ and $y$ only through the
ratio $z= \kk^2/Q_s^2(A,y)$ \cite{SCALING,MT02}. Here, $Q_g(A,y)$
is the ``geometric scaling momentum", which grows faster than the
saturation momentum with $y$, and marks the upper bound of the
geometric scaling region \cite{SCALING} (see also
\cite{MT02,DT02,MP03,MS04}). With a running coupling, the
discussion of the physical regimes becomes slightly more involved,
and will be presented in Sect. \ref{EV:RUN}.

The approximations described in this section will be sufficient to
study the suppression in the ratio ${\mathcal R}_{pA}(\kk,y)$ with
increasing $y$ at generic momenta. On the other hand, they are not
sufficient to describe the flattening of the Cronin peak (see
below).

{\sf iii) $y > 0$ : The general argument for high--$\kk$
suppression (cf. Sect.  \ref{general})}

In Sect. \ref{general} we shall identify and characterize the
general features of the quantum evolution which are responsible
for the suppression of the ratio ${\mathcal R}_{pA}$  at generic
momenta.

The rapid suppression of the ratio ${\mathcal R}_{pA}(\kk,y)$ with
increasing $y$ is shown to be the consequence of the strong
dissymmetry between the quantum evolution of the nucleus and that
of the proton, which in turn reflects the large separation of
scales between the respective saturation momenta. Physically, it
is so because, for the same values of $y$ and $\kk$, the
transverse phase--space available for the evolution, namely
$\rho(A,\kk) \equiv \ln \kk^2/Q_s^2(A)$, is larger for the proton
than for the nucleus (since $Q_s(A) \gg Q_s(p)$). In more
technical terms, the proton evolves faster because of general
properties of the kernel of the BFKL equation, like the convexity
of its eigenvalue $\chi(\gamma)$, which accelerate the evolution
with increasing $\rho$.

Moreover, the {\sf suppression rate} $d \ln{\mathcal R}_{pA}/dy$
is largest at small $y$ and for not so large transverse momenta
(as compared to the nuclear saturation momentum), since in this
regime the difference between the evolution of the proton and that
of the nucleus is most pronounced. This explains the rapid
suppression observed in the early stages of the evolution in the
numerical results in Ref. \cite{Nestor03}: This is the consequence
of the DGLAP \cite{DGLAP} evolution of the proton (as described
here by DLA), while the nucleus evolves comparatively little
(except at extremely large momenta).

The general properties of the evolution equations can also be used
to show that, for fixed $y$ with $\alpha_s y\simge 1$, the ratio
${\mathcal R}_{pA}(\kk,y)$ is {\sf monotonously increasing with
$\kk$}. Thus, the Cronin peak flattens out during the first
$1/\alpha_s$ units of rapidity.

Another interesting consequence of the evolution is a {\sf
reversal in the $A$--dependence of the ratio ${\mathcal R}_{pA}$
when increasing $y$} (see also Refs. \cite{Baier03,KKT}) : Whereas
at $y=0$, the height of the Cronin peak
--- and, more generally, the ratio ${\mathcal R}_{pA}(\kk)$ at generic momenta
around $Q_s(A)$ --- is (logarithmically) {\sf increasing} with
$A$, this tendency is rapidly reversed by the evolution: after
only a small rapidity increase $\alpha_s y \sim 1/\rho_A$,
${\mathcal R}_{pA}(\kk,y)$ becomes a {\sf decreasing} function of
$A$ for any $\kk$. This property could be related to a similar
change of behavior in the centrality dependence of the ratio
$R_{dAu}$ measured at RHIC
\cite{Brahms-data,RHIC-dAu-forward,RHIC-dAu-forward-other}.

{\sf iv) The flattening of the Cronin peak (cf. Sect.
\ref{CRevolve})}

The evolution of the proton {\sf alone} produces a rather uniform
suppression in the ${\mathcal R}_{pA}$--ratio at all momenta.
Thus, by itself, this cannot wash out any structure present in the
initial conditions, like the Cronin peak. Therefore, the
disappearance of the peak is necessarily related to the evolution
of the nucleus.

In order to study this phenomenon, we shall use the Kovchegov
equation to compute the evolution of the nuclear gluon
distribution in the first step $\Delta y$ in rapidity, with
$\alpha_s \Delta y\ll 1$. This calculation allows us to follow the
evolution of the peak at least up to the rapidity $y_0$, with
$\alpha_s y_0 \sim (\ln^2 \rho_A)/\rho_A \ll 1$, where the height
of the peak becomes of order one. As the calculation shows, the
effect of the evolution is to {\sf generate power law tails which
progressively replace the original exponential tail of the gluon
distribution at saturation}. Because of that, the peak flattens
out, and moves up to larger momenta.

The flattening is related to the difference in the nuclear
evolution at momenta below and, respectively, above the saturation
momentum $Q_s(A,y)$. For $\kk < Q_s(A,y)$, the gluons in the
nucleus are saturated, and evolve only slowly. For $\kk >
Q_s(A,y)$, the nucleus is in the linear, BFKL, regime, and the
corresponding gluon distribution increases much faster (although
not so fast as the proton distribution at the same $\kk$). Because
of this dissymmetry, the Cronin peak gets tilted up, and flattens
out very fast. Although we cannot control analytically this
evolution until the complete disappearance of the peak, we shall
check that the peak has flattened out when $\alpha_s y \sim 1$.

{\sf v) High--$\kk$ suppression: the detailed picture
 (cf. Sects. \ref{CRevolve} and \ref{HIGHPT})}

In Sect. \ref{HIGHPT}, we shall give the detailed picture of the
phenomenon of ``high--$\kk$ suppression'' in the kinematic plane
$y-\ln\kk^2$. (Previously, a similar analysis will be presented in
Sect. \ref{CRevolve}, but only for the evolution along the nuclear
saturation line $\kk= Q_s(A,y)$.) In fact, the suppression will be
seen to occur at all momenta, and not only at those which are
``high'' in the sense of the present analysis (and which are such
that $\kk\gg Q_s(A,y)$). Still, the suppression is stronger for
momenta above the proton saturation scale $Q_s(p,y)$
--- since this is where the proton evolves faster --- but below the nuclear
geometric scale $Q_g(A,y)$, since for even larger momenta, both
the proton and the nucleus are in the DLA regime, and their
respective evolutions almost compensate in the ratio ${\mathcal
R}_{pA}(\kk,y)$. Thus, not surprisingly, the ratio is found to
asymptotically approach one from the below when increasing $\kk$
at fixed $y$. But since $Q_g(A,y)$ increases rapidly with $y$, it
is clear that, for $\alpha_s y \simge 1$, a strong suppression
will be visible at all non--asymptotic momenta.

The explicit analysis in Sect. \ref{HIGHPT} will confirm the
general trend of the evolution anticipated in Sect. \ref{general}:
For generic momenta $\kk$ (such that the nucleus is either at
saturation, or in the BFKL regime), the suppression in the ratio
${\mathcal R}_{pA}(\kk,y)$ is very fast in the early stages of the
evolution, when the proton is in the DLA regime, but with
increasing $y$ it slows down,  and eventually ${\mathcal R}_{pA}$
stabilizes at a very small value, proportional to an inverse power
of $A^{1/3}$. Besides, for fixed $y$ (with $\alpha_s y \simge 1$),
the ratio is monotonously increasing with $\kk$, from a very small
value $\sim 1/A^{1/3}$ for $\kk\sim \Lambda_{\rm QCD}$ to
${\mathcal R}_{pA}\simeq 1$ when $\kk\to \infty$.

\begin{figure}[htb]
  \centerline{
  \epsfsize=0.7\textwidth
  \epsfbox{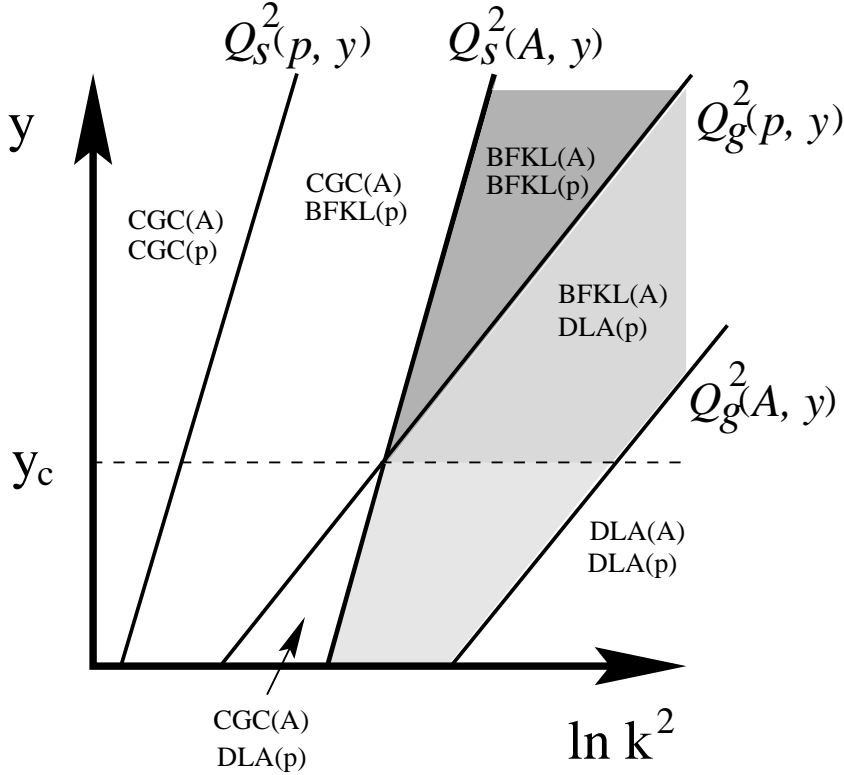}
  }
 \caption[]{{\sl Physical regimes for evolution in the kinematic
plane $y-\ln\kk^2$ (for fixed coupling, for definiteness; see Fig.
\ref{EVOL-MAP-run} for the corresponding picture with a running
coupling).} {\small Both the saturation momentum $Q_s(A)$ and the
`geometric scale' momentum $Q_g(y)$ rise exponentially with $y$
(for both the proton and the nucleus), and thus are represented by
straight lines in the logarithmic scale of the plot. As visible on
this plot, the geometric scale rises faster (its logarithmic slope
is roughly twice as large as that of the saturation momentum).}}
 \label{EVOL-MAP}
\vspace*{0.5cm}
\end{figure}

In order to describe this evolution in more detail, it is
convenient to follow a line which is parallel to the saturation
line in the kinematic plane $y-\ln\kk^2$ (see
Fig.~\ref{EVOL-MAP}). That is, we simultaneously increase $y$ and
$\kk$ in such a way to keep the ratio $z\equiv \kk^2/Q_s^2(A,y)$
fixed. If $z=\mathcal{O}(1)$, then when increasing $y$ from zero,
the proton starts in the DLA regime, because $Q_g(p,y)\ll
Q_s(A,y)$ at sufficiently small $y$. But with increasing $y$,
$Q_g(p,y)$ grows faster than $Q_s(A,y)$, so at some `critical'
value $y_c$ the proton changes from the DLA to the BFKL regime.
The value of $y_c$ is however different for fixed or running
coupling, and so is also the evolution for larger rapidities $y >
y_c$, as we explain now:

(a) With {\sf fixed coupling}, $\alpha_s y_c \sim\rho_A$, and most
of the suppression is achieved already for $y\simle y_c$. For $y
> y_c$, the suppression slows down significantly because the BFKL
evolutions of the proton and of the nucleus almost cancel in the
ratio (\ref{Rdef}) (due to the fact that the respective saturation
momenta evolve in the same way with $y$ \cite{AM03}). Thus, at
large $y$, the ratio ${\mathcal R}_{pA}(z,y)$ decreases only
slowly, and eventually stabilizes, when $\alpha_s y \simge
\rho_A^2$, at a small value ${\mathcal R}_{pA} \sim
1/(A^{1/3}\rho_A)^{1-\gamma}$, which shows a weak dependence on
$z$. Here, $\gamma \approx 0.63$ is the BFKL saddle point (or
`anomalous dimension') near saturation \cite{SCALING,MT02}.

 (b) With {\sf running coupling}, $y_c$ is parametrically larger,
$y_c \propto \rho_A^2$, but the suppression is pursued within a
significant range of $y$ above $y_c$, because in that range the
proton and nuclear saturation momenta evolve differently with $y$
\cite{AM03} : The proton saturation momentum $Q_s(p,y)$ grows
initially faster, and catches up with that of the nucleus when $y
\sim \rho_A^4$.  For larger $y$, $Q_s(A,y) \approx Q_s(p,y)$, and
the ratio stabilizes at the constant value ${\mathcal R}_{pA}
\simeq 1/A^{1/3}$ which is  even smaller than the corresponding
limit for fixed coupling. In fact, this limiting value is simply
the factor of $A^{-1/3}$ introduced by hand in Eq.~(\ref{Rdef}).
This reflects the {\sf universality} of the gluon distribution
which, within a wide range of $\kk$, depends upon the hadron
species only via the saturation momentum.

It should be also stressed that, although the suppression in
${\mathcal R}_{pA}$ is eventually stronger with running coupling,
the evolution leading to this suppression is {\sf slower} in that
case, in the sense that it takes a considerably larger interval of
rapidity before the final limiting value is reached. Besides, all
the intermediate stages of the evolution (like changing from the
DLA to the BFKL regime with increasing $y$ at fixed $z$) require a
longer rapidity evolution with a running coupling than with a
fixed one.

To conclude this introductive discussion, let us emphasize that
all the phenomena discussed so far --- the Cronin peak in the
initial condition, its disappearance with increasing $y$, and the
high--$\kk$ suppression --- are {\sf hallmarks of saturation,
which however reflect different aspects of saturation, and of the
quantum evolution towards saturation} : The Cronin peak reflects
{\sf classical saturation}, i.e., the saturation via
non--linearities in the classical field equations, whereas the
color sources in these equations (the `valence quarks') are
uncorrelated. (Alternatively, this reflects {\sf incoherent},
Glauber--like, multiple scattering.) This is, of course, the
content of the MV model, but it could also be a reasonable
approximation for a large nucleus at not so high energies, where
the effects of quantum evolution are negligible. Furthermore, the
driving force towards both high--$\kk$ suppression and the
flattening of the Cronin peak is the linear evolution (either
DGLAP, or BFKL). But notwithstanding this, these phenomena are
still a signal of saturation, since they occur only because of the
{\sf mismatch between various regimes of evolution, which is made
possible by saturation} : For the flattening of the peak, this is
the mismatch between the nuclear evolution below and above the
saturation scale, whereas for the high--$\kk$ suppression, this is
the mismatch between the (linear) evolutions of the proton and of
the nucleus, which originates in the difference between the
respective saturation momenta.

\section{Cronin effect in the initial conditions}
\setcounter{equation}{0} \label{CRONINMV}
\subsection{Generalities}

By the ``unintegrated gluon distribution'' we shall more precisely
understand in what follows the {\sf gluon occupation factor},
i.e., the number of gluons of given spin and color per unit
phase--space in a nucleus with atomic number $A$:
\be\label{phidef}
\varphi_A(\kk,y)\,\equiv\,\frac{(2\pi)^3}{2(N_c^2-1)}\,
\,\frac{{\rm d} N_A}{\rmd y {\rm d}^2k_\perp{\rm
d}^2b_\perp}\,.\ee Here, $y$ denotes the rapidity (related to the
longitudinal momentum fraction $x$ of the gluons via $y=\ln
(1/x)$), $k_\perp$ is the transverse momentum, and $b_\perp$ is
the gluon position in transverse space. For simplicity, we shall
consider a hadron which is homogeneous in the transverse plane,
and we shall suppress the $b_\perp$--dependence in all the
formulae. Within light--cone quantized QCD, the gluon occupation
factor (\ref{phidef}) is related to a gauge--invariant 2--point
function of the color fields, which can be computed within the CGC
effective theory \cite{RGE,CGCreviews}. More precisely, the
evolution of this quantity with $y$ can be computed within
perturbative QCD (to the accuracy of the effective theory), but
the initial conditions at $y=0$ remain non--perturbative, and
require a model. It turns out that the choice of this model has a
strong influence on the physical problems that we would like to
address, especially as far as the existence of a Cronin peak
\cite{CroninExp} is concerned.

Physically, the initial conditions that we are interested in
correspond to a large nucleus ($A\gg 1$) in a regime of
intermediate energies. The initial energy should be high enough
for the gluons with the smallest values of $x$ to be coherent with
each other, but low enough for the quantum effects in the gluon
distribution to be negligible. These conditions require $x_0
\simle 1/A^{1/3}$, but at the same time $\alpha_s\ln 1/x_0\ll 1$,
where $x_0$ denotes the longitudinal fraction of the slowest
gluons within the nucleus at the initial energy. For instance, for
a gold nucleus and with $\alpha_s \approx 0.2$, these requirements
can be satisfied by choosing $x_0 \sim 10^{-1}$. (From now on, the
rapidity variable will be understood to represent the difference
from this original rapidity: $y=\ln (x_0/x)$.) Under these
conditions, the gluon distribution in the nucleus is simply the
result of classical radiation from the valence quarks. But if $A$
is large enough --- which is what we assume here ---, the
resulting gluon density can be still very high, and thus favor
non--linear effects which lead to {\sf gluon saturation} at
sufficiently low $\kk$,
below some characteristic 
scale $Q_s(A)$.

A simple model which encompasses these physical conditions is the
McLerran--Venugopalan (MV) model \cite{MV,CGCreviews}, in which
the total color charge in the nucleus is the incoherent sum of the
individual color charges of the valence quarks, and the
corresponding  saturation momentum scales like $Q_s(A) \sim
A^{1/6}$ (since $Q_s^2(A)$ is proportional to the color charge
squared per unit transverse area). If $A$ is large enough, this
scale is hard ($Q_s^2(A)\gg\Lambda^2_{\rm QCD}$), and the matter
made of the gluons (the CGC) is weakly coupled.

In what follows we shall adopt the MV model as our initial
condition at $y=0$. The remaining part of this section is devoted
to a study of the gluon distribution in this model, with emphasis
on the Cronin peak, and, more generally, on the role of
non--linear effects in rearranging the gluons in momentum space.
Although this model has been extensively studied in the literature
\cite{MV,K96,JKMW97,KM98,CGCreviews} (in particular, in relation
with the Cronin peak \cite{GJ02,Baier03,JNV,KKT,Nestor03}), our
analysis below will bring some new results and conceptual
clarifications, and will reveal some novel, and perhaps
surprising, features, which to our knowledge have escaped to
previous investigations.

Specifically, in Sect. \ref{MVmodel}, we shall provide a complete,
analytic, study of the gluon distribution in the MV model, that we
shall generalize on this occasion to include running coupling
effects, in such a way to be compatible with the quantum evolution
to be discussed later. The results obtained in this analysis will
permit us to clarify the conditions for the emergence of the peak,
and explicitly compute properties like the location of the peak
and its magnitude (in Sect. \ref{CroninMV}). Then, in Sect.
\ref{SUM-RULE}, we shall reexamine a global argument in favor of
the Cronin peak, based on a sum--rule \cite{KKT}, that we shall
use in order to better understand the origin of the gluons which
make up the condensate.

A rather surprising feature which will emerge from this analysis
refers to the redistribution of gluons in momentum space under the
influence of the non--linear effects : Whereas it has been since
long appreciated that the effect of the repulsive interactions is
to push the gluons towards the modes at larger momenta, and thus
provide a spectrum which is infrared--safe
\cite{JKMW97,KM98,CGCreviews}, it has not been recognized so far
that the dominant part of this spectrum --- the one which is
parametrically enhanced for large $A$, and provides a plateau of
order $1/\alpha_s$ at saturation ($\kk \simle Q_s(A)$)
--- is actually {\sf compact}, i.e., it falls off exponentially
with increasing $\kk$ above $Q_s(A)$. In fact, as we shall see, it
is precisely this exponential decrease in $\varphi_A(\kk)$ for
momenta just above $Q_s(A)$ which explains why a pronounced peak
appears in the ratio ${\mathcal R}_{pA}$ when computed in the MV
model.

To conclude this general discussion, let us mention two
alternative definitions for the unintegrated gluon distribution
which are used in the literature. These definitions have in common
the fact that they relate the gluon distribution to a scattering
operator: the scattering amplitude $\mathcal{N}_A(r_\perp,y)$ for
a color dipole with transverse size $r_\perp$ which scatters of a
hadronic target with atomic number $A$ at relative rapidity $y$.
Under suitable approximations, this quantity obeys a closed,
non--linear, evolution equation (a generalization of the BFKL
equation), originally derived by Kovchegov \cite{K}.

Specifically, in Ref. \cite{SCALING}, the following definition has
been proposed (the color dipole is taken to be made off two
gluons):
 \be\label{phiN} \varphi_A(k_\perp,y)\,\equiv\,
    \int \frac{d^2 r_\perp}{\pi r^2_\perp}\,
   {\rm e}^{-i k_\perp \cdot r_\perp}\,
    \frac{\mathcal{N}_A(r_\perp,y)}{\alpha_s N_c}\,.
    \ee
This definition has no deep motivation, but is simply based on the
analogy with a formula for the true occupation factor
(\ref{phidef}) which holds within the MV model (see Sect.
\ref{MVmodel} below). In fact, the quantities in Eqs.~(\ref{phiN})
and (\ref{phidef}) are very similar to each other: They coincide
with each other in the linear regime at high momenta $\kk\gg
Q_s(A,y)$, where they both obey the BFKL equation, and they also
show a similar behavior at low momenta $\kk\ll Q_s(A,y)$, so they
differ, at most, in the transition region towards saturation. The
approximations that we shall develop in this paper are not
sensitive to the details of this transition region, nor to the
differences \cite{IM03,MS04} between the Kovchegov equation and
the general, functional, evolution equation for the CGC, so all
the results that we shall obtain apply literally to both
definitions. Thus, our results are directly comparable to the
numerical calculations in Ref. \cite{Nestor03}, which are based on
Eq.~(\ref{phiN}) together with the Kovchegov equation.

A different definition, with a deeper physical motivation, has
been introduced in Ref. \cite{Braun}, and reads
 \be\label{hdef}
h_A(k_\perp,y)\,\equiv\,\frac{1}{4}\,\kk^2\grad^2_k
 \varphi_A(k_\perp,y)\,=\,\kk^2
    \int \frac{d^2 r_\perp}{4\pi}\,
   {\rm e}^{-i k_\perp \cdot r_\perp}\,
    \frac{\mathcal{S}_A(r_\perp,y)}{\alpha_s N_c}\,,
    \ee where in this context $\varphi_A(k_\perp,y)$ is the
function defined in Eq.~(\ref{phiN}), and
$\mathcal{S}_A(r_\perp,y) \equiv 1 -\mathcal{N}_A(r_\perp,y)$ is
the $S$--matrix element for dipole--hadron scattering. The
quantity (\ref{hdef}) enters a factorized formula for the
cross--section for gluon production in proton--nucleus collisions
\cite{KM98,KT02,DM01,BGV04}, and as such it is directly relevant
for the phenomenology of d--Au collisions at RHIC. Note that its
interpretation as a `gluon distribution' is only conventional
(this is based on an analogy with the $\kk$--factorization which
holds in the linear regime at not so high energies), and should
not give rise to confusion: There is a priori no reason why the
(canonical) gluon distribution should enter the calculation of
observables for high energy scattering. Indeed, in the
high--density environment at high energy, the scattering operators
are non--linear in the color field in the target (the non--linear
effects describe {\sf multiple scattering}), so, unlike what
happens at low energy, they are not simply proportional to the
2--point function which defines the gluon distribution.

The results that we shall obtain for $\varphi_A$ later in this
paper do not directly apply to $h_A$, although they could be
easily translated for it (at least, in the case of a fixed
coupling), by using $h_A \propto k^2\grad^2_k \varphi_A$. The
numerical analysis in Ref. \cite{Nestor03}, which considers both
definitions (\ref{phiN}) and (\ref{hdef}), shows that the ratio
${\mathcal R}_{pA}$  is qualitatively the same when computed with
any of these two definitions. But important quantitative
differences persist, especially at low momenta, where the
functions $\varphi_A$ and $h_A$ are very different. In view of
this, it would be interesting to repeat for $h_A$, and also for
the convolution yielding the cross--section for gluon production,
the analysis that we shall give below in this paper for
$\varphi_A$. A potential difficulty that we foresee with such an
analysis is the fact that the generalization of Eq.~(\ref{hdef})
to the case of a running coupling is ambiguous (unlike for the
other definitions, Eqs.~(\ref{phidef}) and (\ref{phiN}), for which
a natural generalization exists; see below).

\subsection{The McLerran--Venugopalan model: Fixed \& running coupling}
\label{MVmodel}

The color sources which generate the small--$x$ gluons in the
McLerran--Venugopalan (MV) model \cite{MV,CGCreviews} are the $3A$
valence quarks (from the $A$ nucleons), which are assumed to be
uncorrelated with each other except for the long--range
correlations associated with confinement. The gluon occupation
factor $\varphi_A(k_\perp)\equiv \varphi_A(k_\perp,y=0)$ at
momenta $\kk\gg \Lambda_{\rm QCD}$ is then obtained as
\cite{JKMW97,KM98}  \be\label{phiMV} \varphi_A(\kk)\,=\, \!\int \!
d^2r_\perp \,{\rm e}^{-ik_\perp\cdot r_\perp}\,\, \frac{1-{\rm
exp}\Big\{-\frac{1}{4}\, r_\perp^2 Q_A^2 \ln{4\over
r_\perp^2\Lambda^2}\Big\}}{\pi \alpha_s N_cr_\perp^2 }\,,\,\,\ee
where $\Lambda$ is a non--perturbative scale of order
$\Lambda_{\rm QCD}$ (the only trace of confinement), $N_c=3$ is
the number of colors, and $Q^2_A=\alpha_s N_c \mu_A\propto
A^{1/3}$ is proportional to the color charge squared $\mu_A$ of
the valence quarks  per unit transverse area. The factor of 4
within $\ln(4/r_\perp^2\Lambda^2)$ is only a matter of choice.
The integration in Eq.~(\ref{phiMV}) must be restricted to
$r_\perp < 2/\Lambda$, for consistency with the approximations
leading to this formula \cite{CGCreviews}, and also to avoid that
the logarithm in the exponent changes sign. But as long as $\kk\gg
\Lambda$, the value of the integral is very little sensitive to
the precise value of the upper cutoff.

Since the MV model is a classical approximation, Eq.~(\ref{phiMV})
is a priori written for a fixed coupling $\alpha_s$. Later, we
shall consider quantum evolution with a running coupling, and at
that stage we shall also need a generalization of
Eq.~(\ref{phiMV}) which includes running coupling effects to
one--loop order (i.e., for a running coupling $\alpha_s(Q^2)\equiv
b_0/\ln(Q^2/\Lambda_{\rm QCD}^2)$). The generalization that we
shall consider reads (from now on, we shall not distinguish
between the scales $\Lambda$ and $\Lambda_{QCD}$)
 \be\label{phiMVrun} \varphi_A(\kk)\,=\, \!\int \! d^2r_\perp
\,{\rm e}^{-ik_\perp\cdot r_\perp}\,\, \frac{1-{\rm
e}^{-\frac{1}{4}r_\perp^2 Q_A^2}} {\pi b_0 N_cr_\perp^2
}\,\ln{4\over r_\perp^2\Lambda^2}\,\,,\ee where $Q^2_A\equiv b_0
N_c \mu_A$ has now a slightly different meaning as compared to the
fixed coupling case, but is still proportional to $A^{1/3}$.
Formally, Eq.~(\ref{phiMVrun}) is obtained from Eq.~(\ref{phiMV})
by replacing in the latter $\alpha_s\rightarrow
\alpha_s(4/r_\perp^2)$ within the denominator of the integrand,
and also within $Q^2_A$. In contrast to Eq.~(\ref{phiMV}), there
is no need for an upper cutoff in Eq.~(\ref{phiMVrun}): the
integral is convergent as written, and for any $\kk\gg \Lambda$ it
is dominated by perturbative sizes $r_\perp \ll 1/\Lambda$.

Since the choice of a running in the absence of a complete
one--loop quantum calculation is a priori ambiguous (especially in
the presence of several scales, like in  Eq.~(\ref{phiMV})), it is
important to justify in more detail our choice leading to
Eq.~(\ref{phiMVrun}). Note first that (a) the coupling $\alpha_s$
which is explicit in Eq.~(\ref{phiMV}) (in the denominator) and
the one which is implicit in the definition of $Q^2_A$
have the same origin \cite{JKMW97,CGCreviews}, and thus need to be
renormalized in the same way, and (b) the quantity  $1-{\rm
exp}\Big\{-\frac{1}{4}\, r_\perp^2 Q_A^2 \ln{4\over
r_\perp^2\Lambda^2}\Big\}$ in the numerator of Eq.~(\ref{phiMV})
can be recognized as the scattering amplitude for a color dipole
of size $r_\perp$ which scatters of the nucleus. More precisely,
this is the dipole made of the two primary gluons radiated by the
valence quarks (the gluons which define the distribution), and the
scattering amplitude describes the color precession of these
primary gluons when propagating through the color field of the
nucleus (see \cite{CGCreviews} for details). As well known, a
small dipole couples predominantly to gluons with momenta
$\kk^2\simle 1/ r_\perp^2$, so it is indeed natural to choose the
size of the dipole as the argument for the running of the
corresponding coupling constant; this is also the choice made in
other studies of the evolution of the dipole scattering amplitude
with increasing energy \cite{Motyka,MT02,DT02,LL01,RW03}.

There is one more subtle point about our choice in
Eq.~(\ref{phiMVrun}) : as a density of color charge, the quantity
$\mu_A$ involves itself a factor of $\alpha_s$, that we have
implicitly treated as a constant in our arguments above.
Specifically, for $A\times N_c$ valence quarks which are
homogeneously distributed within the nuclear disk with radius
$R_A$, one obtains \cite{CGCreviews} \be\label{muA} \mu_A
\,=\,\frac{2\alpha_s A}{R_A^2}\,.\ee Treating $\alpha_s$ in
Eq.~(\ref{muA}) as a constant amounts to neglecting quantum
corrections to the distribution of the valence quarks, which is in
the spirit of the MV model. One sees that, in writing
Eq.~(\ref{phiMVrun}), we have treated differently the vertices
describing the radiation of classical color fields from the
valence quarks, and those describing the scattering of the color
dipole off these classical fields. This is justified for the
present purposes because these vertices play different roles in
the subsequent evolution with increasing $y$: Whereas the valence
quarks act simply as sources for the primary gluons at $y=0$, and
as such they can be viewed as classical colored particles, the
quantum evolution proceeds via gluon radiation from the primary
gluons, and the vertices describing this radiation are of the same
type as those describing the scattering of the color dipole in
Eq.~(\ref{phiMV}). Thus dressing {\sf just} these vertices is the
minimal way to render Eq.~(\ref{phiMV}) consistent with the
quantum evolution which includes running coupling effects.

The emergence of the dipole scattering amplitude in
Eqs.~(\ref{phiMV}) and (\ref{phiMVrun}) illustrates the deep
connection between {\sf unitarization} effects in scattering
processes at high energy and {\sf saturation} effects in the
nuclear gluon distribution: Both types of effects arise from
non--linearities associated with strong gluon fields, which here
are encoded in the exponential terms\footnote{Physically, these
exponentials represent the $S$--matrix for dipole--nucleus
scattering.} in the integrands of Eqs.~(\ref{phiMV}) and
(\ref{phiMVrun}). The transition from the linear regime to the
non--linear one takes 
takes place when the exponent is of order one. More precisely, we
shall define the {\sf saturation momentum} $Q_s(A)$ such that the
exponent is equal to one when $r=2/Q_s(A)$. For running coupling,
Eq.~(\ref{phiMVrun}) immediately implies $Q_s^2(A)= Q_A^2\,\sim
A^{1/3}$. For fixed coupling, the saturation scale turns out to be
larger than the scale $Q_A$ introduced by the color sources
 \be\label{QsatMV}
Q_s^2(A)\,=\,Q_A^2\,\ln\frac{Q_s^2(A)}{\Lambda^2} \,\,\sim\,
A^{1/3}\,\ln A^{1/3}\,,\ee and this difference will be seen to
have important consequences. Loosely speaking, the ratio
 \be\label{rhoA}
\rho_A\,\equiv \,\frac{Q_s^2(A)}{Q_A^2}\,=\,
\ln\frac{Q_s^2(A)}{\Lambda^2}\,\,\sim\,\ln A^{1/3}\,,\ee will play
the same role in the fixed coupling case as the inverse of the
coupling constant in the case of a running coupling. In fact, for
a running coupling, we shall define similarly $\rho_A \equiv \ln
({Q_A^2}/{\Lambda^2})$, and then it is clear that $1/\rho_A$ is
essentially the same as the coupling evaluated at the saturation
momentum : $\alpha_s(Q_A^2) = b_0/\rho_A$. Since the MV model
makes sense only for weak coupling, it is justified to treat
$\rho_A$ as a large parameter, $\rho_A\gg 1$, which we shall do in
what follows. It is interesting to note in this context that, for
a gold nucleus at typical RHIC energies one expects
$Q_s^2(A)\simeq 2$ GeV$^2$, which together with $\Lambda \simeq
200$ MeV, yields $\rho_A\simeq \ln 50 \simeq 4$.

It is straightforward to compute the dominant behavior of the
gluon distribution at asymptotically large ($k_{\perp} \gg
Q_s(A)$) or small ($k_{\perp} \ll Q_s(A)$, with $\kk\gg
\Lambda$ though) momenta. One finds : 

\noindent {\sf (a) Fixed coupling:} \be\label{asympfix}
\varphi_A(\kk)\,\approx\, \left\{ \begin{array} {c@{\quad\rm
for\quad}l}
 \frac{1}{\alpha_s N_c}\,\frac{Q_A^2}{k_{\perp}^2}\,, & k_{\perp} \gg Q_s(A) \\
\frac{1}{\alpha_s N_c}\,\ln \frac{Q_s^2(A)}{k_{\perp}^2}\,, &
k_{\perp} \ll Q_s(A)
\end{array}
\right. \ee {\sf (a) Running coupling:} \be\label{asymprun}
\varphi_A(\kk)\,\approx\, \left\{ \begin{array} {c@{\quad\rm
for\quad}l}
 \frac{1}{b_0 N_c}\,\frac{Q_A^2}{k_{\perp}^2}\,, & k_{\perp} \gg Q_A \\
\frac{1}{N_c}\left\langle \frac{1}{\alpha_s}\right\rangle \ln
\frac{Q^2_A}{k_{\perp}^2}\,, & k_{\perp} \ll Q_A
\end{array}
\right. \ee The average inverse coupling constant which enters the
last equation is defined as  \be\label{avalpha} \left\langle
\frac{1}{\alpha_s}\right\rangle \,\equiv\,\frac{1}{2b_0}\left( \ln
\frac{Q^2_A}{\Lambda^2} +\ln
\frac{\kk^2}{\Lambda^2}\right)\,=\,\frac{1}{b_0}\,\ln \frac{\kk
Q_A}{\Lambda^2}\,.\ee

The high--momentum behavior in these equations, which is obtained
after linearizing the exponential terms Eqs.~(\ref{phiMV}) and
(\ref{phiMVrun}), is recognized as the bremsstrahlung spectrum due
to radiation from independent color sources. This would be the
spectrum at any $\kk$ in the absence of non--linear effects in the
dynamics of the radiated gluons. But already at high--$\kk$, this
spectrum receives non--linear corrections which are suppressed by
powers of ${Q^2_s(A)}/{k_{\perp}^2}$ (`higher--twist' effects; see
below). The dominant behavior at low momenta is obtained after
neglecting the exponential terms Eqs.~(\ref{phiMV}) and
(\ref{phiMVrun}), and is correct to leading logarithmic accuracy;
that is, the first corrective term would be a constant term under
the logarithm.

As we shall see in Sect. \ref{CroninMV}, the asymptotic
expressions above are already sufficient to demonstrate the
existence of a peak  in the ratio ${\mathcal R}_{pA}$. But in
order to study the properties of this peak, we also need the gluon
occupation factor at generic (intermediate) momenta, which given
the simplicity of the MV model can be computed analytically.

In fact, in the case of a running coupling, we have been able to
evaluate the integral in Eq.~(\ref{phiMVrun}) in analytic form,
with a result which is displayed in the Appendix, and which will
be also discussed below. For fixed coupling, we do not have such
an exact result, but we have found a convenient series
representation of the integral in Eq.~(\ref{phiMV}), which
converges very fast, and renders the physical interpretation of
the result transparent.

Let us consider the fixed coupling case first. After rewriting the
exponent in the integrand of Eq.~(\ref{phiMV}) as (cf.
Eq.~(\ref{QsatMV}) and (\ref{rhoA}))
    \be \frac{1}{4}\, r_\perp^2
Q_A^2 \ln{4\over r_\perp^2\Lambda^2}\,=\,t\left( 1 + \frac{\ln
1/t}{\rho_A}\right),\qquad t \equiv \frac{r_\perp^2
Q_s^2(A)}{4}\,,\ee one can decompose the integral into two pieces
as
    \be\label{phiMVdec} \varphi_A(\kk)&=& \int d^2r_\perp \,{\rm
e}^{-ik_\perp\cdot r_\perp}\,\left\{\frac{1-{\rm
e}^{-\frac{1}{4}\, r_\perp^2 Q_s^2(A)}}{\pi \alpha_s N_c r_\perp^2
}\, + \, \frac{{\rm e}^{-\frac{1}{4}\, r_\perp^2 Q_s^2(A)} }{\pi
\alpha_s N_cr_\perp^2 } \left[1 - {\rm exp}\left(- t \,\frac{\ln
1/t}{\rho_A}\right)\right]\right\} \nn &=& \varphi_A^{\rm
    sat}(\kk) + \varphi_A^{\rm twist}(\kk)\,,\ee
 where the first,
`saturating', piece can be explicitly evaluated, and reads
\be\label{phisat} \varphi_A^{\rm sat}(\kk) \,=\, \frac{1}{\alpha_s
N_c} \,\Gamma(0,z)\,,\ee while the second piece --- the sum of
all--`twist' contributions (see below) --- can be evaluated as a
series expansion, obtained by expanding the exponential within the
square brackets in powers of $(t \ln 1/t)/\rho_A$ as
\be\label{phiMVexp} \varphi_A^{\rm twist}(\kk)\,=\,-
\frac{1}{\alpha_s N_c\rho_A}\int  dt\, J_0(\sqrt{4zt}\,)\,{\rm
e}^{-t}\sum_{n=1}^{n_{\rm max}}
\left(\frac{t}{\rho_A}\right)^{n-1}\frac{\ln^n t}{n!}\,.\ee In
these equations, $z\equiv \kk^2/Q_s^2(A)$, $\Gamma(0,z)$ is the
incomplete Gamma function: \be \Gamma(0,z) =
\int_z^\infty\,\frac{{\rm e}^{-t}}{t}\,dt\,,\ee $n_{\rm max}\sim
{\rm e}^{\rho_A} = {Q_s^2(A)}/{\Lambda^2}\gg 1$, and the
truncation of the series in Eq.~(\ref{phiMVexp}) at $n\le n_{\rm
max}$ reflects the upper cutoff $r_{\rm max}=2/\Lambda$ which is
implicit in the integral over $r_\perp$ in the first line of
Eq.~(\ref{phiMVdec}). (The corresponding integral giving
$\varphi_A^{\rm sat}$, Eq.~(\ref{phisat}), has been extended up to
infinity, since rapidly convergent.) Without a truncation, the
series would be asymptotically divergent, but the divergent
behavior would start to manifest itself only at very large orders,
beyond $n_{\rm max}$. On the other hand, when $\rho_A\gg 1$ the
truncated series is rapidly convergent for any $z$, and therefore
$\varphi_A(\kk)$ can be computed even analytically with high
accuracy (see the Appendix for details).

The first important observation about the decomposition in
Eq.~(\ref{phiMVdec}) is that the second piece, $\varphi_A^{\rm
twist}$, is multiplied by an overall factor $1/\rho_A$ (this is
manifest on Eq.~(\ref{phiMVexp})), and thus is parametrically
suppressed at large $A$. This implies that, in the saturation
region at $z\simle 1$, the gluon distribution is dominated by the
first piece, $\varphi_A^{\rm sat}$. In particular, $\varphi_A^{\rm
sat}$ captures the leading behavior at low momenta $\kk\ll
Q_s(A)$, cf. Eq.~(\ref{asympfix}): indeed, $\Gamma(0,z)\approx \ln
1/z$ for $z\ll 1$, whereas $\varphi_A^{\rm twist}$ is analytic
near $z = 0$. In fact, if one remembers that $1/\rho_A$ can be
effectively identified with $\alpha_s$ (cf. the discussion after
Eq.~(\ref{rhoA})), it becomes clear that the
$1/\alpha_s$--enhancement of the gluon distribution, which is the
hallmark of saturation, is associated solely with $\varphi_A^{\rm
sat}$ : For $z\simle 1$, the latter provides a plateau with a
height of $\mathcal{O}(1/\alpha_s)$, whereas $\varphi_A^{\rm
twist}$ brings only a small correction, of $\mathcal{O}(1)$.

The second important observation is that the saturating piece
$\varphi_A^{\rm sat}$ is {\sf compact}, in the sense that it
vanishes exponentially at momenta outside the saturation region
(since $\Gamma(0,z)\approx {\rm e}^{-z}/z$ for $z\gg 1$). By
contrast, it can be checked that, for $z\gg 1$, the function
$\varphi_A^{\rm twist}$ can be expanded in powers of $1/z$ (up to
logarithmic corrections), thus generating the `twist expansion' of
the MV model. It should be stressed that each of the terms in the
truncated series in Eq.~(\ref{phiMVexp}) contains terms to all
orders in the twist expansion: the first term ($n=1$) includes the
bremsstrahlung spectrum $\propto 1/z$, together with an infinite
series of `higher--twist' terms (i.e., terms of order $1/z^2$, or
higher), the second term ($n=2$) starts at order $1/z^2$, etc. The
first few terms in the twist expansion are exhibited in the
Appendix.

We see that the gluon distribution of a large nucleus in the MV
model naturally decouples into two pieces, one which controls the
spectrum at saturation, and another one which controls the tail of
the distribution at high momenta.  It is natural to interpret the
first piece, $\varphi_A^{\rm sat}$, as the {\sf occupation factor
in the color glass condensate}. The fact that this distribution
appears to be compact is probably just specific to the MV model,
since, as we shall see, this feature is washed out by the quantum
evolution (cf. Sect. \ref{FLAT}). The previous considerations also
suggest that the twist terms contribute only little to the physics
of the Cronin peak in the MV model. This will be confirmed by the
analysis in the next subsection.

The corresponding discussion in the case of a running coupling is
even simpler, since the corresponding `twist' piece involves only
one term, instead of a series of terms. Also, some arguments
become more transparent since, with a running coupling,
$1/\alpha_s(Q_s^2(A))$ and $\rho_A/b_0$ are explicitly identified.
Specifically, Eq.~(\ref{phiMVrun})
can be decomposed as: \be\label{phiMVrunexp} \varphi_A(\kk)&=&
\varphi_A^{\rm sat}(\kk) + \varphi_A^{\rm twist}(\kk)\,,\nn
\varphi_A^{\rm sat}(z)&=& \frac{1}{b_0 N_c}\,\rho_A
\,\Gamma(0,z)\,,\qquad \varphi_A^{\rm twist}(z)=\int  dt
J_0(\sqrt{4zt}\,)\,\frac{1-{\rm e}^{-t}}{b_0 N_c t}
\ln\frac{1}{t}\,,\ee where the saturating piece is explicitly
enhanced by the factor $\rho_A$ (that is, it is of
$\mathcal{O}(1/\alpha_s)$), while $\varphi_A^{\rm twist}$ is of
$\mathcal{O}(1)$ for $z\sim 1$, and  can be expanded in powers of
$1/z$ for $z\gg 1$ (thus generating the twist expansion). The only
noticeable difference with respect to the fixed coupling case is
that, now, $\varphi_A^{\rm twist}(z)$ is not analytic near $z\to
0$, but rather yields a large {\sf negative} contribution
$-(1/2b_0 N_c)\ln^2z$ in that limit, which contributes to the
dominant low--momentum behavior exhibited in Eq.~(\ref{asymprun}).
Since, however, the overall contribution remains {\sf positive}
for any $\kk \ge \Lambda$, it is clear that $\varphi_A^{\rm
sat}(\kk)$ is still the dominant contribution at any $\Lambda \le
\kk \le Q_A$. The integral giving $\varphi_A^{\rm twist}$ is
explicitly evaluated in the Appendix, where its twist expansion
will be also considered.

\begin{figure}[htb]
\begin{center}
\includegraphics[scale=1.33]{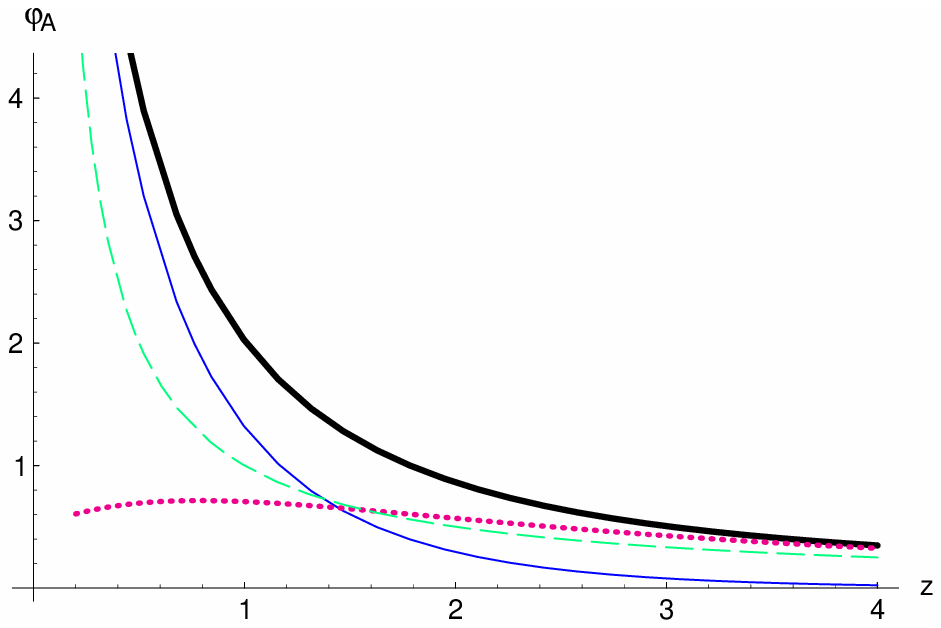}
\includegraphics[scale=1.33]{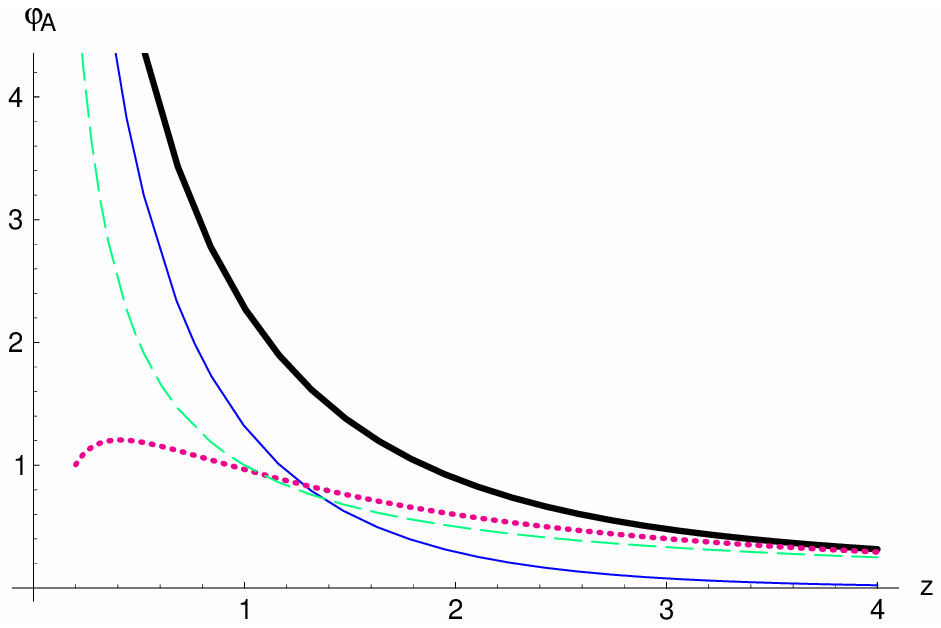}
    \caption{\label{phi} {\sl The gluon occupation factor
    $\varphi_A(z)$ as a function of the scaled momentum variable
    $z=k^2/Q_s^2(A)$ in the MV model with either fixed (figure
    above) or running (below) coupling and $\rho_A=6$.} {\small
    With respect to the text definitions, we plot the quantities
    $\rho_A\alpha_s N_c\times\varphi_A$ for fixed coupling, and
    $b_0 N_c\times\varphi_A$ for running coupling. The black
    (thick) line corresponds to $\varphi_A(z)$; the blue (solid)
    line shows the saturation contribution $\varphi^{\rm
    sat}_A(z)$; the magenta (dotted) line shows the twist
    contribution $\varphi^{\rm twist}_A(z)$; the green (dashed)
    line represents the bremsstrahlung spectrum.}}
\end{center}
\end{figure}

\begin{figure}[]
\begin{center}
\includegraphics[scale=1.33]{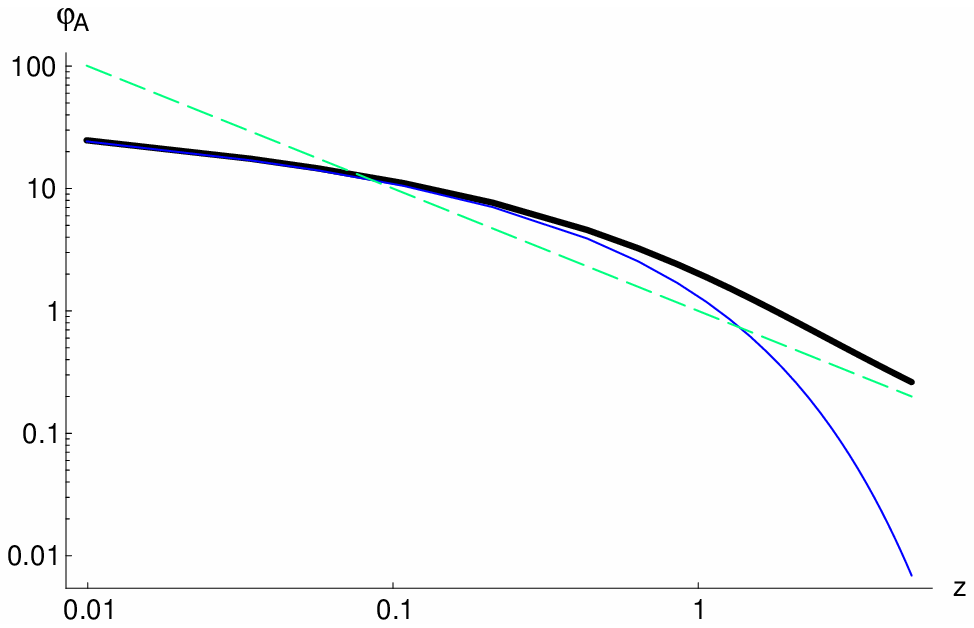}
\vspace*{0.5cm}
\includegraphics[scale=1.33]{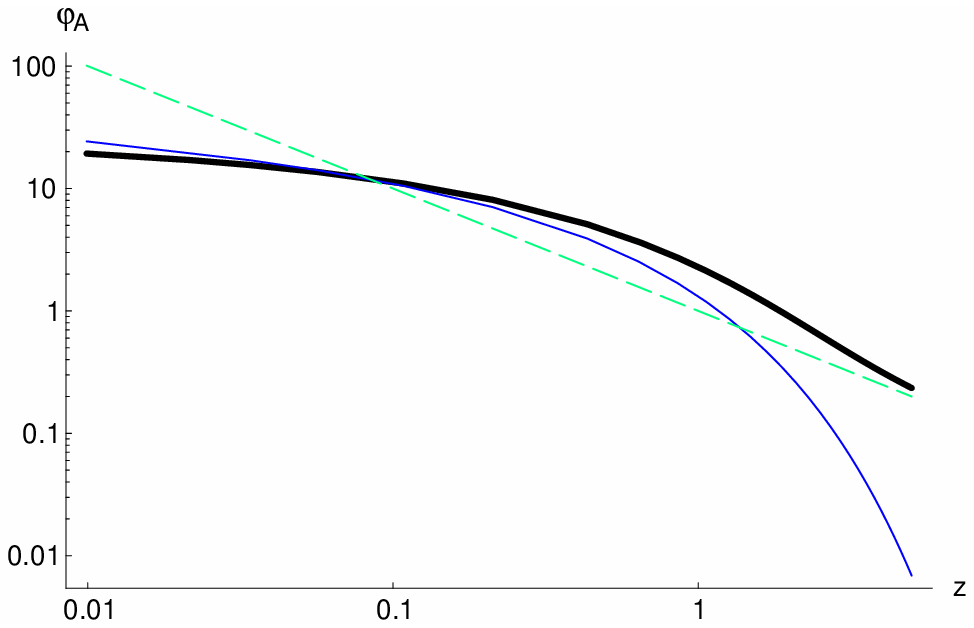}
    \caption{\label{logphi} {\sl Logarithmic plot of the gluon
    occupation factor $\varphi_A(z)$ as a function of the scaled
    momentum variable $z=k^2/Q_s^2(A)$ in the MV model with either
    fixed (figure above) or running (below) coupling and
    $\rho_A=6$.} {\small The plotted quantities are rescaled as
    explained in the caption to Fig. \ref{phi}. The black (thick)
    line corresponds to the gluon occupation factor
    $\varphi_A(z)$; the blue (solid) line shows the saturation
    contribution $\varphi^{\rm sat}_A(z)$; the green (dashed) line
    represents the bremsstrahlung spectrum.}}
\end{center}
\end{figure}

The previous considerations are illustrated in Figs. \ref{phi} and
\ref{logphi}, which exhibit the gluon occupation factor
$\varphi_A(\kk)$, and its various contributions $\varphi_A^{\rm
sat}(\kk)$ and $\varphi_A^{\rm twist}(\kk)$, in the MV model with
fixed and running coupling, in either linear scale (Fig.
\ref{phi}), or log--log scale in a wider range of momenta (Fig.
\ref{logphi}).

\subsection{Cronin effect in the McLerran--Venugopalan model}
\label{CroninMV}

To form the ratio ${\mathcal R}_{pA}$, Eq.~(\ref{Rdef}), one also
needs the gluon distribution in the proton, $\varphi_p(\kk,y)$.
For $y=0$ and $\kk\gg \Lambda$, the proton is in the perturbative
(i.e., linear) region, and will be described by  the
bremsstrahlung spectrum in Eqs.~(\ref{asympfix}) or
(\ref{asymprun}) (for fixed and running coupling, respectively)
with $Q_A^2 \rightarrow Q_p^2$, and $Q_p = {\mathcal O}
(\Lambda)$. For instance, \be\label{phip0}
\varphi_p(\kk)\,=\,\frac{1}{\alpha_s
N_c}\,\frac{Q_p^2}{k_{\perp}^2}\,\Theta (\kk - Q_p)\,\qquad{\rm
(fixed\,\,coupling)}.\ee We shall assume the following relation
between the two scales: \be\label{A13} Q_A^2\,=\,
A^{1/3}\,Q_p^2,\ee which holds formally if one extrapolates the MV
model (a priori valid for a large nucleus) down to $A=1$.

Note that the quantity in the denominator of Eq.~(\ref{Rdef}),
namely $A^{1/3}\varphi_p(\kk)$, coincides with the nuclear gluon
distribution in the high momentum limit (i.e., the bremsstrahlung
spectrum in Eqs.~(\ref{asympfix}) or (\ref{asymprun})). Thus,
within the MV model at least, the ratio ${\mathcal R}_{pA}(\kk)$
is also a measure of the deviation of the actual nuclear gluon
distribution from the corresponding prediction of linear
perturbation theory. As already discussed, this deviation is
associated with non--linear effects in the gluon dynamics, in
particular with saturation. We immediately conclude that
${\mathcal R}_{pA}(\kk)$ must approach one at high $\kk$:
 \be\label{RMVasymp} {\mathcal R}_{pA}(\kk)\simeq 1\qquad{\rm
for}\qquad \kk\gg Q_s(A).\ee By using this condition together with
very general properties of saturation\footnote{Of course, all the
properties to be discussed below are manifest on the expressions
(\ref{phisat}) and (\ref{phiMVrunexp}) for $\varphi_A^{\rm
sat}(\kk)$, but here we would like to construct our argument
without relying on the explicit formulae for $\varphi_A$ that we
have obtained previously, in order to emphasize the general
conditions required by the existence of the peak.}, we can deduce
the existence of a Cronin peak without any detailed calculation:

Consider the fixed coupling case, for definiteness. The basic
consequence of saturation is that, for momenta $\kk\simle Q_s(A)$,
the nuclear occupation factor develops a plateau with the height
of $\mathcal{O}(1/\alpha_s)$; that is, $\varphi_A(\kk)\approx
(1/{\alpha_s N_c})\bar\varphi(z)$, where $z\equiv \kk^2/
Q_s^2(A)$, and the function $\bar\varphi(z)$ is slowly varying
when $z < 1$, and is of $\mathcal{O}(1)$ when $z\sim 1$. This
function depends upon $\kk$ and $A$ only through $z$ because it is
dimensionless, and $Q_s(A)$ is the only scale in the problem other
than $\kk$ (since at saturation the spectrum cannot be sensitive
to the non--perturbative scale $\Lambda$). We deduce that:
\be\label{CRarg} {\mathcal R}_{pA}(\kk) \,\approx
\,\frac{\kk^2}{Q_A^2}\,\bar\varphi(z)\,= \,
\rho_A\,z\,\bar\varphi(z)\,,\qquad{\rm for}\qquad \kk\simle
Q_s(A).\ee Since $\bar\varphi(z)$ is only logarithmically
divergent as $z\to 0$, it is clear that the ratio ${\mathcal
R}_{pA}(\kk)$ is much smaller than one for $\kk$ small enough
(e.g., it is of $\mathcal{O}(A^{-1/3})$ when $\kk\sim Q_p$), but
it is  of $\mathcal{O}(\rho_A)$ --- and thus strictly larger than
one --- for $\kk = Q_s(A)$. This behavior at small $\kk$, together
with the asymptotic behavior (\ref{RMVasymp}) at large $\kk$,
immediately imply that the ratio ${\mathcal R}_{pA}$ must have a
maximum at some intermediate value of $\kk$.

This general argument does not tell us where is the maximum
actually located, and not even whether there is a single maximum,
or several. But it is highly probable, and will be confirmed by
the explicit calculations below, that there is indeed only one
maximum, which is located near $Q_s(A)$ and thus has a height of
$\mathcal{O}(\rho_A)$ : Indeed, there is no intrinsic scale larger
than $Q_s(A)$ in the problem, and for $\kk < Q_s(A)$ the ratio is
still increasing with $\kk$, as manifest on Eq.~(\ref{CRarg}).

Note the specific way how the large factor $\rho_A$ has entered
the calculation in Eq.~(\ref{CRarg}) : this is due to the mismatch
(\ref{QsatMV}) between the scale $Q_A^2$ associated with the color
sources and the saturation scale $Q_s^2(A)$ generated by the
non--linear gluon dynamics. It is easy to check that a similar
enhancement occurs also for a running coupling, although in that
case the argument is more direct: the factor $\rho_A$ is
introduced by the inverse coupling in the saturation condition
$\varphi_A(\kk= Q_A)\approx \kappa/{\alpha_s(Q_A^2) N_c}$, with
$\kappa \sim \mathcal{O}(1)$.

We conclude that the existence of the Cronin peak is a consequence
of the fact that the gluon distribution at saturation is larger,
by a factor of $\rho_A\sim \ln A^{1/3}$, than the na\"{\i}ve
extrapolation of the bremsstrahlung spectrum down to $\kk\sim
Q_s(A)$. This logarithmic enhancement is due to non--linear
effects which cause the gluon occupation factor to increase faster
with $1/\kk^2$ than predicted by linear perturbation theory,
before eventually saturating to a value of order $1/\alpha_s$ at
$\kk \simle Q_s(A)$.

We now rely on the explicit formulae for $\varphi_A$ established
in the previous subsection to give a complete description of the
peak. For more clarity, we start with the large--$A$ limit, in
which the gluon occupation factor for momenta around $Q_s(A)$ is
given by the saturating piece $\varphi_A^{\rm sat}$, up to
corrections of $\mathcal{O}(1/\rho_A)$. Using $\varphi_A^{\rm
sat}(\kk)$ from either Eq.~(\ref{phisat}) or
Eq.~(\ref{phiMVrunexp}), one finds the same expression for the
ratio ${\mathcal R}_{pA}$ with both fixed and running coupling,
namely: \be\label{RAfixed} {\mathcal R}_{pA}(\kk)\,\approx
\,z\,\Gamma(0,z)\,\rho_A\,\qquad{\rm for}\qquad \kk\sim Q_s(A)
\,.\ee This is of the form anticipated in Eq.~(\ref{CRarg}), and
thus is of order $\rho_A$ when $z\sim 1$ ($\Gamma(0,1)=
0.219...$). It can be checked that the function $f(z)\equiv
z\,\Gamma(0,z)$ has a maximum at $z_0\approx 0.435$, with the peak
value $f(z_0)\approx 0.281$. Moreover, this peak is very
pronounced, since $f(z)$ is almost linearly increasing below the
peak, but exponentially decreasing above it.

It is then straightforward to add the contribution of the twist
terms in Eq.~(\ref{phiMVexp}) or (\ref{phiMVrunexp}), and compute
their effect on the location and the magnitude of the peak in an
expansion in powers of  $1/\rho_A$. One finds (with ${\mathcal
R}_{\rm max}(A)\equiv {\mathcal R}_{pA}(z_0)$) :
\be\label{RmaxMVf} z_0&=&
0.435\,+\,\frac{0.882}{\rho_A}\,+\,\frac{2.122}{\rho_A^2}\,+\,{\mathcal
O}\big(\rho_A^{-3}\big),\nn {\mathcal R}_{\rm
max}(A)&=&0.281\,\rho_A
\,+\,0.300\,+\,\frac{0.294}{\rho_A}\,+\,{\mathcal
O}\big(\rho_A^{-2}\big), \ee in fixed coupling case, and
respectively \be\label{RmaxMVr} z_0&=&
0.435\,+\,\frac{1.382}{\rho_A}\,+\,\frac{2.038}{\rho_A^2}\,+\,{\mathcal
O}\big(\rho_A^{-3}\big),\nn {\mathcal R}_{\rm
max}(A)&=&0.281\,\rho_A
\,+\,0.524\,+\,\frac{0.804}{\rho_A}\,+\,{\mathcal
O}\big(\rho_A^{-2}\big), \ee in the running coupling case.

Finally, for both fixed and running coupling one can compare the
results obtained above in the $1/\rho_A$ expansion to the
corresponding exact results, and the agreement is good even for
$\rho_A=4$ (which, we recall, is a realistic value at RHIC). For
example, with running coupling and $\rho_A=4$, the exact
calculation yields $z_0=0.89$ and ${\mathcal R}_{{\rm max}}=1.85$,
while Eq.~(\ref{RmaxMVr}) predicts $z_0=0.91$ and ${\mathcal
R}_{{\rm max}}=1.85$.

\begin{figure}[]
\begin{center}
\includegraphics[scale=1.33]{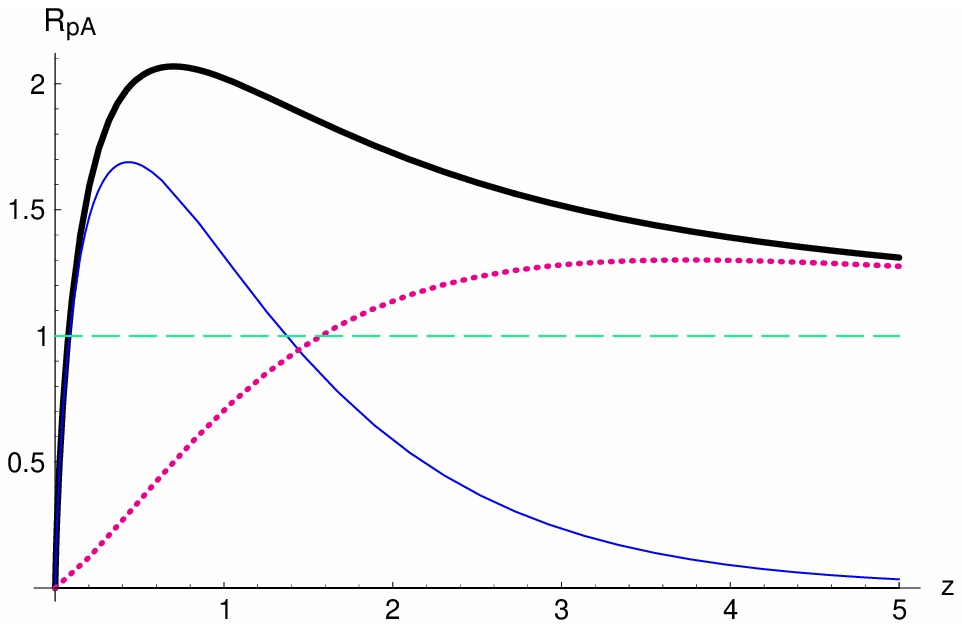}
\includegraphics[scale=1.33]{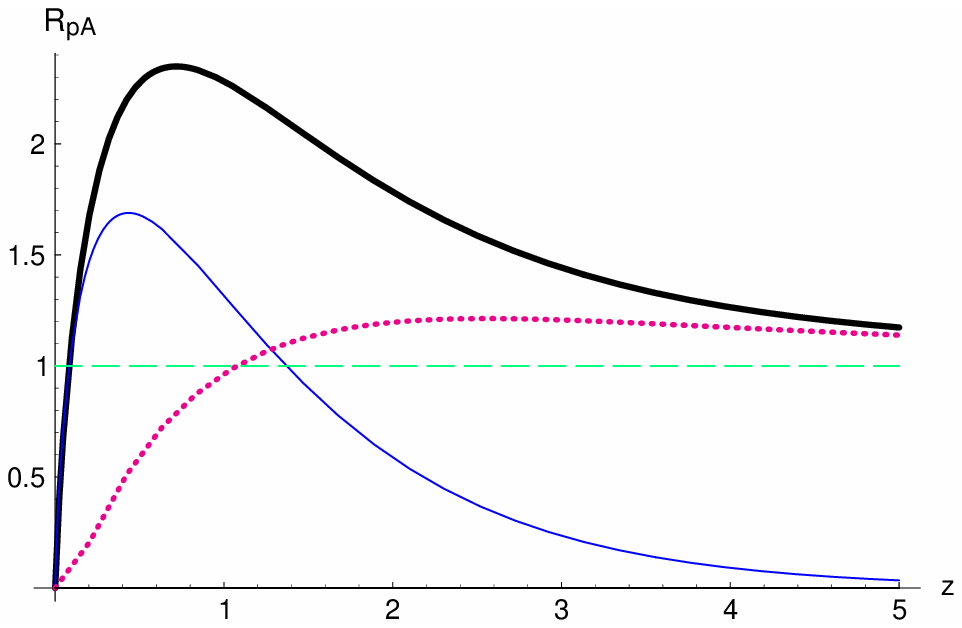}
    \caption{\label{RpA} {\sl The Cronin ratio
    $\mathcal{R}_{pA}(z)$ as a function of the scaled momentum
    variable $z=k^2/Q_s^2(A)$ in the fixed (above) and running
    (below) coupling McLerran-Venugopalan model for $\rho_A=6$.}
    {\small The black (thick) line corresponds the ratio
    $\mathcal{R}_{pA}(z)$; the blue (solid) line shows the
    saturation contribution $\mathcal{R}^{\rm sat}_{pA}(z)$; the
    magenta (dotted) line shows the twist contribution
    $\mathcal{R}^{\rm twist}_{pA}(z)$.}}\vspace*{.5cm}
\end{center}
\end{figure}

In Fig. \ref{RpA}, where we show the Cronin ratio in the MV model
with both fixed and running coupling, we also represent the
individual contributions separated in the r.h.s. of
Eq.~(\ref{phiMVdec}) (or (\ref{phiMVrunexp})), to better emphasize
that the `twist' effects give only a small contribution in the
region of the peak. As obvious on this figure, the effect of the
twist corrections is to flatten the peak a little bit, and also to
move it towards larger momenta. Still, a well pronounced peak
emerges because, even for $\rho_A$ as small as 4, the dominant
contribution to $\varphi_A(\kk)$ for $\kk$ around $Q_s(A)$ is
still given by the compact, `saturating', piece $\varphi_A^{\rm
sat}$. Accordingly, the nuclear occupation factor shows an
exponential fall--off for momenta just above $Q_s(A)$. It can be
easily checked that the behavior changes from an exponential to a
power law around $z\sim \ln\rho_A$.

\subsection{A sum rule  and its consequences}
\label{SUM-RULE}

In this subsection we shall discuss an alternative, global,
argument in favor of the existence of the Cronin peak in the MV
model, which sheds more light on the role of non--linear effects
in the nuclear gluon distribution. This argument, due to Kharzeev,
Kovchegov, and Tuchin \cite{KKT}, is specific to the MV model ---
it is based on a sum--rule which reflects the basic property of
this model that color sources are uncorrelated ---, and cannot be
extended to $y>0$ (since the quantum evolution introduces
correlations among the color sources; see below). But even within
the MV model, the use of this argument is quite subtle, because of
possible complications with ultraviolet divergences. In fact, as
we shall explain below, the  original formulation of the relevant
sum--rule in Ref. \cite{KKT} is not mathematically rigorous, which
may have given rise to confusion (especially, in relation with the
generalization of this sum--rule to $y>0$; see the discussion in
Ref. \cite{Nestor03}). In what follows, we shall properly restate
the argument, and then present a stronger version of it, based on
partial sum--rules, which provides a more detailed information on
the distribution of gluon at saturation.

Specifically, the argument in Ref. \cite{KKT} is based on the
following `sum--rule' (the IR cutoff $\Lambda$ is strictly needed
only for the proton) :
 \be\label{sumright}  \int_{\Lambda^2}\,\frac{d^2\kk}{\pi}\,
 \Big\{\varphi_A(\kk) -
A^{1/3}\varphi_p(\kk)\Big\}\,=\,0\,,\ee which holds indeed within
the MV model, for both fixed and running coupling, as we shall
demonstrate shortly. From this relation, one can infer the
existence of a Cronin peak through the following reasoning: Since
at low momenta $\kk\ll Q_s(A)$ we have $\varphi_A(\kk) <
A^{1/3}\varphi_p(\kk)$ due to gluon saturation in the nucleus,
whereas at large momenta $\kk\gg Q_s(A)$ the two functions
approach the same limit, one concludes that $\varphi_A(\kk)$ must
be larger than $A^{1/3}\varphi_p(\kk)$ at some intermediate
momenta, in order to give rise to the same integrated
distribution. One reason for the debate around this argument in
the literature \cite{Nestor03} is that in Ref. \cite{KKT} the
sum--rule has been abusively written as: \be\label{sumwrong}
\int_{\Lambda^2}\,\frac{d^2\kk}{\pi}\,
 \varphi_A(\kk) \,=\,
A^{1/3}\int_{\Lambda^2}\,\frac{d^2\kk}{\pi}\, \varphi_p(\kk)\,.\ee
At a first sight, this might look identical to
Eq.~(\ref{sumright}), but in reality Eq.~(\ref{sumwrong}) is
meaningless, since the expressions on both sides are
logarithmically divergent in the ultraviolet. If, moreover, one
introduces an upper cutoff $Q^2$ to eliminate these divergences,
then the ensuing, finite, expressions do not coincide with each
other for any finite $Q^2$ (see below). The would--be divergent
pieces (i.e., the contributions proportional to $\ln Q^2$) are
indeed the same for both integrals --- because $\varphi_A(\kk)
\simeq
 A^{1/3}\varphi_p(\kk)$ for large $\kk\gg Q_s(A)$
---, but by itself this property only tells us that the integral
in the left hand side of Eq.~(\ref{sumright}) is well defined, but
not also that the value of this integral is  actually zero. In
other terms, the equality of the divergent pieces in
Eq.~(\ref{sumwrong}) carries no information about the behavior of
the integrands at intermediate momenta, and thus cannot guarantee
the existence of the Cronin peak \cite{Nestor03}.

Still, as we shall prove below, the difference ${\mathcal
G}_A(Q^2)\,-\,A^{1/3}{\mathcal G}_p(Q^2)$ vanishes like $1/Q^2$
when $Q^2\to\infty$, so the  sum--rule (\ref{sumright}) holds
indeed as written. We have defined here: \be\label{xGdef}
{\mathcal G}_A(Q^2)\,\equiv\,\int^{Q^2}_{\Lambda^2}
 \frac{d^2\kk}{\pi}\,\varphi_A(\kk)\,,\ee (together with a similar
expression for the proton) which, up to a global factor $\pi
R_A^2\times (N_c^2-1)/(2\pi)^2$ (cf. Eq.~(\ref{phidef})), is the
integrated gluon distribution in the MV model, i.e., the total
number of gluons localized within a transverse area $1/Q^2$.

 Our proof of Eq.~(\ref{sumright}) is very similar to the original
one in Ref. \cite{KKT}, namely it uses the Fourier representation
of the integrand (cf. Eqs.~(\ref{phiMV})--(\ref{phiMVrun})) to
easily perform the integral over $\kk$, and thus produce a
$\delta$--function $\delta^{(2)}(r_\perp)$. Then, the sum--rule
immediately follows because the integrand vanishes when
$r_\perp\to 0$. For instance, in the case of a running coupling,
Eq.~(\ref{phiMVrun}), we write (we omit unnecessary constant
factors) : \be\label{sumproof} \int\,\frac{d^2\kk}{\pi}\int \!
d^2r_\perp \,{\rm e}^{-ik_\perp\cdot r_\perp}\,\left\{
\frac{1-{\rm e}^{-\frac{1}{4}\,r_\perp^2 Q_A^2}} {r_\perp^2 }-
\frac{Q_A^2}{4} \right\} \,\ln{4\over r_\perp^2\Lambda^2}\nn
\propto\,\lim_{r_\perp\to 0}\left\{ \frac{1-{\rm
e}^{-\frac{1}{4}\,r_\perp^2 Q_A^2}} {r_\perp^2 }- \frac{Q_A^2}{4}
\right\} \,\ln{4\over r_\perp^2\Lambda^2} \,=\,0,\ee which
vanishes as $r_\perp^2\ln r_\perp^2$ when $r_\perp\to 0$.

\begin{figure}[]
\begin{center}
\includegraphics[scale=1.33]{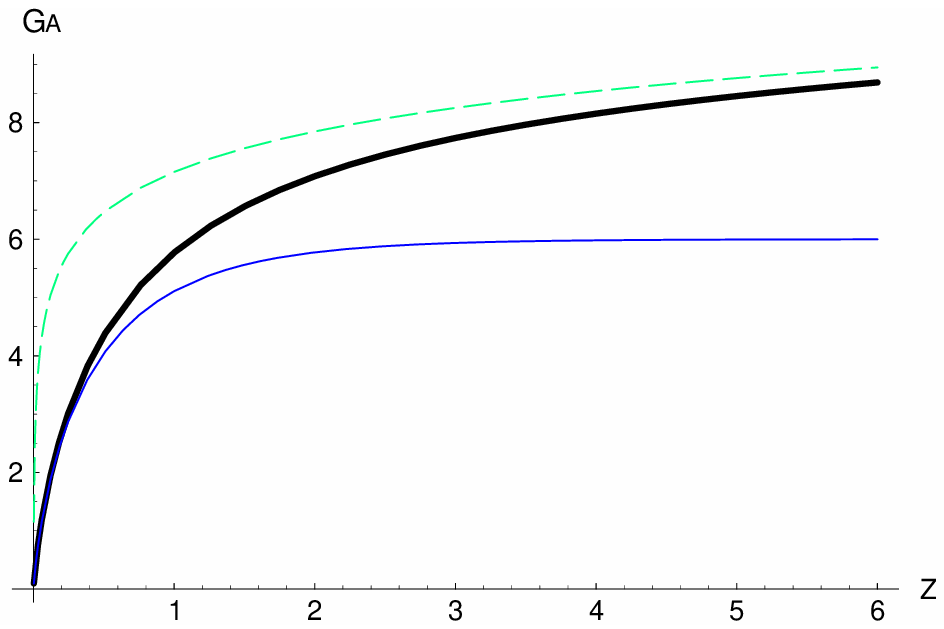}
\includegraphics[scale=1.33]{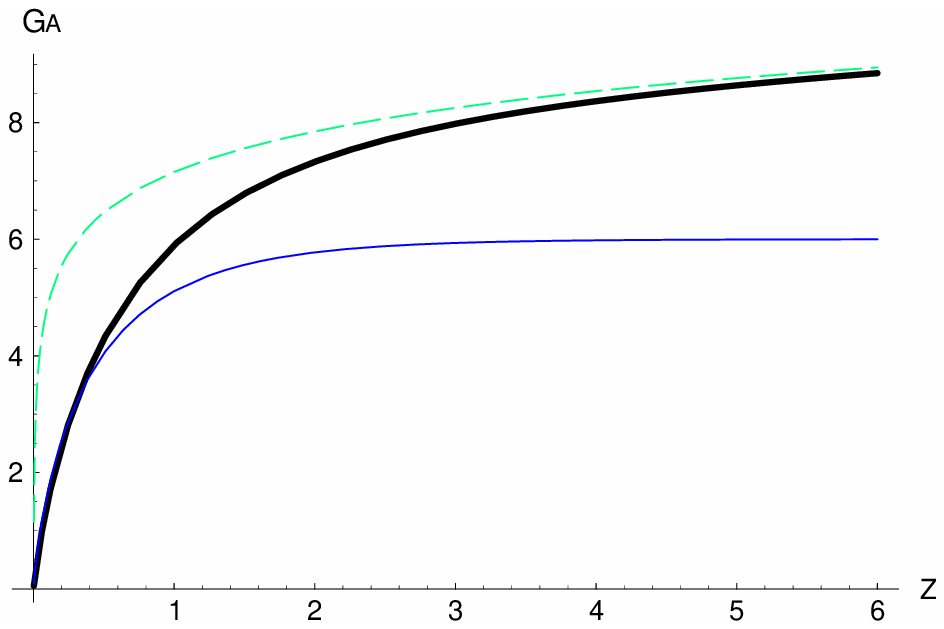}
    \caption{\label{intphi} {\sl The integrated gluon distribution
    function $\mathcal{G}_A(Z)$ as a function of the scaled
    momentum variable $Z=Q^2/Q_s^2(A)$ in fixed (above) and
    running (below) coupling McLerran-Venugopalan model for
    $\rho_A=6$.} {\small With respect to the text definitions, we
    plot the quantities $[\rho_A\alpha_s N_c/Q_s^2(A)]\times
    \mathcal{G}_A$ for fixed coupling, and $[b_0
    N_c/Q_s^2(A)]\times\mathcal{G}_A$ for running coupling. The
    black (thick) line corresponds to the integrated gluon
    distribution function $\mathcal{G}_A(Z)$; the blue (solid)
    line shows the saturation contribution $\mathcal{G}^{\rm
    sat}_A(Z)$; the green (dashed) line represents the
    bremsstrahlung distribution function.}}
\end{center}
\end{figure}

If the above integral over $\kk$ is restricted to some finite, but
large $Q^2$, with $Q^2\gg Q_A^2$, then instead of the
$\delta$--function $\delta^{(2)}(r_\perp)$ one generates a smooth
distribution in $r_\perp$ which is peaked at $r_\perp= 0$, with
the peak height proportional to $Q^2$, and the width of the peak
of order $1/Q^2$. When integrated over such a distribution, the
`higher--twist' terms in Eq.~(\ref{sumproof}) (i.e., the higher
order terms in the expansion of the exponential) give
contributions which vanish as powers of $1/Q^2$ when $Q^2
\to\infty$. The dominant higher--twist contribution, of order
$1/Q^2$, turns out to be {\sf negative} (this is explicitly
computed in the Appendix). Thus, for a finite, but large, $Q^2$,
the total number of gluons with transverse area $1/Q^2$ is {\sf
smaller} in the MV spectrum of a nucleus than in the corresponding
bremsstrahlung spectrum (i.e., than it would be in the absence of
non--linear effects). Specifically, one can use the results in the
Appendix to deduce that
 \be\label{xGA-P} {\mathcal
G}_{A}(Q^2)\,-\,{\mathcal G}_{BS}(Q^2)\,\simeq\,
 - \,\frac{1}{b_0 N_c}\,\frac{Q_s^4(A)}{2 Q^2} \qquad {\rm
for}\qquad Q^2 \gg Q_s^2(A)\,,\ee with ${\mathcal
G}_{BS}(Q^2)\equiv A^{1/3}{\mathcal G}_p(Q^2)$. This is
illustrated in Fig. \ref{intphi}.

The sum--rule (\ref{sumright}) can be understood as the
consequence of two basic properties, out of which one is generic
--- the fact that the non--linear effects fall off as inverse powers
of $\kk^2$  in the high--momentum limit ---, and the other one is
specific to the MV model: the fact that the color sources (the
valence quarks) are {\sf uncorrelated}. Specifically, the second
property implies that, at high momenta $\kk\gg Q_s(A)$, the gluon
distributions produced by the $3A$ valence quarks sum up
incoherently, so that the leading--twist terms cancel exactly in
the difference $\varphi_A(\kk) - A^{1/3}\varphi_p(\kk)$. Moreover,
by the first property, the remaining terms in this difference
involve higher powers of $1/\kk^2$, which explains
Eq.~(\ref{xGA-P}), and thus the sum--rule (\ref{sumright}).

The above considerations also help us to understand why the sum
rule is bound to fail after taking the quantum evolution with $y$
into account: At $y > 0$, the color sources are predominantly
gluons which are themselves products of radiation, and thus are
correlated with each other (unlike the valence quarks) even in the
leading--twist approximation. Because of that, the gluon
occupation factors in the nucleus and in the proton are not simply
proportional to each other, not even at high $\kk$. This can be
easily verified on Eqs.~(\ref{DLAp0}) and (\ref{DLAprun}), which
give the dominant behavior at large momenta for $y>0$. For
instance, for fixed coupling, Eq.~(\ref{DLAp0}) implies (we omit
the trivial factor $1/\alpha_s N_c$) : \be\label{diffDLA}
\varphi_A(\kk,y) - A^{1/3}\varphi_p(\kk,y)&\approx&
\,\frac{Q_A^2}{k_{\perp}^2}\,\left\{ {\rm e}^{\sqrt{4\bar\alpha_s
y \rho(A,\kk)}} -
{\rm e}^{\sqrt{4\bar\alpha_s y \rho(p,\kk)}}\right\}\\
&\approx&
 - \rho_A\,\frac{Q_A^2}{k_{\perp}^2}\, \sqrt{\frac{\bar\alpha_s
y}{\rho(A,\kk)}}\, {\rm e}^{\sqrt{4\bar\alpha_s y \rho(A,\kk)}}
\qquad{\rm for}\quad \kk\gg Q_s(A,y)\,,\nonumber\ee where $
\rho(A,\kk)\equiv\ln \kk^2/Q_s^2(A)$, $\rho(p,\kk) \equiv\ln
\kk^2/Q_p^2 = \rho(A,\kk) + \rho_A$, and in writing the second
line we have kept only the leading term at large $\rho(A,\kk)$.
Although the expression above does vanish as $\kk\to\infty$, it is
clear that it is not integrable in the sense of
Eq.~(\ref{sumright}). The very fact that the sum--rule fails to
apply at $y > 0$ does not preclude the existence of the Cronin
peak: The sum--rule is only a {\sf sufficient} condition for the
existence of the peak, but not also a {\sf necessary} one. And
indeed the peak does persist (although its height is rapidly
decreasing) in the early stages of the evolution, as we shall show
in Sect. \ref{CRevolve}. This is also seen in the numerical
results in Ref. \cite{Nestor03}.

Returning to the MV model, where the sum--rule  (\ref{sumright})
{\sf is} valid, it is instructive to understand its physical
implications in more detail: Eq.~(\ref{sumright}) tells us that
the net result of the non--linear effects in the gluon dynamics is
merely a {\sf redistribution} of the gluons in the transverse
momentum space. As it should be clear from the previous arguments,
and also from the plot in  Fig. \ref{logphi}, this redistribution
is associated with repulsive interactions which push the gluons
towards the modes with larger momenta. In Fig. \ref{logphi} one
can see that, whereas at low momenta $\kk < Q_c(A)$ the gluon
occupation factor is larger for the bremsstrahlung spectrum than
for the MV spectrum, at higher momenta $\kk > Q_c(A)$ the opposite
situation occurs. Here, $Q_c(A)$ is defined by\footnote{The
subsequent formulae in this subsection are written for the case of
a fixed coupling, for definiteness. It can be easily checked that
analog formulae, leading to similar conclusions, apply also for
the MV model with a running coupling.} (see Eq.~(\ref{asympfix})):
\be \frac{Q_A^2}{k_{\perp}^2}\,\sim \,\ln
\frac{Q_s^2(A)}{k_{\perp}^2}\qquad {\rm for}\qquad \kk\sim
Q_c(A),\ee
 which implies :
 \be\label{Qcdef}
 Q_c^2(A) \,\simeq\,\frac{Q_A^2}{\ln\frac{Q_s^2(A)}{Q_c^2(A)}}
\,\simeq\,\frac{Q_A^2}{\ln\frac{Q_s^2(A)}{Q_A^2}}
\,\simeq\,\frac{Q_s^2(A)}{\rho_A\ln \rho_A}\,,\ee where we have
also used $Q_s^2(A) = Q_A^2 \rho_A$, cf. Eq.~(\ref{QsatMV}).
It is easily checked that $\Lambda^2\ll Q_c^2(A)\ll Q_s^2(A)$
(since $\rho_A\gg 1$).

Thus, the effect of the interactions is to move the gluons up in
transverse momentum, from below $Q_c(A)$ to above it. This
explains, in particular, why when counting the gluons up to a
finite (large) $Q^2$, one necessarily finds a smaller number for
the MV spectrum than for the bremsstrahlung one, cf.
Eq.~(\ref{xGA-P}). But as we shall explain now, at large $\kk\gg
Q_s(A)$ the effects of this redistribution are only tiny: {\sf
Most of the gluons which are in excess in the bremsstrahlung
spectrum at $\kk < Q_c(A)$ are found in  the MV spectrum in the
saturation region at $\kk\simle Q_s(A)$.} In particular, the
enhancement factor $1/\alpha_s \sim \rho_A$ characteristic of
saturation finds its origin  in the phase--space for
bremsstrahlung radiation at momenta $\Lambda < \kk < Q_c(A)$. This
factor is large because of the infrared sensitivity of the
bremsstrahlung spectrum.

A simple way to see this is to notice that the total number of
gluons in the region  $\Lambda < \kk < Q_s(A)$ is essentially the
same for  the bremsstrahlung spectrum and for the MV spectrum.
Indeed, we have first :
 \be\label{GBSsat} {\mathcal G}_{BS}(Q^2_s(A))\,=\,
\int^{Q^2_s(A)}_{\Lambda^2}\,\frac{d^2\kk}{\pi}\,
\frac{1}{\alpha_s N_c}\, \frac{Q_A^2}{\kk^2}
 \,=\,\frac{Q_A^2}{\alpha_s N_c}\,\ln\frac{Q^2_s(A)}{\Lambda^2}
\,=\,\frac{Q_s^2(A)}{\alpha_s N_c}\,,\ee where the saturation
momentum $Q_s(A)$ has been reconstructed as $Q_s^2(A) = Q_A^2
\rho_A$, cf. Eq.~(\ref{QsatMV}). Note that the large factor
$\rho_A= \ln {Q^2_s(A)}/{\Lambda^2}$ has been generated from the
phase--space for the bremsstrahlung gluons at $\Lambda < \kk <
Q_s(A)$. For the MV spectrum, we shall define ${\mathcal
G}_{A}(Q^2_s(A))$ as the total number of gluons contained in the
saturating component of the spectrum $\varphi^{\rm sat}_A(\kk)$.
This is appropriate since $\varphi_A(\kk)\approx \varphi^{\rm
sat}_A(\kk)$ for any $\kk < Q_s(A)$, whereas at larger momenta
$\varphi^{\rm sat}_A(\kk)$ falls off exponentially. One then
obtains:
 \be\label{GMVsat} {\mathcal G}_{A}(Q^2_s(A))\,=\,
\int\,\frac{d^2\kk}{\pi}\, \frac{1}{\alpha_s N_c}\,\Gamma\big(0,
\kk^2/Q^2_s(A)\big)\,=\,\frac{Q_s^2(A)}{\alpha_s N_c}\,.\ee (The
simplest way to obtain this result is to notice that the
unrestricted integral of $\varphi^{\rm sat}_A(\kk)$ is tantamount
to letting $r_\perp\to 0$ in the integral representation of
$\varphi^{\rm sat}_A(\kk)$ in  Eq.~(\ref{phiMVdec}).) As
anticipated, the final expressions in Eqs.~(\ref{GBSsat}) and
(\ref{GMVsat}) are identical. But, of course, the way how these
gluons are distributed in $\kk$ is very different in the two
cases: For the bremsstrahlung spectrum, which is infrared
divergent, most of these gluons are located at low momenta $\kk <
Q_c(A)$, while for the MV spectrum they are rather uniformly
distributed at all momenta up to $Q_s(A)$. For instance
 (see Eq.~(\ref{Qcdef})):
\be\label{GBSI} \hspace*{-0.6cm} {\mathcal G}_{BS}(Q^2_c(A))\,=\,
\int^{Q^2_c(A)}_{\Lambda^2}\,\frac{d^2\kk}{\pi}\,
\frac{1}{\alpha_s N_c}\, \frac{Q_A^2}{\kk^2}
 \,=\,\frac{Q_A^2}{\alpha_s N_c}\,\ln\frac{Q^2_c(A)}{\Lambda^2}
\,\approx\,\frac{Q_A^2}{\alpha_s N_c}\,\left({\rho_A} -
{\ln\rho_A} \right)
\,,\ee is almost the same as the result in Eq.~(\ref{GBSsat}) (in
particular, it includes already the enhancement factor $\rho_A$
due to the logarithmic phase--space\footnote{Although
$Q^2_c(A)\approx Q^2_s(A)/\rho_A \ll Q^2_s(A)$, the logarithmic
phase--space at
 $\Lambda < \kk < Q_c(A)$ (namely ${\rho_A} - {\ln\rho_A}$) is
almost the same as that for the whole interval $\Lambda < \kk <
Q_s(A)$ (which is equal to ${\rho_A}$).}). This confirms the fact
that the color glass condensate in the MV model is formed by
redistributing the `infrared' gluons at $\kk < Q_c(A)$ in the
bremsstrahlung spectrum.

\section{Non--linear gluon evolution in the Color Glass Condensate}
\setcounter{equation}{0} \label{EVOLUTION}

Within the effective theory for the CGC \cite{RGE}, the evolution
of the gluon distribution with $y$ is not described by a single,
closed, equation, but rather by an infinite hierarchy of coupled
equations, which relate $n$--point functions with arbitrary $n\ge
2$ (by itself, $\varphi(\kk,y)$ corresponds to a 2--point
function), and are compactly summarized in a {\it functional}
evolution equation, the JIMWLK equation. Fortunately,  this
complicated evolution simplifies considerably in interesting
limiting cases, which are tractable through analytic
approximations \cite{SAT,GAUSS}.

At low momenta $\kk \simle Q_s(A,y)$, the gluons form a condensate
with occupation factors of order $1/\alpha_s$, and the dynamics is
highly non--linear. Still, one can rely on mean field
approximations to deduce a formula for $\varphi(\kk,y)$ valid in
the low momentum limit $\kk \ll Q_s(A,y)$
\cite{SAT,GAUSS,AM99,LT99,AM02}, which will be presented in Sect.
\ref{NONLINEAR} below.

At high momenta $\kk \gg Q_s(A,y)$, the gluon density is low, and
non--linear effects in the evolution become negligible. Then, the
general equations for the evolution of the gluon distribution boil
down to a single, linear, equation: the BFKL equation \cite{BFKL}.
The latter controls also the approach towards saturation from the
above (at least, approximately), and thus determines the
saturation momentum $Q_s(A,y)$ \cite{GLR,SCALING,MT02,DT02,MP03}.

For a fixed coupling, we shall discuss in Sect. \ref{LINEAR} the
solution to the BFKL equation in the saddle point approximation
and with saturation boundary conditions
\cite{SCALING,MT02,DT02,MP03} (see also \cite{MS04}). Our
derivation of this solution will be quite schematic, and only
intended to emphasize a subtle point --- namely, the fact that it
is the saturation scale in the initial conditions, $Q_s(A)$, which
sets the infrared cutoff for the transverse phase--space available
for evolution \cite{AM03} ---, and to prepare the ground for a
general discussion of high--$\kk$ suppression through quantum
evolution, to be presented in Sect. \ref{general}.

For a running coupling, with the scale for running set by the
gluon transverse momentum $\kk$, we shall follow the analysis in
Refs. \cite{MT02,DT02,MP03}, from which we shall simply quote here
the relevant results, and discuss their range of validity.

\subsection{Non--linear evolution at low $\kk$ :
 Saturation and universality}
\label{NONLINEAR}

In the high density regime deeply at saturation ($\kk \ll
Q_s(A,y)$), the color fields are strong ($A^i\sim 1/g$) and the
dynamics is fully non--linear, which makes it difficult to use
standard perturbative techniques. On the other hand, these
conditions are favorable to the use of mean field approximations
\cite{SAT,GAUSS}, which enable us to determine the $y-$ and
$\kk-$dependencies of the gluon occupation factor. Similar results
are also obtained from studies of the unitarization effects in
dipole--hadron scattering \cite{AM99,LT99,AM02}.

Remarkably, it turns out that the functional form of the gluon
occupation factor at saturation is {\sf universal} \cite{SAT}: For
$\kk \ll Q_s(A,y)$, $\varphi_A(\kk,y)$ is independent of either
the initial condition at $y=0$, or the details of the evolution
leading to saturation, except for the corresponding dependencies
of the saturation scale itself. Moreover, this universal form
happens to be the same as in the MV model (cf. the expressions in
the second lines of Eqs.~(\ref{asympfix}) and (\ref{asymprun})),
which is quite non--trivial, and to some extent surprising, since
the physical conditions leading to this functional form are very
different in the two cases: Whereas in the MV model the color
sources are uncorrelated, and the logarithmic behavior at low
$\kk$ is merely the result of non--linear effects in the classical
equations of motions for the color fields
\cite{JKMW97,CGCreviews}, in the effective theory at small $x$ the
color sources are strongly correlated, in such a way to ensure
color neutrality over a transverse size $\sim 1/Q_s(A,y)$
\cite{SAT,AM02,GAUSS}, and the logarithmic behavior at low $\kk$
is already encoded in the correlations of the color sources (i.e.,
it holds independently of the presence of non--linear terms in the
classical field equations; the latter affect  only the overall
normalization of the gluon distribution, which in this non--linear
regime is anyway not under control).

Specifically, for fixed coupling, one finds \cite{AM99,SAT,GAUSS}
: \be\label{phiAsat} \varphi_A(\kk,y)\,\approx\,
 \frac{a_0}{\alpha_s N_c}\,\ln \frac{Q_s^2(A,y)}{k_{\perp}^2}\,,
\qquad  {\rm (fixed\,\,coupling)} ,\ee whereas for running
coupling (with the scale for running set by the gluon momentum
$\kk$) one obtains \cite{AM02,CGCreviews}.
 \be\label{phiAsatrun0}
\varphi_A(\kk,y)&\approx&
 \frac{a_0}{N_c}\,\left\langle \frac{1}{\alpha_s}\right\rangle\,
\ln \frac{Q_s^2(A,y)}{k_{\perp}^2}\,, \qquad {\rm
(running\,\,coupling)},\ee with (compare to Eq.~(\ref{avalpha})) :
 \be\left\langle
\frac{1}{\alpha_s}\right\rangle \equiv \frac{1}{2b_0}\left\{\ln
\frac{Q_s^2(A,y)}{\Lambda^2} + \ln\frac
{k_{\perp}^2}{\Lambda^2}\right\}.\ee
  The overall factor $a_0$ in these equations is a number of
order one which cannot be computed analytically (and which needs
not be the same for fixed and running coupling).

Note that, in the fixed coupling case, the gluon distribution at
saturation depends upon the two kinematical variables $\kk$ and
$y$ only via the ratio $z \equiv k_{\perp}^2/Q_s^2(A,y)$. This
property, known as {\sf geometric scaling} \cite{geometric},
reflects the fact that the saturation momentum is the only
intrinsic scale at saturation. As manifest on
Eq.~(\ref{phiAsatrun0}), the running of the coupling introduces a
second intrinsic scale, namely $\Lambda$, and this is a source of
geometric scaling {\sf violations}, which are however under
control.

\subsection{Linear evolution at high $\kk$ : Fixed coupling}
\label{LINEAR}

 In the low density regime at high momenta $\kk \gg
Q_s(A,y)$, the dynamics is linear, and the nuclear gluon
occupation factor  can be obtained (to the present accuracy) by
solving the BFKL equation \cite{BFKL} with initial conditions at
$y=0$ provided by the MV model (cf.
Eq.~(\ref{phiMV})--(\ref{phiMVrun})) and with an `absorptive'
boundary condition at $\kk \sim Q_s(A,y)$ \cite{MT02,DT02} which
mimics the non--linear effects in the approach towards saturation.
(Similar results have been recently obtained \cite{MP03} through
direct studies of the non--linear Kovchegov equation \cite{B,K}.)
In particular, the saturation scale itself is obtained by
requiring $\varphi_A(\kk,y)$ to become of order $1/\alpha_s$ at
saturation (up to subleading terms of ${\mathcal O}(1)$) :
 \be\label{SATcond}
\varphi_A(\kk,y)\,\simeq\,\frac{\kappa}{\alpha_s(Q_s^2(A,y))
N_c}\,\qquad {\rm for} \qquad \kk\sim Q_s(A,y)\,.\ee ($\kappa$ is
a number of order one.)

When $y$ and/or $\kk$ are relatively large (see below), an
approximate analytic solution can be obtained, via a saddle point
approximation. Here we shall outline only a few steps in the
construction of this solution (see Refs. \cite{SCALING,MT02} for
details).


For fixed coupling, the solution to the BFKL equation for
$\varphi_A(\kk,y)$ can be written in Mellin form as:
\BQ\label{Mellin} \varphi_A(\kk,y)\,=\, \int_{C}
\frac{d\gamma}{2\pi i} \left(\frac{Q_0^2}{\kk^2}\right)^{\gamma}
 {\rm e}^{\bar\alpha_s y \chi(\gamma)}\,
\widetilde{\varphi_A}(\gamma; Q_0), \EQ where
$\bar\alpha_s=\alpha_s N_c/\pi$,
$\chi(\gamma)=2\psi(1)-\psi(\gamma)-\psi(1-\gamma)$ with
$\psi(\gamma)\equiv d \ln \Gamma(\gamma)/d\gamma$ and complex
$\gamma$ is the eigenvalue of the BFKL kernel,
$\widetilde{\varphi_A}(\gamma; Q_0)$ is the Mellin transform of
the initial condition $\varphi_A(\kk)$, Eq.~(\ref{phiMV}):
\be\label{invMellin} \widetilde{\varphi_A}(\gamma; Q_0)\,=\,
\int_0^\infty\frac{d\kk^2}{\kk^2}\,\left(\frac{\kk^2}{Q_0^2}\right)^{\gamma}
\varphi_A(\kk),\ee and $Q_0$ is an arbitrary reference scale
introduced for dimensional reasons, and which drops out in the
complete result, as obvious on the above equations. Given the
behavior of the initial distribution $\varphi_A(\kk)$ in various
limits, cf. Eq.~(\ref{asympfix}), it can be checked that the
integral over $\kk^2$ in Eq.~(\ref{invMellin}) is absolutely
convergent for $0 < {\rm Re} \,\gamma < 1$. Thus the contour for
the complex integration in Eq.~(\ref{Mellin}) can be chosen as $C
=\left\{\gamma=\gamma_1 + i \gamma_2\,;-\infty< \gamma_2
<\infty\right\}$, with a fixed $\gamma_1$ in the range $0 <
\gamma_1 < 1$.

The same study of the convergence properties of the integral in
Eq.~(\ref{invMellin}) tells us that {\sf this integral is
dominated by momenta of the order of the initial saturation
momentum} $Q_s(A)$: Indeed, for large $\kk \gg Q_s(A)$,
$\varphi_A(\kk) \propto 1/\kk^2$, and the integral is saturated by
momenta of the order of the lower cutoff $Q_s(A)$. Furthermore,
for low momenta $\kk \ll Q_s(A)$, the initial distribution
saturates: $\varphi_A(\kk) \propto \ln(Q_s^2/\kk^2)$, so the
integral is now dominated by momenta near the upper cutoff, i.e.,
by $\kk \sim Q_s(A)$ once again. Thus, after performing the
integral over $\kk$, $Q_s(A)$ replaces $Q_0$ as the {\sf natural}
reference scale in the Mellin representation of the solution
$\varphi_A(\kk,y)$. Then, Eq.~(\ref{Mellin}) can be rewritten as:
\BQ\label{Mellin1} \varphi_A(\kk,y)\,=\,\frac{1}{\alpha_s N_c}
\int_{C} \frac{d\gamma}{2\pi i}
\left(\frac{Q_s^2(A)}{\kk^2}\right)^{\gamma} \, {\rm
e}^{\bar\alpha_s y \chi(\gamma)}\,\, \widetilde{\varphi_A}(\gamma;
\rho_A), \EQ where any reference to the arbitrary scale $Q_0$ has
disappeared. Note that, for more clarity, we have extracted a
factor $1/\alpha_s N_c$ out of the (Mellin transform of) the
initial condition $\widetilde{\varphi_A}(\gamma; \rho_A)$. The
latter still depends upon $Q_s(A)$, but only logarithmically, via
$\rho_A \equiv \ln(Q_s^2(A)/\Lambda^2)$. The precise form of the
function $\widetilde{\varphi_A}(\gamma; \rho_A)$ will not be
needed in what follows. Rather, we shall use its approximate form
which is obtained when $\varphi_A(\kk)$ in Eq.~(\ref{invMellin})
is replaced by the simplest interpolation between the asymptotic
behaviors shown in Eq.~(\ref{asympfix}). This reads :
\be\label{phinit} \widetilde{\varphi_A}(\gamma; \rho_A)\,=\,
 - \frac{\Gamma(\gamma) \psi(1-\gamma)}{\rho_A}\, +\,
  \frac{\Gamma(\gamma)}{\gamma}\,+ \cdot \cdot \cdot
\,\approx\,\frac{1}{\rho_A}
\frac{1}{1-\gamma}\,+\,\frac{1}{\gamma^2}\,.\ee The first term in
the last (approximate) equality, which has a pole at $\gamma=1$
and is suppressed as $1/\rho_A$, comes from the high--$\kk$
behavior (the bremsstrahlung spectrum in Eqs.~(\ref{asympfix})),
while the second term, with a double pole at $\gamma=0$, comes
from the saturating behavior at low $\kk$.

\subsubsection{The saddle point approximation}
\begin{figure}[]
\begin{center}
\includegraphics[scale=1.33]{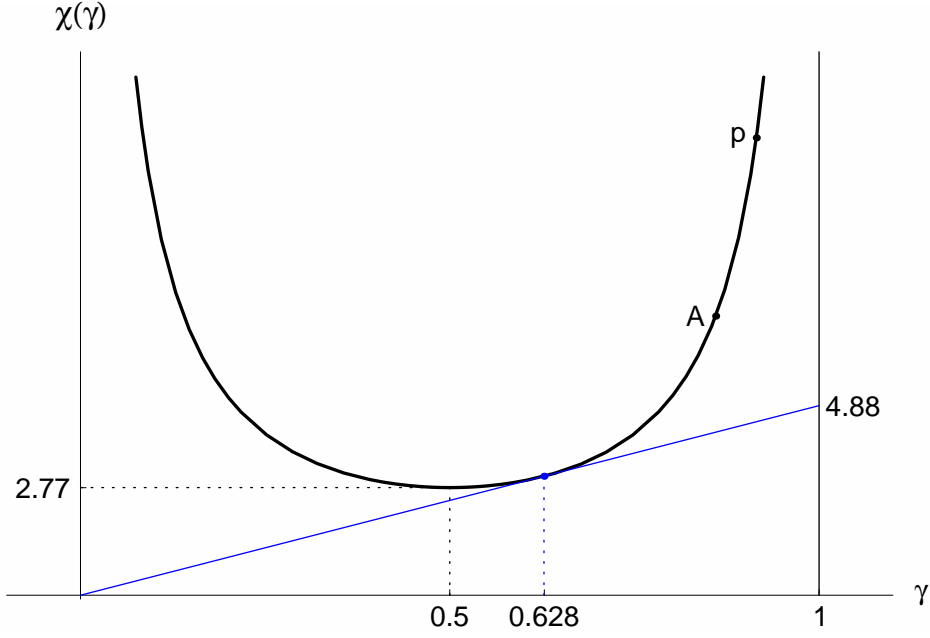}
    \caption{\label{CHI}
    {\sl The BFKL eigenvalue $\chi(\gamma)$ and the graphical
    solution to the saturation problem}
    {\small The value $\gamma_s
    = 0.628$ corresponds to the saturation saddle point and the value
    $\chi(\gamma_s)/\gamma_s = 4.88$ determines the asymptotic energy
    dependence of the saturation momentum (for comparison, the ``hard
    pomeron'' saddle point $\gamma_{\mathbb P}  =1/2$ and its
    intercept $\omega_{\mathbb P} = 4 \ln 2 = 2.77$ are shown). The
    points {\sf p} and {\sf A} correspond to the proton and the
    nucleus respectively, for the same transverse momentum and
    rapidity, when they are far above the saturation scale.}}
\vspace*{.5cm}
\end{center}
\end{figure}

At this stage, it becomes natural to introduce the logarithmic
momentum variable $\rho\equiv \ln(\kk^2/Q_s^2(A)) = \rho(A,\kk)$,
and notice that, when either $\bar\alpha_s y$, or $\rho$, or both,
are large, the integral in Eq.~(\ref{Mellin1}) can be evaluated
via a {\sf saddle point approximation}. Specifically, if one
writes: \be\label{Mellin2} \varphi_A(\kk,y)\,=\,\frac{1}{\alpha_s
N_c} \int_{C} \frac{d\gamma}{2\pi i}\, \,{\rm e}^{\bar\alpha_s y
F(\gamma,{\rm r})}\, \,\widetilde{\varphi_A}(\gamma; \rho_A)\,,\ee
where: \be\label{RFdef} {\rm r}\equiv \frac{\rho}{\bar\alpha_s
y},\qquad F(\gamma,{\rm r})=-\gamma {\rm r} + \chi(\gamma), \ee
then the saddle point $\gamma_0\equiv \gamma_0({\rm r})$ satisfies
the condition:
 \BQ\label{saddle}
\left. \frac{\del F}{\del \gamma}(\gamma,{\rm
r})\right\vert_{\gamma= \gamma_0({\rm r})} = -{\rm r} +
\chi'(\gamma_0({\rm r}))=0, \EQ which shows that $\gamma_0({\rm
r})$ is a real number, in between 0 and 1. The behavior of the
function $\chi(\gamma)$ in this range is illustrated in Fig.
\ref{CHI}.

After expanding around the saddle point and performing the
Gaussian integral over the small fluctuations, one obtains :
\be\label{SPA} \varphi_A(\kk,y)\,\approx\,\frac{1}{\alpha_s N_c}\,
\frac{{\rm e}^{\bar\alpha_s y F(\gamma_0,{\rm r})}
}{\sqrt{2\pi\bar\alpha_s y\chi''(\gamma_0)}}\,
\widetilde{\varphi_A}(\gamma_0; \rho_A)\,.\ee This is a reasonable
approximation to Eq.~(\ref{Mellin1}) provided $\bar\alpha_s y
\simge 1$, $\rho \simge 1$, and at least one of these quantities
is much larger than one.

When ${\rm r}$ varies from 0 to $\infty$, the function
$\gamma_0({\rm r})$ is monotonously increasing, and spans the
range from $\gamma_0(0)=1/2$ to $\gamma_0(\infty)=1$. Note,
however, that {\sf physically} the ratio ${\rm r}$ cannot be
smaller than the value ${\rm r}_s\equiv \rho_s(y)/\bar\alpha_s y$
(with $\rho_s(y)\equiv \ln[Q_s^2(A,y)/Q_s^2(A)]$) which
corresponds to the saturation momentum. Indeed, the linear
approximation is justified only as long as $\kk \gg Q_s(A,y)$, or
$\rho > \rho_s(y)$.

To compute the saturation momentum, and the corresponding saddle
point $\gamma_s\equiv \gamma_0({\rm r}_s)$, we shall use the
saturation boundary condition (\ref{SATcond}), together with the
approximation (\ref{SPA}) for $\varphi_A(\kk,y)$, which for this
purpose is extrapolated down to $\kk \sim Q_s(A,y)$. It can be
justified that this is a legitimate procedure for the calculation
of the energy dependence of $Q_s(A,y)$, although not necessarily
also for its absolute normalization \cite{SCALING,MT02}. To the
accuracy of interest here, Eqs.~(\ref{SPA}) and (\ref{SATcond})
imply just that the exponent in Eq.~(\ref{SPA}) must vanish on the
saturation line: \BQ\label{sat_cond} F(\gamma_0({\rm r}),{\rm
r})\Big\vert_{{\rm r}={\rm r}_s} \equiv \, -\,\gamma_s {\rm r}_s
+\chi(\gamma_s)=0, \EQ which shows that ${\rm r}_s$ is a pure
number, to be denoted as $c$ in what follows. This number together
with $\gamma_s$ are determined by Eqs.~(\ref{saddle}) and
(\ref{sat_cond}) as:
 $\gamma_s = \chi(\gamma_s)/\chi'(\gamma_s)
=  0.627...$ and $c\equiv {\rm r}_s = \chi'(\gamma_s)= 4.883....$
\cite{GLR,AM99,SCALING,MT02}.

Note that in writing the saturation condition in the form
(\ref{sat_cond}) we have neglected the effect of the Gaussian
fluctuations around the saddle point. The reason is that this
effect is modified by the absorptive boundary condition at
saturation \cite{MT02} (that we neglect here), and thus cannot be
correctly computed in the present approximations. In what follows,
we shall often ignore such slowly varying prefactors, but only
include the dominant behavior at high energy. To the same
accuracy, the nuclear saturation momentum reads: \be\label{Qsfix0}
Q_s^2(A,y)\,\simeq\,Q_s^2(A)\, {\rm e}^{c\bar\alpha_s y}
\,, \qquad c\,=\,\,4.883....\ee This expression appears to satisfy
the initial condition $Q_s^2(A,y=0)=Q_s^2(A)$, but this is only
formal, since the calculation above is valid only for relatively
large $y$, with $\alpha_s y > 1$, so the extrapolation to $y\to 0$
is not justified. In fact, the energy dependence of $Q_s^2(A,y)$
displayed in Eq.~(\ref{Qsfix0}) is only the first term in an
asymptotic expansion at large $y$, out of which also the second
term \cite{MT02,DT02} and the third one \cite{MP03} are presently
known.

For practical purposes, the saddle point approximation (\ref{SPA})
is still quite involved, since the function $\gamma_0({\rm r})$ is
known only implicitly, as the solution to Eq.~(\ref{saddle}).
However, important simplifications leading to fully explicit
formulae can be performed either for very large momenta, $\rho \gg
\bar\alpha_s y$ --- this corresponds to the so--called
``double--logarithmic accuracy'', or DLA, regime ---, or for
momenta above, but relatively close to the saturation scale, where
one can expand Eq.~(\ref{SPA}) around the saturation saddle point.
Below, we shall refer to the latter regime as the ``BFKL
regime''\footnote{Note that this is not the {\sf usual} BFKL
regime associated with the  saddle point $\gamma_{\mathbb P}=1/2$
and corresponding to the high energy limit $\alpha_s y\gg \rho$.
Whereas $\gamma_{\mathbb P}=1/2$ describes the evolution with $y$
at fixed (and relatively low) $\kk$, $\gamma_s=0.63$ corresponds
rather to an evolution where, when increasing $y$, $\kk$ is
correspondingly increased, in such a way that the condition
$\kk^2> Q_s^2(A,Y)$ (or $\rho > \rho_s(Y)$)  remains satisfied.}.

Let us briefly describe here the modifications which should be
brought into the previous formulae in order to adapt them to the
case of a proton. Recall that there is no saturation scale (i.e.,
no analog of the nuclear scale $Q_s(A)$) in the proton at $y=0$.
The infrared cutoff is rather set by the scale $Q_p$ which is soft
and reflects non--perturbative physics (cf. Eq.~(\ref{phip0})).
This is therefore also the scale which enters the formulae for the
linear evolution of $\varphi_p(\kk,y)$ with increasing $y$. E.g.,
the proton analog of Eq.~(\ref{Mellin1}) is obtained by replacing
$Q_s(A)\rightarrow Q_p$ and taking an initial condition
$\widetilde{\varphi_p}(\gamma) = 1/(1-\gamma)$ (compare to
Eq.~(\ref{phinit})). In particular, the variable $\rho$ for the
proton is defined as $\rho(p,\kk)\equiv \ln (\kk^2/Q_p^2)$. Of
course, for sufficiently large $y$, the gluon density in the
proton will become large, and a (hard) saturation scale $Q_s(p,y)$
will be eventually generated. The dominant $y$--dependence of this
scale at large $y$ is described again by Eq.~(\ref{Qsfix0}), but
with $Q_s(A)\rightarrow Q_p$. More generally, all the formulae to
be obtained below in this section can be translated to the case of
a proton by  replacing everywhere $Q_s(A)\rightarrow Q_p$.

\subsubsection{DLA regime}

If $\rho \gg \alpha_s y$, the saddle point is close to one, and
only the first term (the one having a pole at $\gamma =1$) must be
retained in the initial condition (\ref{phinit}). A
straightforward calculation yields $\gamma_0({\rm r})\approx 1 -
1/\sqrt{{\rm r}}$, and :
 \be\label{DLAp0}
\varphi_A(\kk,y)\,\simeq\,\frac{1}{\alpha_s
N_c}\,\frac{Q_A^2}{k_{\perp}^2}\,
\frac{{\rm e}^{\sqrt{4\bar\alpha_s y \rho}}}
{\sqrt{2\pi\sqrt{4\bar\alpha_s y \rho}}} \,,\ee where the factor
$Q_A^2$ has been reconstructed as $Q_s^2(A)/\rho_A$. This
approximation is valid for $\bar\alpha_s y \rho \gg 1$, and as
long as this condition is satisfied, Eq.~(\ref{DLAp0}) can be even
used for relatively small $y$, with $\bar\alpha_s y\ll 1$. Indeed,
in this regime Eq.~(\ref{DLAp0}) can be obtained from the DGLAP
equation \cite{DGLAP}, which provides a resummation based on the
assumptions that $\alpha_s  \rho \ge 1$, but $\alpha_s y\ll 1$.

\subsubsection{BFKL regime: weak universality and approximate scaling}
\label{BFKLFIX}

 When $\rho$ is larger than $\rho_s(y)\equiv
\ln(Q_s^2(A,y)/Q_s^2(A))= c \bar\alpha_s y $, but not {\sf much}
larger, one can obtain an explicit form for $\varphi_A(\kk,y)$ by
expanding the general saddle point approximation, Eq.~(\ref{SPA}),
in powers of ${\rm r} - {\rm r}_s$ (or, equivalently, in powers of
$(\rho -\rho_s)/\rho_s$). To second order in this expansion, one
obtains \cite{SCALING,MT02,DT02,MP03}
 \be\label{BFKLfix} \hspace*{-.5cm} \varphi_A(\kk,y)\approx
 \frac{\kappa_0}{\alpha_s N_c}
\left\{\rho-\rho_s+{\Delta} \right\} {\rm e}^{-\gamma_s
(\rho-\rho_s)} \exp\left\{-\frac{(\rho-\rho_s)^2}{2\beta
\bar\alpha_s y} \right\},\ee or, more explicitly [recall that
$\rho-\rho_s = \ln ({k_{\perp}^2}/{Q_s^2(A,y)})$] :
 \be\label{BFKLp0} \hspace*{-.4cm} \varphi_A(\kk,y)\approx
 \frac{\kappa_0}{\alpha_s N_c}
\left\{\ln \frac{k_{\perp}^2}{Q_s^2(A,y)}+{\Delta}
\right\}\left(\frac{Q_s^2(A,y)}{k_{\perp}^2}\right)^{\gamma_s}
\exp\left\{-\frac{1}{2\beta \bar\alpha_s y}\ln^2
\frac{k_{\perp}^2}{Q_s^2(A,y)} \right\}\ee
 In these equations, $\gamma_s\simeq 0.63$ appears
as an ``anomalous dimension'' (recall that, at very high $\kk$,
the actual power of $1/{k_{\perp}^2}$ was one, both in the initial
condition at $y=0$, and in the DLA approximation (\ref{DLAp0}) at
$y > 0$). Furthermore, 
$\beta \equiv \chi''(\gamma_s)= 48.518...$ plays the role of a
diffusion coefficient for the diffusion in $\ln
({k_{\perp}^2}/{Q_s^2})$.

The constant numbers $\kappa_0$ and $\Delta$ in Eq.~(\ref{BFKLp0})
are not known\footnote{But it seems natural to impose $\Delta\ge
1/\gamma$, to avoid that Eq.~(\ref{BFKLp0}) become a decreasing
function of $1/{k_{\perp}^2}$ when approaching $Q_s(A,y)$ from the
above.}, as they depend upon the detailed matching onto the
(unknown) solution in the transition region towards saturation. In
particular, Eqs.~(\ref{BFKLp0}) and (\ref{SATcond}) need not match
with each other, since there is no reason why Eq.~(\ref{BFKLp0})
should remain correct down to $Q_s(A,y)$. But in writing
Eq.~(\ref{BFKLp0}) we have nevertheless assumed that, when
extrapolated to $\kk\sim Q_s(A,y)$, {\sf the BFKL solution becomes
parametrically of the same order as the actual solution in the
transition region} (namely, they are both of order $1/\alpha_s$).
This may look surprising in view of the previous experience with
the MV model where we have seen that, when extrapolated down to
$Q_s(A)$, the twist piece $\varphi_A^{\rm twist}$ remains
parametrically smaller (by a factor $\rho_A$) than the actual
distribution at saturation $\varphi_A^{\rm sat}$. The two
functions match with each other only at the (parametrically
larger) momenta $\kk^2\sim (\ln\rho_A) Q_s^2(A)$. Still, as we
shall see through a more detailed analysis in Sect. \ref{FLAT},
the evolution with $y$ is such that a power--law, `twist', tail
develops in the gluon distribution at momenta just above
$Q_s(A,y)$, and this tail is predominantly generated via radiation
from those gluons which were originally at saturation (i.e., which
at $y=0$ were included in $\varphi_A^{\rm sat}$). Thus, for
$\alpha_s y\simge 1$, the twist contributions generated by the
evolution are {\sf not} parametrically suppressed with respect to
the saturating distribution anymore. 

 An important property of Eq.~(\ref{BFKLp0}), that we shall refer
to as {\sf weak universality} (as opposed to the {\sf strong
universality} that has been found at saturation \cite{SAT}) is the
fact that the functional form of this expression, seen as a
function of $z\equiv \kk^2/Q_s^2(A,y)$ and $y$, is independent of
the initial condition at $y=0$ (and, in particular, of $A$), but
rather is fully determined by the quantum (BFKL) evolution
together with the saturation boundary condition: $\varphi_A(\kk,y)
= \varphi(z,y)$, with $\varphi$ an universal function. The initial
condition affects only the value of the
saturation momentum, cf. 
Eq.~(\ref{Qsfix0}) \cite{AM03}.

The factor linear in $\rho -\rho_s$ in the right hand side of Eq.~
(\ref{BFKLfix}), or (\ref{BFKLp0}), does not come from the
expansion of Eq.~(\ref{SPA}), but rather from the absorptive
boundary condition at saturation \cite{MT02,MP03}, that we have
ignored so far. The reason why this factor cannot be neglected (in
contrast, e.g., with the prefactor in the expression
(\ref{Qsfix0}) for the saturation scale) is that this factor
introduces a dependence upon $\rho$ which is not suppressed at
large $y$. The other factors in Eq.~ (\ref{BFKLfix}), which
involve $\rho -\rho_s$ in the exponent, do come from the expansion
of Eq.~(\ref{SPA}). A priori, one would expect this expansion to
be valid so long as $(\rho -\rho_s)/\rho_s \ll 1$, since this is
the ``small parameter". However, since $\beta$ is a
relatively large number, 
the second order term in the expansion, quadratic in $\rho
-\rho_s$, remains small as compared to the linear term even for
$\rho -\rho_s\sim \rho_s$. We conclude that Eq.~ (\ref{BFKLfix})
(or (\ref{BFKLp0})) is a good approximation for any $\rho$ within
the following range  \cite{SCALING} :
 \be\label{swindow0} 0 < \rho - \rho_s(y)
\simle \rho_s(y), \,\ee
often referred to as the ``extended scaling window", since within
this range the BFKL solution approximately preserves the geometric
scaling property characteristic of saturation \cite{SCALING,MT02}:
More precisely, Eq.~(\ref{BFKLfix}) shows scaling so long as
 $\rho - \rho_s(y)\ll \sqrt{2\beta \bar\alpha_s y}$,
whereas for larger $\rho$, the scaling is 
violated by the diffusion term. Both the scaling property, and its
violation, seem to be necessary to understand the small--$x$ data
at HERA \cite{geometric,GBW99,IIM03}.

Note finally that the existence of a relatively large window for
geometric scaling {\sf above} the saturation momentum is a
consequence of the evolution\footnote{For instance, there is no
such a window in the initial conditions at $y=0$, where the
property of geometric scaling can be defined with respect to the
dependencies upon $\kk$ and $A$. As manifest on
Eq.~(\ref{phiMVdec})--(\ref{phiMVexp}), it is only the saturating
distribution which is just function of $z=\kk^2/Q_s^2(A)$; the
twist terms have additional, explicit, dependencies upon $A$ (and
$\Lambda$), through factors of $1/\rho_A$.}, so it takes some
finite amount of rapidity (typically, $\bar\alpha_s y\sim 1$)
until this behavior is actually reached.

\subsubsection{The geometric scaling line : Approximately matching BFKL
and DLA}

The approximate solutions that we have found so far apply in
different kinematical regions --- $\rho \gg \alpha_s y$ for the
DLA formula (\ref{DLAp0}), and $\rho - \rho_s \simle\rho_s$ for
the BFKL expression (\ref{BFKLfix}) ---, which have no overlap
with each other. So, a priori, there seems to be no reason why
these expressions could be matched onto each other. Still, if one
chooses $\rho=2\rho_s(y) = 2c \bar\alpha_s y$, and one estimates
both the DLA and the BFKL expressions along this line, then one
finds that the dominant (exponential) behavior in $y$ is very
similar, although not exactly the same, for both approximations.
This is in agreement with an observation \cite{SCALING} that the
{\sf geometric scaling line} $\rho=2\rho_s(y)$ is approximately
the borderline between the regions dominated by the saturation
saddle point and, respectively, by the DLA saddle point in the $y
-\ln \kk^2$ plane.

This opens the possibility to achieve at least an {\sf approximate
matching} between the two limiting expressions, by appropriately
tuning the borderline between these two regime. Of course, there
is no fundamental reason in favor of this matching, but this will
be convenient in practice, as it will allow us to avoid
exponentially large discontinuities when studying the ratio
${\mathcal R}_{pA}$ in different regimes. Specifically, we shall
tune the upper limit in the window (\ref{swindow0}) in such a way
that the leading, exponential, behavior along that line be exactly
the same for the DLA and BFKL approximations with fixed coupling.
A straightforward calculation shows that Eq.~(\ref{swindow0}) must
be replaced by
 \be\label{swindow1} 0 < \rho- \rho_s(y)
\le \nu \rho_s(y)\,,\ee or, equivalently, \be\label{swindow}
 Q_s^2(A,y)\,\, \ll\,\, \kk^2\,\,<\,\,Q_g^2(A,y)\equiv
Q_s^2(A,y)\left( \frac{Q_s^2(A,y)}{Q^2_s(A)}\right)^\nu\,,\ee with
$\nu\approx 1.708$ determined by solving the following equation:
 \be
 c[(1-\gamma_s)\nu +1] - \sqrt{4c(\nu+1)} - \frac{(\nu
c)^2}{2\beta}\, =\,0\,.\ee As anticipated, $\nu$ is a number of
order one.

\subsection{Linear evolution at high $\kk$ : Running coupling}
\label{EV:RUN}

Since the (leading--order) BFKL equation corresponds to a
fixed--coupling approximation, the inclusion of the running is a
priori ambiguous. Physical arguments, together with recent
experience \cite{Salam99,DT02} with the next--to--leading order
BFKL equation \cite{NLBFKL}, suggest that it should be appropriate
to use a coupling which is running with the transverse momentum
$\kk$ of the gluon [i.e., $\alpha_s\equiv\alpha_s(\kk^2)$]. This
running has been already used in obtaining
Eq.~(\ref{phiAsatrun0}), valid in the saturation region. With such
a running, the BFKL equation becomes difficult to solve in
general, even formally, so it is interesting to keep in mind that,
in order to study the high--energy limit\footnote{In the present
context, this is the limit in which, with increasing $y$, $\rho$
is increased as well, in such a way to remain in the linear regime
at $\rho
>\rho_s(y)$.}, one can also use the running with the saturation
momentum [i.e., $\alpha_s\equiv\alpha_s(Q_s^2(A,y))$]. Indeed,
with increasing $y$, the relative separation from the saturation
line $(\rho -\rho_s(y))/\rho_s(y)$ becomes smaller and smaller,
and we can approximate, e.g. (with $b\equiv b_0 N_c/\pi =
{12N_c}/{(11N_c -2N_f)}$),
    \be\label{alphaexp} \bar\alpha_s(\kk^2) \,\equiv\,\frac{b}{\ln
(\kk^2/\Lambda^2)} \,\approx\,\bar\alpha_s(Q_s^2(A,y))
\,\left(1 -\frac{\rho-\rho_s(y)}{\tau_A(y)}\right)\,.\ee In this
equation $\tau_A(y) = \ln (Q_s^2(A,y)/\Lambda^2) =
\rho_s(y)+\rho_A$ (see Eq.~(\ref{Qsrun}) below for a more precise
definition). By using just the first term in this expansion, i.e.,
$\bar\alpha_s(Q_s^2(A,y))$, one obtains the correct dominant
behavior at high energy for both the saturation momentum and the
gluon distribution \cite{SCALING,MT02,DT02,MP03}. However,
interesting subleading effects at high energy (like the
``diffusion term" in the gluon distribution) are not correctly
given by this approximation, whose applicability at finite $y$ is
restricted to a rather narrow strip $(\rho -\rho_s)/\rho_s\ll 1$
above the saturation line. To keep control on such subleading
effects, we shall follow Refs. \cite{MT02,DT02,MP03} and use the
whole expression in the r.h.s. of Eq.~(\ref{alphaexp}) when
studying the ``BFKL regime" precursory of saturation. Of course,
the resulting approximation will be still limited to $(\rho
-\rho_s)/\rho_s\ll 1$ --- because of the expansion performed in
Eq.~(\ref{alphaexp}), and also of the other approximations needed
when solving the (corresponding version of) BFKL equation
\cite{MT02,DT02} ---, but as we shall see, the accessible window
is considerably wider with this improved running.

On the other hand, for much larger momenta, such that $\rho
-\rho_s\simge \rho_s$, the running effects cannot be expanded
anymore, and the actual running coupling $\bar\alpha_s(\kk^2)$ has
to be used. But in this regime, the BFKL equation itself boils
down to the DLA equation, which is much simpler, and whose exact
solution is known also with running coupling.

As it should be clear from the previous considerations, the
strategy for performing approximations is now different from the
fixed coupling case: Rather than solving first the general BFKL
equation and then simplifying the result, it is more convenient to
perform the approximations on the equation itself, and then solve
a simpler equation. Since the approximations that we shall
consider are different for the BFKL and the DLA regimes, our
analysis will not cover the whole linear regime at
$\rho>\rho_s(y)$ : the intermediate region between the BFKL and
DLA regimes remains out of control.

But before we present the general results, let us use the
simplified running $\bar\alpha_s(Q_s^2(A,y))$ to give a rapid
derivation of the leading energy dependence of the saturation
momentum at large $y$. With this running, the only modification in
the previous, fixed--coupling, analysis refers to the replacement
$\bar\alpha_s y \rightarrow h_A(y)$ in the exponent of the
integrand in Eq.~(\ref{Mellin}) (and the subsequent formulae).
Here, $h_A(y)$ is defined as the solution to $dh_A/ dy =
\alpha_s(Q_s^2(A,y))$ with $h_A(0)=0$. Then calculations entirely
similar to those described in the previous subsection lead to the
condition ${\rm r}_s\equiv \rho_s(y)/h_A(y) = c$, where
$c=4.883...$ and $\rho_s(y) \equiv \ln(Q_s^2(A,y)/Q_A^2)$, as
before. This condition together with the equation defining
$h_A(y)$ form a coupled system of equations for the functions
$h_A(y)$ and $Q_s^2(A,y)$, whose solution is conveniently written
as \cite{SCALING,MT02,AM03}:
   \be\label{Qsrun} Q_s^2(A,y)\,\simeq\,\Lambda^2 \,{\rm
e}^{\tau_A(y)}\,,\qquad\tau_A(y)\,\equiv \, \sqrt{2c b y+
\rho_A^2}\,.\ee For completeness, we note that $c h_A(y) =
\tau_A(y) - \rho_A$.

For relatively small $y$, such that $2c b y\ll \rho_A^2$, one can
expand $\tau_A(y) \simeq \rho_A + (b cy/\rho_A)$. Then,
Eq.~(\ref{Qsrun}) reduces to the same expression as for fixed
coupling, cf. Eq.~(\ref{Qsfix0}), but with $\bar\alpha_s$
evaluated at the initial saturation scale (i.e.,
$\bar\alpha_s\equiv \bar\alpha_s(Q_A^2) = b/\rho_A$) \cite{DT02}.

But with increasing $y$, the dependence on $A$ becomes weaker and
weaker\footnote{Choosing $\rho_A=4$ (a reasonable value for the
RHIC phenomenology) and using $2c b y\sim 10 y$, one finds that
the running effects become important --- in the sense that the two
terms under the square root in Eq.~(\ref{Qsrun}) become comparable
--- already for $y$ of order one.} \cite{AM03}
 : \be\label{QsrunY} Q_s^2(A,y)\,\simeq\,\Lambda^2
\,{\rm e}^{\sqrt{2c b y}}\,\exp\left\{\frac{\rho_A^2}{2\sqrt{2c b
y}}\right\} \qquad{\rm for} \qquad 2c b y\,\gg\, \rho_A^2\,.\ee By
using similar formulae (with $\rho_A\rightarrow \rho_p\sim 1$)
also for the proton, it becomes clear that, unlike in the fixed
coupling case, where the ratio $Q_s^2(A,y)/Q_s^2(p,y) =
Q_s^2(A)/Q_s^2(p)$ is independent of $y$ and thus fixed by the
initial conditions, in the case of a running coupling, this ratio
decreases with $y$, and converges to one at large $y$ \cite{AM03}
: \be\label{QsrunAp}
\frac{Q_s^2(A,y)}{Q_s^2(p,y)}\,\simeq\,\,\exp\left\{\frac{\rho_A^2-\rho_p^2}
{2\sqrt{2c b y}}\right\}\,\simeq\, 1 \qquad{\rm for} \qquad 2c b
y\,\simge\, \rho_A^4\,.\ee Thus, due to running coupling effects
in the evolution, the initial difference between the proton and
the nucleus is washed out after a rapidity evolution $cby \sim
\rho_A^4$.

\subsubsection{DLA regime}

With a running coupling $\bar\alpha_s(\kk^2)$, the solution to the
DLA equation --- the simplified equation valid at high $\kk^2$,
which is a common limit of the BFKL and DGLAP equations --- is
well known, even exactly. For consistency with the other
approximations that we have considered so far, we shall only use
the corresponding saddle point approximation, valid when $4 b y
\eta\gg 1$, with $\eta\sim\ln \rho$ (see below for a precise
definition). This reads: \be\label{DLAprun}
\varphi_A(\kk,y)\,\simeq\,\frac{1}{b_0
 N_c}\,\frac{Q_A^2}{k_{\perp}^2}\,
 \frac{{\rm e}^{\sqrt{4 b y \eta}}} {\sqrt{2\pi\sqrt{4 b y
\eta}}}\,,\qquad \eta\equiv\, \ln
\left[\frac{\ln\frac{\kk^2}{\Lambda^2}}{\ln
\frac{Q_A^2}{\Lambda^2}}\right] \,.\ee We expect this to be a good
approximation so long as  $\rho -\rho_s > \rho_s$, a condition
that in practice we shall use under the stronger form $\rho
-\rho_s> \tau_A$.

\subsubsection{BFKL regime}

The solution to the  BFKL equation with the running coupling
expanded as in Eq.~(\ref{alphaexp}) has been obtained in
\cite{MT02} (see also  Refs. \cite{DT02,MP03}) for the case of the
dipole scattering amplitude. In what follows we shall adapt this
solution to the case of the gluon distribution, explain its
properties, and carefully establish its validity limits (thus
going beyond the original discussion in Refs.
\cite{MT02,DT02,MP03})).

When applied to the gluon occupation factor, the solution in Refs.
\cite{MT02,DT02,MP03}) reads
\begin{equation}\label{BFKLrun}
    \varphi_A(\kk,y)\,=\,\frac{\kappa_0}{\alpha_s(k^2_{\perp}) N_c}\,
   \tau_A^{1/3}\, z^{-\gamma_s}\,{\rm Ai}
    \left (\xi_1 +\frac{\ln z + \Delta}{D\,\tau_A^{1/3}}\right)
    \exp \left( -\frac{2}{3\beta} \frac{\ln^2 z}{\tau_A} \right),
\end{equation}
with the following notations: $\kappa_0$ is a pure constant of
$\mathcal{O}(1)$, $\tau_A(y)$ has been defined in
Eq.~(\ref{Qsrun}), $\gamma_s=0.628$ is the saturation saddle
point, $z \equiv {\kk^2}/{Q_s^2(A,y)}$, and therefore $\ln z = \ln
({\kk^2}/{Q_s^2}) = \rho-\rho_s$. Furthermore,  ${\rm Ai}(x)$ is
the Airy function,  $\xi_1 = -2.33...$ is its first (rightmost)
zero, $\Delta$ is a constant that we are not able to determine,
    \be D=[\chi''(\gamma_s)/(2\,\chi(\gamma_s))]^{1/3} =
1.99...,\ee
 is the ``anomalous'' diffusion coefficient, and
$\beta=\chi''(\gamma_s)=48.5...$ is the ``standard'' one (however,
notice the 3/2 factor in the Gaussian in Eq.~(\ref{BFKLrun}),
instead of a 2 in Eq.~(\ref{BFKLfix})).
 Note that, in the present context, we find it more
natural to measure the transverse momenta with respect to
$\Lambda$, rather than with respect to the original saturation
scale (as we did for fixed coupling).

To the same approximation, the saturation momentum is given by
\begin{equation}\label{20satmom}
    Q_s^2(A,y)=\Lambda^2
    \exp \left(\tau_A - \frac{3 |\xi_1|}{4 D}\,\tau_A^{1/3} +\Delta
    \right).
\end{equation}
Since obtained after high energy approximations,
Eqs.~(\ref{BFKLrun}) and (\ref{20satmom}) apply only for $y$ large
enough, such that $b y/\rho_A \gtrsim 1$. (This is the usual
condition $\bar\alpha_s y \gtrsim 1$ for the validity of the BFKL
approximation, where $\bar\alpha_s$ is now evaluated at the
original saturation momentum $Q_A$.) As expected,
Eqs.~(\ref{20satmom}) and (\ref{Qsrun}) show the same dominant
behavior at large $y$.

The validity range of Eq.~(\ref{BFKLrun}) in $z$ is determined by
the various approximations performed in deriving this result
\cite{MT02}: First, there is the condition $\ln z \ll\tau_A$,
which has two sources: {\textsf (i)} the expansion of the BFKL
$\chi$ function around the saturation saddle--point\footnote{By
itself, this expansion could be justified up to $\ln z \sim
\tau_A$, because of the relative large value of the coefficient
$\beta$.} (the analog of the saddle--point approximation in the
fixed coupling case), and {\textsf(ii)} the expansion of the
running coupling around the saturation line, cf.
Eq.~(\ref{alphaexp}). Second, there is a more restrictive
condition, related to the further approximations performed when
solving the ``diffusion equation'' (Eq.~(74) in Ref. \cite{MT02})
obtained after expanding the $\chi$ function. This condition leads
to the following window for the applicability of
Eq.~(\ref{BFKLrun}) :
\begin{equation}\label{swindowrun}
   0 \,<\, \ln z \equiv
\rho-\rho_s\, \ll \frac{3\gamma_s D^2 }{\sqrt{5}}\,
\,\tau_A^{2/3}\,,
\end{equation}
to be referred as the `BFKL regime' in the presence of a running
coupling. This window includes several interesting kinematical
regimes, as we discuss now:

Using the expansion of the Airy function around its zero,
\begin{equation}\label{20Aismall}
    {\rm Ai}(\xi_1 + x)=
    {\rm Ai}'(\xi_1) \left( x + \frac{\xi_1}{6}\, x^3 +
    \cdot \cdot \cdot \right)
    \qquad {\rm for} \,\, x \ll 1,
\end{equation}
one can deduce an approximation for $\varphi$ valid near to the
saturation line, namely,
\begin{equation}\label{20phiscale}
    \varphi_A(\kk,y)\,\approx\, \frac{\kappa_0}{\alpha_s(k^2_{\perp}) N_c}\, \,
    z^{-\gamma_s} \left( \ln z + \Delta  \right),
\end{equation}
where trivial constants have been absorbed in the unknown
multiplicative factor $\kappa_0$. Except for the mild scaling
violations brought by the factor $1/\alpha_s(k_{\perp})$, this
expression is a scaling function, i.e., a function of
$z=\kk^2/Q_s^2(A,y)$. Eq.(\ref{20phiscale}) is valid so long as
the cubic term in (\ref{20Aismall}) is smaller than the linear
one; this condition implies $\kk\ll Q_g(A,y)$ where:
\begin{equation}\label{20geom}
    Q_g^2(A,y) \approx Q_s^2(A,y)\,
    \exp \left( \sqrt{\frac{6}{|\xi_1|}}\, {D\tau_A^{1/3}} \right).
\end{equation}
Thus, it is natural to choose $Q_g(A,y)$ as the upper boundary for
the region of geometric scaling in the present case. Note that,
with running coupling, this region is only a small part of the
BFKL regime defined by Eq.~(\ref{swindowrun}). Since $\tau_A(y)$
is monotonously increasing with $A$ and $y$, and approaches an
$A$--independent value when $y\to\infty$, it is clear that
\begin{equation}
     Q_g^2(A_1,y) > Q_g^2(A_2,y) \qquad {\rm for}\,\, A_1>A_2,
\end{equation}
and the two scales will approach each other at very large $y$.
Note also that the expression in Eq.(\ref{20phiscale}) is a
universal function, in the sense of Sect. \ref{BFKLFIX}.

Consider also the region outside the scaling domain, at $\ln z
\gtrsim\tau_A^{1/3}$. Using the asymptotic form of the Airy
function,
\begin{equation}\label{20Ailarge}
    {\rm Ai}(x)\,\approx\,
    \frac{1}{2 \sqrt{\pi}}\,
    \frac{1}{x^{1/4}}\,
    \exp \left( -\frac{2}{3}\,x^{3/2}  \right)
    \qquad {\rm for} \,\, x\gg 1,
\end{equation}
one immediately finds
\begin{equation}\label{20philarge}
    \varphi_A(\kk,y)\,\approx\,\frac{\kappa_0}{\alpha_s(k^2_{\perp}) N_c}\,
    \frac{\tau_A^{5/12}}{\ln^{1/4} z}
    \exp \left(
    -\gamma_s \ln z
    -\frac{2}{3 D^{3/2}} \frac{\ln^{3/2} z}{\tau_A^{1/2}}
    -\frac{2}{3 \beta} \frac{\ln^2 z}{\tau_A} \right),
\end{equation}
where $\kappa_0$ has been redefined in order to absorb some
trivial constants, and $\Delta$ has been dropped since it is small
compared to $\ln z$ in the region of interest. Note the appearance
of  diffusion terms of two types --- from the Gaussian
Eq.~(\ref{BFKLrun}), and from the asymptotic expansion of the Airy
function ---, which both lead to {\sf scaling violations}. There
is also a mild violation from the prefactors.

Moreover, at variance with the fixed coupling case (cf.
Eq.~(\ref{BFKLfix})), the scaling violations in
Eq.~(\ref{20philarge}) are synonymous of {\sf universality
violations} : indeed, the ``diffusion time" $\tau_A(y)$ (which
replaces the $\bar\alpha_s y$ of the fixed coupling case) is now
explicitly dependent on $A$. But for sufficiently large $y$, the
$A$--dependence of $\tau_A(y)$ becomes very weak, as we have seen,
and universality gets restored when $ 2c b y \gg \rho_A^2$. For
even larger $y$, such that $ 2c b y \gg \rho_A^4$, the saturation
momentum becomes independent of $A$, cf. Eq.~(\ref{QsrunAp}), so
the gluon occupation factor --- which in this regime is a
universal function of $z$ and $y$ --- is then the same in the
proton and in the nucleus for any $\kk$ in the BFKL regime.

\section{High--$k_\perp$ suppression from quantum evolution:
The general argument}\label{general}\setcounter{equation}{0}

Before we embark ourselves in a systematic analysis of the ratio
${\mathcal R}_{pA}(\kk,y)$, let us present here a general argument
explaining why the quantum evolution leads to a suppression in
this ratio at generic momenta. The argument can be succinctly
formulated as follows\footnote{We would like to thank Al Mueller
for helping us clarifying this general argument.}: {\sf The
suppression of the ratio ${\mathcal R}_{pA}(\kk,y)$ with
increasing $y$ is due to the different evolution rates for the
gluon distribution in the nucleus and in the proton: the proton
distribution evolves faster because (a) the corresponding
saturation momentum in the initial conditions is much smaller
($Q_p\ll Q_s(A)$), and (b) the convexity of the function
$\chi(\gamma)$ (the eigenvalue of the BFKL kernel) accelerates the
evolution with increasing $\rho$.} In particular, the difference
between the proton and the nuclear evolution rates is particularly
pronounced at small $y$ (when the proton is in the DLA regime for
the interesting values of $\kk$), which explains the rapid
suppression seen in the early stages of the evolution
\cite{Nestor03}.

Moreover, similar arguments will allow us to conclude that, after
some initial evolution $\bar\alpha_s y\simge 1$, and for
transverse momenta in the proton perturbative region ($\kk\gg
Q_s(p,y)$), {\sf the ratio ${\mathcal R}_{pA}(\kk,y)$ is
monotonously increasing with $\kk$, and a decreasing function of
$A$}. The monotonic behavior with $\kk$ shows that the Cronin peak
has already disappeared for such values of $y$. As for the
$A$--dependence, this can be related to the experimental
observation that the central--to--peripheral ratio $R_{CP}$ for
(pseudo)rapidity $\eta>1$ is smaller for central collisions than
for more peripheral ones
\cite{Brahms-data,RHIC-dAu-forward,RHIC-dAu-forward-other}.
(Peripheral collisions involve less nucleons than the central
ones, thus effectively corresponding to smaller values of $A$.)
This tendency is opposite to that identified at $y=0$ in our
theoretical calculations (cf.
Eqs.~(\ref{RmaxMVf})--(\ref{RmaxMVr})), and also at $\eta=0$ in
the experimental results at RHIC
\cite{Brahms-data,RHIC-dAu-forward,RHIC-dAu-forward-other}, where
the amplitude of the Cronin peak appears to be increasing with
$A$.

In this context, it is interesting to notice that the quantity
$\alpha$  (the logarithmic slope in the parametrization
$R_{CP}\sim e^{\alpha\eta}$) which has been measured at BRAHMS
\cite{Brahms-data} corresponds in our approach to the right hand
side of either Eq.~(\ref{dRdyfin}), or Eq.~(\ref{dRdysat}), for
the nucleus in the linear regime, or in the saturation regime,
respectively. With this interpretation, $\alpha$ is dominated by
the linear evolution of the {\sf proton}, and is not the same as
the saturation exponent (in contrast to a suggestion in Ref.
\cite{Brahms-data}).

The general arguments to be presented in this section apply for
the case of fixed coupling alone: for a running coupling, we have
used different approximations in the BFKL and DLA regimes, so,
based on these approximations, one cannot construct a general
argument valid in the whole perturbative region. But in Sect.
\ref{HIGHPT} we shall check through explicit calculations that all
the conclusions about the general behavior of the ratio ${\mathcal
R}_{pA}(\kk,y)$ to be established below in this section remain
true for a running coupling.

\subsection{A general argument on the $y$--dependence}

The $y$--dependence of the ratio in Eq.~(\ref{Rdef}) is encoded
in: \beq\label{dRdy} \frac{d\ln {\mathcal
R}_{pA}}{dy}\,=\,\frac{d\ln \varphi_A}{dy} \,
-\,\frac{d\ln\varphi_p}{dy}\,.\eeq
 We shall study separately the
linear and the saturation regimes for the nucleus. The proton will
be always in the linear regime, as described by the BFKL solution
in the saddle point approximation, Eq.~(\ref{SPA}).

\subsubsection{Nucleus in the linear regime ($\kk \gg Q_s(A,y)$)}

In this regime, both the proton and the nucleus are described by
Eq.~(\ref{SPA}), which implies: \be \label{lnphi}
\ln\varphi_A(\kk,y)\,\approx\, \bar\alpha_s y F(\gamma_0({\rm
r}_A),{\rm r}_A)\,,\ee where we have neglected the slowly varying
prefactors. Here, we have introduced the more specific notation:
\be {\rm r}_A(\kk, y) \,\equiv\, \frac{\rho(A,\kk)}{\bar\alpha_s
y}\,=\,\frac{\ln\kk^2/Q_s^2(A)} {\bar\alpha_s y}\,,\ee (with
$Q_s(A)\rightarrow Q_p$ for the proton) to emphasize that
quantities like $\rho\equiv \rho(A,\kk)$ and ${\rm r}_A$ are
different for the proton and the nucleus, even when considered for
the same values of the kinematical variables $\kk$ and $y$. Since
$Q_s(A) \gg Q_p$ then, clearly, \be\label{ineqpA}
\rho(A,\kk)\,<\,\rho(p,\kk),\qquad{\rm and}\qquad
{\rm r}_A(\kk, y) < {\rm r}_p(\kk, y)\,.\ee After taking a total
derivative w.r.t. $y$ in Eq.~(\ref{lnphi}), and using the saddle
point condition, Eq.~(\ref{saddle}), together with the definition
(\ref{RFdef}) of the function $F$, one finds : \be \frac{d\ln
\varphi_A}{dy} \,\approx\,\bar\alpha_s\chi (\gamma_A)\,,\ee (with
the simplified notation $\gamma_A\equiv \gamma_0({\rm r}_A)$) and
therefore: \beq\label{dRdyfin} \frac{1}{\bar\alpha_s} \frac{d\ln
{\mathcal R}_{pA}}{dy}\,\approx\,\chi (\gamma_A)
\,-\,\chi(\gamma_p)\,,\ee which is always {\sf negative} (for
$\kk$ and $y$ within its range of validity), as we show now: Since
$\chi(\gamma)$ is a convex function ($\chi''(\gamma) > 0$ for $0 <
\gamma <1$), the function $\gamma_0({\rm r})$ which gives the
saddle point is monotonously increasing: \be \frac{d \gamma_0({\rm
r})}{d{\rm r}}\,=\,\frac{1} {\chi''(\gamma_0({\rm
r}))}\,>\,0\,,\ee (this follows by differentiating the saddle
point condition (\ref{saddle}) w.r.t. ${\rm r}$), which together
with Eq.~(\ref{ineqpA}) implies that $\gamma_A < \gamma_p$, from
which Eq.~(\ref{dRdyfin}) finally follows because $\chi(\gamma)$
is monotonously increasing for any $\gamma > 1/2$ (recall that
$\gamma_A\ge \gamma_s \approx 0.63$). Some of these properties are
manifest on Fig. \ref{CHI}.

Moreover, the suppression rate (\ref{dRdyfin}) is largest in the
early stages of the evolution, but decreases with $y$, and
approaches zero (through negative values) when $y\to\infty$.
Indeed, the difference: \be  {\rm r}_p(\kk, y) - {\rm r}_A(\kk, y)
\,=\, \frac{\rho(p,\kk)-\rho(A,\kk)}{\bar\alpha_s y}\,=\,
\frac{\ln(Q_s^2(A)/Q_p^2)}{\bar\alpha_s y}\,\approx\,
\frac{\rho_A}{\bar\alpha_s y}\,,\ee is independent of $\kk$ and
largest at small $y$, but it decreases with $y$, and vanishes
asymptotically at very large $y$. Therefore, when $y\to\infty$, we
have $\gamma_A - \gamma_p\to 0$, and thus also $\chi (\gamma_A)
-\chi(\gamma_p) \to 0$.

To be more specific, consider two interesting physical regimes:

a) For very large $\kk \gg Q_g(A,y)$, both the proton and the
nucleus are in the DLA regime, where ${\rm r} \gg 1$,
$\gamma_0({\rm r})\approx 1 - 1/\sqrt{{\rm r}}$, $\chi(\gamma_0)
\approx \sqrt{{\rm r}}$, and therefore: \beq\label{dRdyDLA}
\frac{1}{\bar\alpha_s} \frac{d\ln {\mathcal
R}_{pA}}{dy}\,\approx\,\sqrt{\frac{\rho(A,\kk)}{\bar\alpha_s
y}}\,-\,\sqrt{\frac{\rho(p,\kk)}{\bar\alpha_s y}}\,,\ee which
confirms that the rate of variation is largest in the early stages
of the evolution (although Eq.~(\ref{dRdyDLA}) cannot be trusted,
strictly speaking, when $y\to 0$).

b) For $\kk$ and $y$ such that both the nucleus and the proton are
in the BFKL regime (which becomes possible when $y$ is large
enough for $Q_g(p,y) > Q_s(A,y)$ ; see Fig. \ref{EVOL-MAP}), then
one can expand both $\chi(\gamma_A)$ and $\chi(\gamma_p)$ around
$\gamma_s \approx 0.63$. This is entirely similar to the expansion
leading from Eq.~(\ref{SPA}) to Eq.~(\ref{BFKLfix}), and gives:
\beq\label{dRdyBFKL} \frac{1}{\bar\alpha_s} \frac{d\ln {\mathcal
R}_{pA}}{dy}\,\approx\,-\,\frac{c\rho_A}{\beta \bar\alpha_s
y}\left\{1 + \frac{1}{c\bar\alpha_s y}\left(
\ln\frac{\kk^2}{Q_s^2(A,y)} + \frac{\rho_A}{2}\right)
\right\},\eeq which shows that, when increasing $y$ and $\kk$ in
such a way that $\ln (\kk^2/Q_s^2(A,y)) \sim $ const., the rate
eventually becomes independent of $\kk$, and vanishes as $1/y$.

\subsubsection{Nucleus in the saturation regime ($\kk \le
Q_s(A,y)$)}

All we need to know about this regime for the present purposes is
that the nuclear gluon distribution at saturation is only a
function of $z\equiv \kk^2/Q_s^2(A,y)$. To remain in the
saturation region when increasing $y$, one must simultaneously
increase $\kk$ along a line parallel to the saturation line, such
that $z$ remains fixed. Proceeding as before, we have:
 \be \label{lnR}
\ln {\mathcal R}_{pA}(z,y)\,\approx\,\ln\varphi_A(z) -
\bar\alpha_s y \chi(\gamma_p) + \gamma_p\ln\frac{z Q_s^2(A,y)}
{Q_p^2}\, ,\ee where the proton saddle point $\gamma_p$ is now
determined by (cf. Eq.~(\ref{saddle})) : \be \label{saddle1}
\chi'(\gamma_p) = \frac{\ln[{z Q_s^2(A,y)}/
{Q_p^2}\,]}{\bar\alpha_s y}\,=\, \frac{\ln[{z Q_s^2(A)}/
{Q_p^2}\,]} {\bar\alpha_s y} \,+\,\chi'(\gamma_s).\ee In the above
equation, the second equality follows after using
Eq.~(\ref{Qsfix0}) for the saturation momentum and recalling that
$c=\chi'(\gamma_s)\approx 4.88$ with $\gamma_s \approx 0.63$.
Taking a derivative w.r.t. $y$ in Eq.~(\ref{lnR}) and using
Eq.~(\ref{saddle1}) yields:
 \beq\label{dRdysat}
\frac{1}{\bar\alpha_s} \frac{d\ln {\mathcal
R}_{pA}}{dy}\bigg\vert_{z}\,\approx\,\gamma_p\chi'(\gamma_s) -
\chi(\gamma_p)\,, \ee which is negative for any $y$, as is quite
obvious by inspection of Fig. \ref{CHI} : The straightline
$f(\gamma_p) \equiv \gamma_p\chi'(\gamma_s)$ is the line which is
tangent to $\chi(\gamma_p)$ at $\gamma_p=\gamma_s$; but
$\chi(\gamma_p)$ is a convex function, and thus it lies above its
tangent for any
$\gamma_p\ne\gamma_s$. 
Moreover, the suppression rate is very large  at small $y$ (since
$\chi(\gamma_p) \to \infty$ as $\gamma_p\to 1$), but it decreases
to zero when $y\to\infty$ (recall that $\gamma_s\chi'(\gamma_s) =
\chi(\gamma_s)$). We conclude that the ratio ${\mathcal
R}_{pA}(z,y)$ is rapidly decreasing with $y$ in the early stages
of the evolution, and approaches a constant value for very large
$y$.

\subsection{A general argument on the $\kk$--dependence}

We now present a mathematically similar argument which allows us
to characterize the $\kk$ dependence of the ratio ${\mathcal
R}_{pA}(\kk,y)$. We would like to show that, for a fixed rapidity
$y$ with $\bar\alpha_s y \simge 1$ (such that the saddle point
approximation (\ref{SPA}) applies), and for all momenta $\kk \gg
Q_s(p,y)$ (where the proton is in a linear regime), the function
${\mathcal R}_{pA}(\kk,y)$ is monotonously increasing with $\kk$.

Consider first the linear regime for the nucleus, at $\kk \gg
Q_s(A,y)$. Using Eq.~(\ref{lnphi}) and the saddle point condition,
Eq.~(\ref{saddle}), one obtains (e.g., for the proton),
\be\label{dphip} \frac{d\ln
\varphi_p}{d\ln(\kk^2/\Lambda^2)}\,\approx\,-\gamma_p\,,\ee
together with a similar formula for the nucleus, and therefore:
 \beq\label{dRdkfin} \frac{d\ln
{\mathcal R}_{pA}}{d\ln(\kk^2/\Lambda^2)}\,\approx\,\gamma_p
\,-\,\gamma_A\,,\ee which by the previous arguments is strictly
positive and vanishes asymptotically when $\kk\to\infty$. Thus, in
this regime, the function ${\mathcal R}_{pA}(\kk,y)$ is strictly
increasing with $\kk$ for any fixed $y$, and saturates to a
constant value when $\kk\to\infty$. As it will be checked in Sect.
\ref{HIGHPT}, this limiting value is one.

Consider also the case where the nucleus is deeply at saturation
($\kk \ll Q_s(A,y)$). Then,  $\varphi_A$ is only logarithmically
decreasing with $z\equiv \kk^2/Q_s^2(A,y)$ (see
Eq.~(\ref{phiAsat})), while $\varphi_p$ has a faster, power--like,
decrease, with ``anomalous dimension'' $\gamma_p \ge \gamma_s
\simeq 0.63$ (cf. Eq.~(\ref{dphip})). Therefore, in this range
too, the function  ${\mathcal R}_{pA}(\kk,y)$ is increasing with
$\kk$.

The previous analysis does not cover the nuclear transition region
at $\kk \sim Q_s(A,y)$. But the discussion following
Eq.~(\ref{BFKLp0}) shows that, when $\bar\alpha_s y\simge 1$,
there is no large (i.e., parametrically enhanced) mismatch between
the value of $\varphi_A$ at $\kk \sim Q_s(A,y)$,
Eq.~(\ref{SATcond}), and the corresponding extrapolation of the
BFKL solution (\ref{BFKLp0}):
the original mismatch between $\varphi_A^{\rm sat}$ and
$\varphi_A^{\rm twist}$  at $\kk \sim Q_s(A)$ (cf. Sect.
\ref{MVmodel}) has been washed out by the evolution. This
property, together with the above finding that $\varphi_A$ is
monotonously increasing with $\kk$ on both sides of the saturation
line, precludes the existence of a (parametrically enhanced) peak
at $\kk \sim Q_s(A,y)$. That is, the peak must have already
disappeared when $\bar\alpha_s y\simge 1$. The mechanism for this
disappearance will be clarified in Sect. \ref{FLAT}.

\subsection{A general argument on the $A$--dependence}

Finally, let us use similar arguments to study the dependence of
the ratio ${\mathcal R}_{pA}$ upon the atomic number $A$. We shall
find that, after an evolution\footnote{This is a conservative
estimate, necessary for the validity of the general arguments in
this section. But the explicit calculations in Sect. \ref{CR_DLAF}
will reveal that, for $\kk\sim Q_s(A,y)$ at least, ${\mathcal
R}_{pA}$ starts to be a decreasing function of $A$ already after
the shorter evolution $\bar\alpha_s y\sim 1/\rho_A$.}
$\bar\alpha_s y\simge 1$, the ratio becomes a decreasing function
of $A$. We shall consider the $A$--dependence for two cases: (a)
with fixed $\kk$ and $y$ (this corresponds to the real
experimental situation for the measurement of $R_{CP}$ at RHIC)
and (b) with fixed $z\equiv \kk^2/Q_s^2(A,y)$ and $y$ (as
relevant, e.g., for the $A$--dependence of the Cronin peak).

Notice first that the dominant $A$--dependence enters the gluon
occupation factor through the ratio between $\kk^2$ and the
saturation momentum (see, e.g., Eqs.~(\ref{phiAsat}) and
(\ref{SPA}); the factor $\widetilde{\varphi_A}$ in the latter
brings only a subleading, logarithmic, dependence on $A$, whose
contribution to the derivative in Eq.~(\ref{derA}) below is
suppressed by a factor $1/\rho_A$).

(a) With fixed $\kk$ and $y$, $\varphi_p(\kk,y)$ is clearly
independent of $A$, and one can rely on Eq.~(\ref{SPA}) (nucleus
in the linear regime) to write: \BQA\label{derA}
\left.\frac{\del}{\del A}\ln {\mathcal R}_{pA}\right\vert_{\kk,y}
&=&\frac{\del \ln \varphi_A}{\del A}-\frac{\del \ln A^{1/3}}{\del
A} \,=\,\frac{\del \rho}{\del A} \frac{\del \ln \varphi_A}{\del
\rho} -\frac{\del \ln A^{1/3}}{\del A}\NN &\simeq &-\frac{\del
\rho_A}{\del A} (-\gamma_A)-\frac{\del \rho_A}{\del A}
\,=\,-\frac{\del \rho_A}{\del A} (1-\gamma_A)\, < \, 0, \EQA where
we have also used $\rho(A,\kk) \equiv \ln \kk^2/Q_s^2(A) = \ln
\kk^2/\Lambda^2 - \rho_A$ and the relation (\ref{dphip}) written
for the nucleus. The same conclusion holds when the nucleus is
deeply at saturation, since there $\varphi_A$ is only
logarithmically varying with $\kk^2/Q_s^2(A,y)$ (and thus with
$A$). Finally, for $\kk$ around $Q_s(A,y)$ one can rely on the
same matching argument as in the previous discussion of the
$\kk$--dependence to conclude that the ratio  ${\mathcal R}_{pA}$
preserves a monotonic, decreasing, behavior with $A$.

(b) Consider now the situation in which $A$ and $\kk$ are
increased simultaneously, in such a way that $z\equiv
\kk^2/Q_s^2(A,y)$ stays fixed. Then $\rho (A,\kk)\equiv \ln
\kk^2/Q_s^2(A)$ is fixed as well, and the gluon distributions of
the nucleus (seen now as a function of $z$ and $y$) has no extra
dependence on $A$. On the other hand, the distribution of the
proton becomes now dependent on $A$, through $\kk$: Writing the
proton $\rho$--variable as \BQ \rho (p,\kk)\equiv \ln
\frac{\kk^2}{Q_p^2} = \ln z + \ln \frac{Q_s^2(A,y)}{Q_p^2} \simeq
\ln z + c\bar\alpha_s y + \rho_A, \EQ where the last term carries
the dependence on $A$, and using  Eq.~(\ref{dphip}), one deduces
that \BQ \left.\frac{\del \ln \varphi_p}{\del A}\right\vert_{z,y}
=-\gamma_p \,\frac{\del \rho_A}{\del A}\,. \EQ This yields the
expected conclusion that  ${\mathcal R}_{pA}$ is decreasing with
$A$ : \BQ \left.\frac{\del}{\del A}\ln {\mathcal
R}_{pA}\right\vert_{z,y} =-\frac{\del \rho_A}{\del A} (1-\gamma_p)
< 0. \EQ  Note also that the rate of decrease in ${\mathcal
R}_{pA}$ is faster when measured with $\kk$ and $y$ fixed than
with $z$ and $y$ fixed (because $1-\gamma_A>1-\gamma_p$).

\section{The evolution of the Cronin peak with increasing $y$}
 \label{CRevolve}
\setcounter{equation}{0}

The problem of the evolution of the Cronin peak with increasing
$y$ is a delicate one, as it involves the nuclear transition
region towards saturation, on which we have little analytic
control at $y>0$. Moreover, the experience with the MV model
suggests that the existence of the peak and its properties are
tributary to the actual behavior of the nuclear gluon distribution
$\varphi_A(\kk,y)$ at momenta just above the saturation scale: If
the distribution is rapidly decreasing with $\kk$ (say, according
to an exponential law), then a pronounced peak exists; but if its
decrease is only power--like, then there is at most a very flat
peak, if any. In the MV model, the gluon distribution above
$Q_s(A)$ has been found to be the superposition between an
exponential and a power--law tail, with the exponential being the
dominant contribution though, since parametrically enhanced at
large $A$. As we shall see below in this section, the effect of
the evolution is to enhance the power--law contributions, which
for $\alpha_s y\sim 1$ become as large as the exponential one (for
momenta just above $Q_s(A)$). When this happens, the Cronin peak
has completely flattened out.

In order to study this flattening, one needs a more accurate
calculation which follows the non--linear evolution of the nucleus
in the saturation region. To achieve the necessary accuracy while
still preserving an analytic control, we shall perform just one
step in the evolution of the MV model according  to the
non--linear Kovchegov equation \cite{K}. Strictly speaking, this
calculation applies only for rapidities $\alpha_s y\ll 1$, but as
we shall see in Sect. \ref{FLAT}, this is enough to reveal the
mechanism responsible for the flattening of the peak. By
extrapolation, we shall then conclude that the peak disappears
after an evolution $\alpha_s y\sim 1$.

Whereas the flattening of the peak is a rather subtle effect which
has to do with the nuclear evolution, the rapid decrease in the
height of the peak, on the other hand, is a more robust
phenomenon, and also easier to calculate, since this is due solely
to the perturbative evolution of the proton. Because of that, we
shall begin our analysis in this section with a study of the
evolution of the magnitude of the peak with increasing $y$. Since
we expect the peak to follow the nuclear saturation momentum, we
shall consider the evolution of the ratio ${\mathcal
R}_{pA}(\kk,y)$ along the nuclear saturation line $\kk=Q_s(A,y)$.
This is possible within the present approximation because, along
this particular line, the nuclear gluon distribution {\sf is}
known,  cf. Eq.~(\ref{SATcond}).
By itself, this calculation cannot tell us whether a peak actually
exists or not; but so long as the peak exists, it provides us with
a correct estimate for the magnitude of the peak and its
parametric dependencies upon $A$ and $y$. Although, as we shall
argue later, the peak disappears already when $\alpha_s y\sim 1$,
in Sects. \ref{CR_DLAF} and \ref{CR_RUN} we shall follow the
evolution along the nuclear saturation line up to very large $y$.
This will reveal the basic features of the high--$\kk$
suppression, to be more systematically analyzed in Sect.
\ref{HIGHPT}.

\subsection{The suppression of the peak: Fixed coupling}
\label{CR_DLAF}

In this subsection and the following one, we shall need only the
(proton and nuclear) gluon distributions along the nuclear
saturation line $\kk=Q_s(A,y)$. For the nucleus, this is simply a
constant times $1/\alpha_s$, cf. Eq.~(\ref{SATcond}). For the
proton, the corresponding distribution is in a linear regime,
which can be either DLA, or BFKL, depending upon the value of $y$:
With increasing $y$, the proton geometric scale $Q_g(p,y_c)$ rises
faster than the nuclear saturation momentum, so the corresponding
evolution lines in the plane $y-\ln\kk^2$ cross each other at some
rapidity $y_c$ (see Fig. \ref{EVOL-MAP}), where the proton changes
from the DLA to the BFKL regime. The condition $Q_s(A,y_c) =
Q_g(p,y_c)$ together with Eqs.~(\ref{Qsfix0}) and (\ref{swindow})
imply :
 \be\label{ycfixed} c\bar\alpha_s y_c
\,\simeq\,\frac{1}{\nu}\,\ln\frac{Q_s^2(A)}{Q_p^2}\,\simeq\,
\frac{\rho_A}{\nu}\,.\ee

\subsubsection{$y < y_c$ : Proton in the DLA regime}

For $y < y_c$, but such that $\bar\alpha_s y \rho > 1$, one can
use Eq.~(\ref{DLAp0}) with $Q_s(A)\rightarrow
Q_p$ to deduce :
\be\label{RmaxDLAfix} {\mathcal R}_{\rm sat}(A,y)\equiv {\mathcal
R}_{pA}(\kk=Q_s(A,y),y) &\sim& \frac{Q_s^2(A,y)}{A^{1/3}Q_p^2}
\,\,{\rm e}^{-\sqrt{4\bar\alpha_s y \rho}} \nn &\sim& \rho_A \,\,
{\rm exp}\Big\{c\bar\alpha_s y-\sqrt{4\bar\alpha_s y
(\rho_A+c\bar\alpha_s y)}\,\Big\}, \ee where we have used
Eqs.~(\ref{QsatMV}), (\ref{A13}) and (\ref{Qsfix0}) to write:
 \be \frac{Q_s^2(A,y)}{A^{1/3}Q_p^2}\,= \, \frac{Q_s^2(A,y)}{Q_A^2}
 \,= \,{\rm e}^{c\bar\alpha_s y}\,\frac{Q_s^2(A)}{Q_A^2}
\,=\, {\rm e}^{c\bar\alpha_s y}\,{\rho_A}\,,\ee and also (recall
that $Q_p^2 \sim \Lambda^2$) :
 \be\label{rhofix0}
\rho\,\equiv\,\ln\frac{Q_s^2(A,y)}{Q_p^2}\,= \, c\bar\alpha_s y +
\ln\frac{Q_s^2(A)}{Q_p^2}\simeq \,  c\bar\alpha_s y + \rho_A.\ee
Note that, in writing Eq.~(\ref{RmaxDLAfix}), we have ignored the
slowly varying prefactor in Eq.~(\ref{DLAp0}), since presently we
are only interested in the dominant parametric dependencies upon
$A$ and $y$. When $y\to 0$, Eq.~(\ref{RmaxDLAfix}) is formally
consistent with the corresponding result in the MV model,
Eq.~(\ref{RAfixed}), but one should keep in mind that the results
obtained here cannot be used for very small values of $y$.

Eq.~(\ref{RmaxDLAfix}) offers, in particular, an estimate for the
magnitude of the Cronin peak; indeed, for as long as it survives,
the peak should be located in the vicinity of $Q_s(A,y)$. Note
that, for any $0 < y \le y_c$, the expression within the exponent
in Eq.~(\ref{RmaxDLAfix}) is negative, showing that the magnitude
of the peak is rapidly decreasing with $y$. Since the initial
maximum at $y=0$ was relatively large (${\mathcal R}_{\rm max}(A)
\sim \rho_A$, cf. Eq.~(\ref{RmaxMVf})), it is interesting to check
how fast is the height of the peak decreasing to a value which is
parametrically of order one. The condition ${\mathcal R}_{\rm
sat}(A,y) \sim 1$ for $y\sim y_0$ implies:
 \be\label{Ry0} \bar\alpha_s y_0\,\sim\,\frac{\ln^2
\rho_A}{4\rho_A}\,\sim\, \frac{\big(\ln  \ln A^{1/3}\big)^2}{\ln
A^{1/3}}\,\ll\,\bar\alpha_s y_c\,.\ee This is a small rapidity
interval, but is still within the reach of the saddle point
approximation (\ref{DLAp0}), since $\rho_A \bar\alpha_s y_0 \sim
\ln^2 \rho_A > 1$. In fact, for such a short evolution in $y$, one
can ignore the evolution of the nucleus in the vicinity of the
peak, i.e., one can neglect $c\bar\alpha_s y \ll 1$ next to
$\rho_A$ within Eq.~(\ref{RmaxDLAfix}). This is reassuring since,
when $\alpha_s y \ll 1$, one cannot really trust the BFKL
approximation (\ref{Qsfix0}) for the saturation scale of the
nucleus.

In view of this, it is possible to obtain a more accurate estimate
for the rapidity $y_0$ by using the MV model for the gluon
distribution of the nucleus together with the complete DLA
expression (\ref{DLAp0}) for the proton distribution,  including
the prefactor. This yields: \be\label{y0fix} \bar\alpha_s
y_0\,=\,\frac{1}{4\rho_A}\,
\ln^2\Big[a\rho_A\sqrt{\ln(a\rho_A)}\Big],\qquad a\equiv
\sqrt{2\pi}z_0\Gamma(0,z_0)\simeq 0.706\,.\ee The corrections to
this result (coming  either from the evolution of the nucleus, cf.
Sect. \ref{FLAT}, or from the twist terms in the MV model, that we
have neglected here) are of $\mathcal{O}(\rho_A^{-2})$.

Eq.~(\ref{RmaxDLAfix}) also shows that, after a short rapidity
evolution $\bar\alpha_s y \sim 1/\rho_A$, the magnitude of the
Cronin peak becomes a {\sf decreasing} function of $\rho_A$ (and
thus of $A$), in sharp contrast with the corresponding behavior at
$y=0$, cf. Eq.~(\ref{RmaxMVf}). This behavior is in agreement with
the general arguments about $A$--dependence given in Sect.
\ref{general}.

As anticipated,  the rapid decrease in the height of the Cronin
peak during the early stages of the evolution is to be attributed
solely to the fast evolution of the proton. For larger rapidities,
$y \gg y_0$, the evolution of the nucleus starts to matter as well
(and Eq.~(\ref{Qsfix0}) can be trusted), but as long as $y \le
y_c$ the proton evolution is still faster, so ${\mathcal R}_{\rm
sat}$ keeps decreasing with $y$, as manifest on
Eq.~(\ref{RmaxDLAfix}). In particular, for $y=y_c$ one has:
\be\label{RmaxDLAyc} {\mathcal R}_{\rm sat}(A,y_c)\,\sim
\rho_A\,{\rm e}^{-\kappa\rho_A} \,\sim\,\frac{\ln
A^{1/3}}{A^{\kappa/3}}\,\ll 1,\quad \kappa
=\frac{1}{\nu}\left(\sqrt{4(1+\nu)/c} - 1\right) \approx 0.29\,.
\ee
For even larger $y$, 
Eq.~(\ref{RmaxDLAfix}) would predict an {\sf increase} of the peak
with $y$, but this is incorrect, since for $y>y_c$ this equation
does not apply any longer.

\subsubsection{$y>y_c$ : Proton in the BFKL regime}

For $y>y_c$, the proton enters the scaling window (\ref{swindow}),
where Eq.~(\ref{BFKLp0}) becomes appropriate. We shall shortly
argue that, when this happens, the Cronin peak has already
disappeared; but it is still interesting to follow the ratio
${\mathcal R}_{pA}(\kk,y)$ further up along the nuclear saturation
line. By using Eq.~(\ref{BFKLp0}) with $Q_s(A,y)\rightarrow
Q_s(p,y)$, together with the relations:
 \be\label{QSAPY}\frac{Q_s^2(A,y)}{Q_s^2(p,y)}\,= \, \frac{Q_s^2(A)}{Q_p^2}
 \,= \,A^{1/3}\,{\rho_A}\,,\qquad \ln(A^{1/3}\rho_A) \simeq \rho_A\,,
 \ee one finds (with the simpler notation $\gamma\equiv\gamma_s$)
: \be\label{RmaxBFKLfix} {\mathcal R}_{\rm
sat}(A,y)\,\sim\,\frac{1}{A^{\frac{1-\gamma}{3}}}
\,\frac{1}{\rho_A^{1-\gamma}}\,\exp\left\{\frac{\rho_A^2}{2\beta
\bar\alpha_s y} \right\}\,\,\ll\,1.\ee Note the emergence of the
power of $A$ in the denominator, which provides a strong
suppression factor which is independent of $y$. This power was
missing in the DLA (compare to Eq.~(\ref{RmaxDLAfix})), but
appears here as a consequence of the `anomalous dimension' $\gamma
< 1$ characteristic of the BFKL solution in the vicinity of the
saturation line \cite{SCALING,MT02}.

Note furthermore that Eq.~(\ref{RmaxBFKLfix}) shows only a weak
dependence on $y$, coming from the ``diffusion term" in
Eq.~(\ref{BFKLp0}): the dominant dependencies, which are
exponential, have cancelled in the ratio between the nuclear and
the proton saturation momenta, cf. Eq.~(\ref{QSAPY}). Thus, as
compared to the DLA regime at $y < y_c$, where ${\mathcal R}_{\rm
sat}(A,y)$ is rapidly decreasing with $y$, in the BFKL regime at
$y > y_c$ this ratio almost stabilizes at a very small value,
proportional to an inverse power of $A$.

In particular, when approaching $y_c$ from the above, one finds:
\be\label{RmaxBFKLyc} {\mathcal R}_{\rm
sat}(A,y_c)\,\sim\,\frac{1}{\big( A^{1/3}
\rho_A\big)^{\kappa}}\,\sim\,\frac{1}{\big( A^{1/3} \ln
A^{1/3}\big)^{\kappa}},\ee where $\kappa$ has now been generated
as $\kappa=1-\gamma -\nu c/2\beta$, but this is the same power as
in Eq.~(\ref{RmaxDLAyc}), by construction (because of our matching
between the DLA and BFKL approximations along the geometric
scaling line). The subleading dependencies on $A$, which are
logarithmic, are different in Eqs.~(\ref{RmaxDLAyc}) and
(\ref{RmaxBFKLyc}), but this difference merely reflects our
imperfect matching.

Furthermore, for very large $y$, such that $2\beta\bar\alpha_s
y\gg \rho_A^2$ (in practice, this limit is approached very fast,
since $2\beta \approx 97$ is a large number), the ratio becomes
independent of $y$ : \be\label{RmaxBFKLyy} {\mathcal R}_{\rm
sat}(A,y) \,\sim\,\frac{1}{(A^{{1}/{3}}\rho_A)^{1-\gamma}}\,
\sim\,\frac{1}{(A^{{1}/{3}}\ln A^{1/3})^{1-\gamma}}\qquad{\rm
for}\qquad 2\beta\bar\alpha_s y\gg \rho_A^2.\ee

To summarize, the inverse power of $A$ which characterizes the
suppression increases from $\kappa\approx 0.29$ at the end of the
DLA regime to $1-\gamma\approx 0.37$ at asymptotically large $y$,
which shows that a large fraction of the suppression is actually
built during the DLA phase of the evolution.

\subsection{The suppression of the peak: Running coupling}
\label{CR_RUN}

With  running coupling, the kinematical domains in which our `DLA'
and `BFKL' approximations apply are well separated from each
other, and the intermediate region between the two regimes is not
under control. From Sect. \ref{EV:RUN}, we recall that for very
large $\kk$, such that $\ln z > \tau_p(y)$ (with $z \equiv
\kk^2/Q_s^2(p,y)$), the proton is described by the DLA
approximation (\ref{DLAprun}), while for intermediate $\kk$,
within the range $0<\ln z < {\rm const}\times \tau_p^{2/3}(y)$
(cf. Eq.~(\ref{swindowrun})), the BFKL approximation
(\ref{BFKLrun}) applies. Thus, when the momentum is taken along
the nuclear saturation line ($\kk = Q_s(A,y)$, or $\ln z \simeq
\tau_A(y)- \tau_p(y)$), the proton is in the DLA regime for $y <
y_c$, and in the BFKL regime for $y\simge y_c'$. Here, $2cby_c
\approx \rho_A^2$, and $y_c'$ is determined from the condition
$\tau_A(y) - \tau_p(y) \approx {\rm const}\times \tau_p^{2/3}(y)$
as:
\begin{equation}\label{20ycdp}
    2 c b y_c' \approx
    \left( \frac{\sqrt{5}}{6 \gamma_s D^2} \right)^{6/5}
    \rho_A^{12/5},
\end{equation}
where the constant factor in the r.h.s. has been obtained
according to Eq.~(\ref{swindowrun}). For even larger $y \ge
y_c''$, the proton enters its geometrical scaling region, where
the simpler expression (\ref{20phiscale}) applies; $y_c''$ is
clearly determined by the condition
$Q_s^2(A,y_c'')=Q_g^2(p,y_c'')$, which together with
Eq.~(\ref{20geom}) implies:
\begin{equation}\label{20ycrit}
    2 c b y_c'' \approx
    \left( \frac{|\xi_1|}{24 D^2} \right)^{3/4} \rho_A^3.
\end{equation}
Finally, for $y$ so large that $2 c b y \simge \rho_A^4$, the
proton saturation momentum catches up with that of nucleus, cf.
Eq.~(\ref{QsrunAp}), and then the proton too enters a non--linear
regime. The constant factors appearing in the above equations for
$y_c$, $y_c'$ and $y_c''$ are not really under control, but their
parametric dependencies on $\rho_A$ are correct as written.

At this level, several differences with respect to the fixed
coupling case are already manifest: First, there is an interval in
rapidity that remains out of control, namely
\begin{equation}\label{20nocontrol}
    \rho_A^2 \lesssim 2cb y \lesssim \rho_A^{12/5}.
\end{equation} Second, the characteristic
rapidities $y_c'$ and $y_c''$, at which the proton enters the BFKL
and scaling regimes respectively, are parametrically larger than
the (unique) corresponding scale for fixed coupling, namely $y_c$
in Eq.~(\ref{ycfixed}). This shows that the evolution is {\sf
slower} in the presence of a running coupling. Third, with a
running coupling, the linear (BFKL) approximation for the proton
does not apply up to arbitrarily large $y$, but only so long as $2
c b y \lesssim \rho_A^4$. But for larger $y$, and $\kk =
Q_s(A,y)$, the proton and nuclear distributions coincide with each
other, so the ratio ${\mathcal R}_{\rm sat}$ reduces to the
trivial factor $1/A^{1/3}$.

Other differences between the fixed and running coupling scenarios
will be revealed by the subsequent analysis.

\subsubsection{$y < y_c$ : Proton in the DLA regime}

In this regime, the proton is described by Eq.~(\ref{DLAprun})
where, for $\kk^2= Q_s^2(A,y)$, the variable $\eta$ can be
evaluated as (see also Eq.~(\ref{Qsrun})): \be \eta\,=\, \ln
\ln\frac{Q_s^2(A,y)}{\Lambda^2}\,-\,\ln \ln
\frac{Q_p^2}{\Lambda^2}\,\simeq \,\ln \tau_A(y) \,.\ee By also
using Eq.~(\ref{SATcond}) for the nucleus, one finds :
\be\label{RmaxDLArun}
{\mathcal R}_{\rm sat}(A,y)\,\sim\,  \,\tau_A(y) \,\, {\rm
exp}\Big\{\tau_A(y) -\rho_A -\sqrt{4by \ln\tau_A(y)}\Big\}, \ee
which for $y\to 0$ is of $\mathcal{O}(\rho_A)$, but it decreases
rapidly with $y$, and it becomes of order one already after a
small rapidity evolution $y_0$, with: \be\label{y0run}
 y_0\,\simeq\,\frac{1}{4b}\,\ln
\rho_A\,\sim\,\ln A^{1/3}.\ee The saddle point approximation for
$\varphi_p(\kk,y)$, Eq.~(\ref{DLAprun}), is justified for $y\sim
y_0$ since $\sqrt{4 b y_0\eta} \simeq \sqrt{4 b y_0\ln \rho_A} =
\ln \rho_A >1$. We find a situation similar to the fixed coupling
case (cf. Eq.~(\ref{Ry0})), namely, $y_0$ is so small that the
corresponding evolution of the nucleus can be neglected:
$\tau_A(y_0)-\rho_A \sim \ln \rho_A/ \rho_A\ll 1$, and therefore
$Q_s^2(A,y) \simeq Q_A^2$ for $y\le y_0$. In particular, one can
use the MV model for the nucleus in order to obtain a more
accurate estimate for $y_0$, valid up to corrections of
$\mathcal{O}(\rho_A^{-2})$  (the analog of Eq.~(\ref{y0fix})) :
\be\label{y0runexact}  y_0\,=\,\frac{1}{4 b\ln\rho_A}\,
\ln^2\Big[a\rho_A\sqrt{\ln(a\rho_A)}\Big],\qquad a\equiv
\sqrt{2\pi}z_0\Gamma(0,z_0)\simeq 0.706\,.\ee

When comparing Eqs.~(\ref{Ry0}) and (\ref{y0run}), it appears that
the running coupling estimate for $y_0$ is parametrically larger
than the fixed coupling one in the limit where $A$ is large. This
confirms the fact that the running effects slow down the
evolution. However, one should keep in mind that the rapidity is
naturally measured in units of $1/\alpha_s$, and whereas $\alpha_s
y_0\sim \ln^2\rho_A/\rho_A$ for fixed coupling, we also have
$\alpha_s(Q_A^2) y_0\sim \ln\rho_A/\rho_A$ for running coupling.
Thus, when expressed in natural units, the two values are rather
close to each other, and the fixed coupling estimate could be even
slightly larger.

\subsubsection{$y>y_c'$ : Proton in the BFKL regime}

So long as $2 c b y \ll\rho_A^4$ (with $y>y_c'$ though), the
proton saturation momentum remains considerably smaller than that
of the nucleus, so the proton with $\kk \sim Q_s(A,y)$  is in the
BFKL regime (\ref{swindowrun}), where Eq.~(\ref{BFKLrun}) applies.
Therefore, the ratio $\mathcal{R}_{\rm sat}(A,y)$ is known
analytically within this whole range, up to an overall constant
factor. In order to simplify the discussion, we shall study this
ratio in particular intervals of rapidity, so that
Eq.~(\ref{BFKLrun}) reduces to either Eq.~(\ref{20philarge}) or
Eq.~(\ref{20phiscale}).

\noindent $\bullet$ The interval $\rho_A^{12/5} \lesssim 2 c b y
\lesssim \rho_A^3$.

In this region, $\varphi_p$ is described by Eq.~(\ref{20philarge})
with $\ln z = \ln (Q_s^2(A,y)/Q_s^2(p,y)) \approx \tau_A(y) -
\tau_p(y)$. Since we are in a regime where $2 c b y \gg\rho_A^2$
we can expand $\tau_A(y)=\sqrt{2 c b y +\rho_A^2}$, to obtain
\begin{equation}\label{20logz}
    \ln z = \frac{1}{2}\,
    \frac{\rho_A^2}{\sqrt{2 c b y}}
    + \cdot \cdot \cdot\,.
\end{equation}
(Subleading terms in this expansion, including those arising from
the $y^{1/6}$ contributions to the saturation momentum, cf.
Eq.~(\ref{20satmom}), result in multiplicative factors in
$\mathcal{R}_{\rm sat}(A,y)$ that go to 1 when $\rho_A \gg 1$.)
Then it is straightforward to show that
\begin{equation}\label{RmaxBFKLrun}
    \mathcal{R}_{\rm sat}(A,y) \approx  \frac{1}{A^{1/3}}
    \frac{\sqrt{\rho_A}}{(2 c b y)^{1/3}}
    \exp \left[
    \frac{\gamma_s}{2} \frac{\rho_A^2}{(2 c b y)^{1/2}}
    +\frac{2}{3 (2D)^{3/2}} \frac{\rho_A^3}{2cby}
    +\frac{1}{6 \beta} \frac{\rho_A^4}{(2 c b y)^{3/2}}
    \right].
\end{equation}
Notice that with increasing $y$, the ratio is decreasing. The main
difference with respect to the corresponding result for fixed
coupling, Eq.~(\ref{RmaxBFKLfix}), is that for running coupling
the dominant $y$--dependencies do not compensate in the ratio
${Q_s^2(A,y)}/{Q_s^2(p,y)}$. Thus, at least for not too large $y$,
Eq.~(\ref{RmaxBFKLrun}) preserves a relatively fast decrease with
$y$, which is dominated by the first term in the exponent,
proportional to the ``anomalous dimension". The other terms in the
exponent, due to diffusion effects, are parametrically smaller
than the first term. However, since these terms are exponentiated,
their contribution to $\mathcal{R}_{\rm sat}$ cannot be neglected
in general.

 At the low end of our interval,
i.e. when $2 c b y \simeq \rho_A^{12/5}$, the ratio becomes
\begin{equation}\label{20rsatrho125}
    \mathcal{R}_{\rm sat}(A) \approx
    \frac{\rho_A^{-3/10}}{A^{1/3}}\,
    \exp \left[
    \frac{\gamma_s}{2}  \rho_A^{4/5}
    +\frac{2}{3 (2D)^{3/2}} \rho_A^{3/5}
    +\frac{1}{6 \beta} \rho_A^{2/5}
    \right]
    \qquad {\rm for}\,\, 2 c b y \approx \rho_A^{12/5}.
\end{equation}
Notice that the very leading behavior $A^{-1/3} \sim
\exp(-\rho_A)$, has settled in already at this rapidity, and what
remains in the exponent is fractional powers of $\rho_A$. These
powers (4/5, 3/5 and 2/5) are smaller than 1, so that
$\mathcal{R}_{\rm sat}(A)$ is (much) smaller than 1, but the whole
exponent is positive so that $\mathcal{R}_{\rm sat}(A)$ is (much)
larger than $A^{-1/3}$. \noindent This behavior is not unexpected
though; one would really need to go back to lower rapidities such
that $2 c b y \sim \rho_A^2$, in order to see the power of $A$
deviate from $-1/3$ (cf. Eq.~(\ref{RmaxBFKLrun})).

At the upper end of our interval, i.e. when $2 c b y \simeq
\rho_A^3$, the expression becomes considerably simpler since one
can keep only the first term in the exponent of
Eq.~(\ref{RmaxBFKLrun}). One has
\begin{equation}\label{20rsatrho3}
    \mathcal{R}_{\rm sat}(A) \approx
    \frac{1}{A^{1/3}}
    \,\frac{1}{\sqrt{\rho_A}}
    \exp \left[
    \frac{\gamma_s}{2} \sqrt{\rho_A}  \right]
    \qquad {\rm for} \quad 2 c b y \approx \rho_A^3.
\end{equation}
which clearly shows more suppression than
Eq.~(\ref{20rsatrho125}).

\noindent $\bullet$ The interval $\rho_A^3 \lesssim 2 c b y
\lesssim \rho_A^4 $.\\

\noindent In this region the proton is in the perturbative
geometric scaling region and its gluon occupation factor is given
by Eq.~(\ref{20phiscale}). The Cronin ratio becomes
\begin{equation}\label{20rsatscale}
    \mathcal{R}_{\rm sat}(A,y)
    \approx
    \frac{1}{A^{1/3}}\,
    \left[ \Delta + \frac{1}{2}\,
    \frac{\rho_A^2}{\sqrt{2 c b y}} \right]^{-1}
    \exp \left[
    \frac{\gamma_s}{2}\,
    \frac{\rho_A^2}{\sqrt{2 c b y}}
    \right].
\end{equation}
It is again clear that $\mathcal{R}_{\rm sat}(A,y)$ is a
decreasing function of $y$. At the low end of the interval,
i.e.~when $2 c b y \simeq \rho_A^3$, we obtain again
Eq.~(\ref{20rsatrho3}). On the other hand when we approach the
upper end $\rho_A^4$, we have
\begin{equation}\label{20rsatrho4}
    \mathcal{R}_{\rm sat}(A) \propto
    \frac{1}{A^{1/3}}
    \qquad {\rm for} \,\, 2 c b y \approx \rho_A^4,
\end{equation}
where the constant of proportionality in the last equation cannot
be determined (since the proton approaches a non--linear regime),
but it is a number greater than 1.

\subsubsection{$2  c b y > \rho_A^4$ : Proton at saturation}

 Even though this region is not analytically under
control, since the proton distribution becomes non-linear, it is
clear that, when further increasing $y$, the ratio will be
(slowly) decreasing until it stabilizes, for $2 c b y
\gg\rho_A^4$, at the very small value:
 \be\label{Rmaxlimit} {\mathcal R}_{\rm sat}(A,y)\, \simeq\,
  \frac{1}{A^{1/{3}}} \,\,\qquad {\rm for}\qquad 2 c b y \gg\rho_A^4
\,,\ee which is simply the factor introduced by hand in the
definition (\ref{Rdef}) of  ${\mathcal R}_{pA}$. Note that also
the overall normalization is now under control. This reflects the
fact that, in this limit, the nuclear and proton saturation scales
coincide with each other, cf. Eq.~(\ref{QsrunAp}), so the
corresponding occupation factors at $\kk \sim Q_s(A,y)$ coincide
as well.

Moreover, by the same argument, it is clear that for $2 c b y
\gg\rho_A^4$, the limiting value ${\mathcal R}_{pA}= 1/A^{1/{3}}$
is reached for {\sf any} $\kk \le Q_g(A,y)$  --- this includes the
saturation region, and also the window (\ref{swindowrun}) for the
BFKL regime
--- since in this whole range of momenta the gluon distribution is
{\sf universal} (for sufficiently large $y$), i.e., it depends
upon the specific properties of the hadron at hand only via its
saturation momentum (cf. Sect. \ref{EV:RUN}). This will be further
discussed in Sect. \ref{HIGHPT}.

Let us finally compare the asymptotic regimes for fixed and
running coupling: By inspection of Eqs.~(\ref{RmaxBFKLyy}) and
(\ref{Rmaxlimit}), it becomes clear that the ultimate value of
${\mathcal R}_{\rm sat}$ is much smaller for running coupling, and
it is reached after a much longer evolution. Note, however, that
if one measures the rapidity in units of $1/\alpha_s$, with
$\alpha_s$ evaluated at the nuclear saturation scale $Q_s(A,y)$ as
natural, then for large $y$ :\be
\bar\alpha_s(Q_s^2(A,y))\,\equiv\,
\frac{b}{\ln(Q_s^2(A,y)/\Lambda^2)}\, \simeq\,
  \frac{b}{\sqrt{2c b y}} \,\,\qquad {\rm for}\qquad 2 c b y
  \gg\rho_A^2,\ee and the asymptotic regime in
Eq.~(\ref{Rmaxlimit}) corresponds to $\bar\alpha_s y \gg
\rho_A^2$. This is parametrically of the same order in $A$ as the
evolution required to reach the limit (\ref{RmaxBFKLyy}) in the
fixed coupling case.  Still, the presence of the large factor
$\beta\approx 48$ in the fixed coupling condition $\beta
\bar\alpha_s y \gg \rho_A^2$ makes that, in practice, the lowest
value of ${\mathcal R}_{\rm sat}$ is reached considerably faster
with fixed coupling evolution than with the running coupling one.
In view of this, the running coupling evolution is indeed slower.

\subsection{The flattening of the Cronin peak}
\label{FLAT}

Since restricted to momenta along the saturation line
$\kk=Q_s(A,y)$, the previous analysis could not tell us whether a
maximum actually exists or not, neither describe its detailed
properties, like the position of the peak and its actual shape. To
study this, we need the nuclear gluon distribution for $y>0$ and
generic momenta around $Q_s(A,y)$, that we shall compute in this
section to lowest non-trivial order in $\bar\alpha_s y$, that is,
after a single step in the quantum evolution.  Since, as we shall
see, the quantum evolution and the twist expansion interfere with
each other, it is convenient to choose the same ``small
parameter", namely $1/\rho_A \ll 1$, to control both of them. If
$\Delta Y\equiv \bar{\alpha}_s y$ denotes the rapidity increment
that we shall consider (this satisfies $\Delta Y\ll 1$), then our
objective in what follows will be to construct the nuclear gluon
distribution $\varphi_A(\kk, \Delta Y)$ in such a way that, for
$\Delta Y \sim 1/\rho_A$ and $\kk\simle Q_s(A,y)$, all the terms
to order $1/\rho_A$ are correctly included. This means that when
we use this result to construct the $\mathcal{R}_{pA}$--ratio (and
therefore we multiply by $\rho_A$; see e.g. Eq.~(\ref{RAfixed})),
we just lose contributions that vanish when $\rho_A \gg 1$.

This implies, in particular, that only the terms of
$\mathcal{O}(1)$ and of $\mathcal{O}(1/\rho_A)$ need to be kept in
the initial condition $\varphi_{\rm in}\equiv \varphi_A(\kk,
Y=0)$. (Throughout this subsection, we shall often omit the
subscript $A$ on the various formulae, to simplify writing.) We
thus write: \be\label{10phiini}
    \varphi_{\rm in}=\,
    \varphi_0
    +\frac{1}{\rho_A}\, \varphi_1
    +\mathcal{O} \left(  \rho_A^{-2} \right),\ee where $\varphi_0$
is the same as the saturating piece in Eq.~(\ref{phiMVdec}) (i.e.,
the piece involving the function $\Gamma(0,z)$), which is of
$\mathcal{O}(1)$, while $\varphi_1$ is the term with $n=1$ in the
twist expansion shown in Eq.~(\ref{phiMVexp}) (which, we recall,
includes the bremsstrahlung spectrum together with higher twist
effects to all orders in $1/z$) and is of $\mathcal{O}(1/\rho_A)$.

To study the non--linear evolution with $y$, it is convenient to
use the Kovchegov equation \cite{K}, since this is a closed
equation. This can be written both in momentum space, i.e., as an
equation for $\varphi(\kk, Y)$, and in coordinate space, as an
equation for the dipole scattering amplitude which appears in
Eq.~(\ref{phiN}). In what follows, both forms of this equation
will be used, for convenience. We start by rewriting
Eq.~(\ref{phiN}) as:
\begin{equation}
\varphi(k_\perp,Y)=\label{10phi}
    \frac{1}{\alpha_s N_c}
    \int \frac{d^2 r_\perp}{\pi r^2_\perp}\,
    e^{-i k_\perp \cdot r_\perp}\,
    \mathcal{N}(r_\perp,Y)=
    \frac{2}{\alpha_s N_c}
    \int \frac{dr}{r}\,
    J_0(k r)\,
    \mathcal{N}(r,Y).
\end{equation}
In the following we shall omit the overall factor $1/\alpha_s
N_c$, which we can restore at the end. In terms of
$\mathcal{N}(r,Y)\equiv \mathcal{N}_{xy}(Y)$ (with $r_\perp =
x_\perp - y_\perp$), the Kovchegov equation reads:
\begin{equation}\label{10BK}
    \frac{\partial \mathcal{N}_{xy}}{\partial Y}\,=
    \int \frac {d^2z_{\bot}}{2\pi}\,
    \mathcal{K}_{xyz}
    \left( \mathcal{N}_{xz}+
    \mathcal{N}_{zy}-
    \mathcal{N}_{xy}
    -\mathcal{N}_{xz} \mathcal{N}_{zy}
    \right),
\end{equation}
where $\mathcal{K}_{xyz}$ is the dipole emission kernel :
\be\label{Kdef}
\mathcal{K}_{xyz}\,\equiv\,{(x_\perp-y_\perp)^2\over
(x_\perp-z_\perp)^2(z_\perp-y_\perp)^2}\,.\ee One can transform
the above integration over $z_{\bot}$ into an integration over the
sizes of the emitted dipoles $r_1$ and $r_2$ :
\begin{align}\label{10BKJ}
    \frac{\partial \mathcal{N}(r)}{\partial Y}=&
    \int \frac{dr_1}{r_1} \,\frac{dr_2}{r_2}\,
    r^2\, \ell\, d\ell\,
    J_0(\ell r_1)\,
    J_0(\ell r_2)\,
    J_0(\ell r)\,
    \nonumber \\
    &\times \left[
    \mathcal{N}(r_1)
    +\mathcal{N}(r_2)
    -\mathcal{N}(r)
    -\mathcal{N}(r_1)\,\mathcal{N}(r_2)
    \right].
\end{align}
Through Eq.~(\ref{10phi}), this gives a non--linear equation for
$\varphi(\kk, Y)$, in which the structure of the non--linear term
is actually simpler \cite{K,SCALING} :
\begin{equation}\label{10BKphi}
    \frac{\partial \varphi}{\partial Y}=
    \mathcal{K} \otimes \varphi - \frac{1}{2}\, \varphi^2,
\end{equation}
where now $\mathcal{K}$ is an operator with the BFKL eigenvalue
spectrum, but for our current purposes we prefer to view it as a
multiple integration determined from Eqs.~(\ref{10phi}) and
(\ref{10BKJ}), that is :
\begin{align}\label{10Kphi}
    \mathcal{K} \otimes \varphi =
    2 & \int \frac{dr_1}{r_1} \,\frac{dr_2}{r_2}\,
    r dr\, \ell\, d\ell\,
    J_0(\ell r_1)\,
    J_0(\ell r_2)\,
    J_0(\ell r)\,
    J_0(k r)\
    \nonumber \\
    &\times \left[
    \mathcal{N}(r_1)
    +\mathcal{N}(r_2)
    -\mathcal{N}(r)
    \right].
\end{align}
To obtain the  change $\Delta \varphi$ in $\varphi$ after a small
step in rapidity $\Delta Y \ll 1$, it is enough to iterate the
evolution equation only once. This gives:
\begin{equation}\label{10dphi}
    \Delta \varphi =
    \Delta Y \left(
    \mathcal{K} \otimes \varphi_{\rm in}
    - \frac{\varphi^2_{\rm in}}{2}
    \right)
    +\mathcal{O}\left( (\Delta Y)^2 \right),
\end{equation}
with $\varphi_{\rm in}$ the initial condition. To keep the
discussion simple, we shall imagine that $\Delta Y \sim 1/\rho_A$,
which is the amount of rapidity (up to $\ln \rho_A$ factors) that
gives a suppression of the Cronin peak to a value of order
$\mathcal{O}(1)$, due to the fast evolution of the proton (cf.
Eq.~(\ref{Ry0})). Then we have:
\begin{equation}\label{10dphi2}
    \Delta \varphi =
    \Delta Y \left( \mathcal{K} \otimes \varphi_0
    -\frac{1}{2}\, \varphi_0^2  \right)
    +\mathcal{O} \left( \frac{\Delta Y}{\rho_A}\,,
    (\Delta Y)^2 \right),
\end{equation}
and therefore the evolved gluon distribution is finally written as
\begin{equation}\label{10phi2}
    \varphi=\varphi_0
    +\frac{1}{\rho_A}\, \varphi_1
    + \Delta Y \tilde{\varphi}_0
    +\mathcal{O}
    \left( \frac{1}{\rho_A^2}\,,
    \frac{\Delta Y}{\rho_A}\,,
    (\Delta Y)^2 \right),
\end{equation}
where we have introduced the shorthand notation
\begin{equation}\label{10phitilde}
    \tilde{\varphi}_0=
    \mathcal{K} \otimes \varphi_0
    -\frac{1}{2}\, \varphi_0^2.
\end{equation}
Eq.~(\ref{10phi2}) fulfills our objective in that all the
neglected terms there become of order $1/\rho_A^2$ when $\Delta Y
\sim 1/\rho_A$ (for generic momenta $\kk\sim Q_s(A,y)$).

To compute the action of the BFKL operator on $\varphi_0$ we shall
use Eq.~(\ref{10Kphi}) in which $\mathcal{N}\rightarrow
\mathcal{N}_0\equiv 1 -\exp(-r^2 Q_s^2/4)$. Since the three
separate integrations (corresponding to the three terms in the
square bracket) in (\ref{10Kphi}) are individually ultraviolet
divergent (the divergences only cancel in their sum), we shall
regulate the $r_{1,2}$ integrations by letting
\begin{equation}\label{10reg}
    \int \frac{dr_1}{r_1}\, \frac{dr_2}{r_2}
    \rightarrow
    \lim_{\epsilon \rightarrow 0^{+}}
    \int \frac{dr_1}{r_1^{1-2\epsilon}}\,
    \frac{dr_2}{r_2^{1-2\epsilon}},
\end{equation}
with the limit to be taken at the end of the calculation. After
performing each of the three (multiple) integrations separately,
and adding the respective results, we obtain :
\begin{align}
    \mathcal{K} \otimes \varphi_0=
    \lim_{\epsilon \rightarrow 0^{+}} &\,
    \frac{1}{2}\,
    \frac{\Gamma^2(\epsilon)}{\Gamma^2(1-\epsilon)}\,
    \left( \frac{z}{4} \right)^{-2\epsilon}
    \nonumber \\
    &\times
    \left[
    1-2\, z^{\epsilon} \,\Gamma(1-\epsilon) \,\Phi(\epsilon,1,-z)
    +z^{2\epsilon}\, \Gamma(1-2\epsilon)\, \Phi(2\epsilon,1,-z)
    \right],
\end{align}
where, as usual, we have used the scaled momentum variable
$z=k^2/Q_s^2(A)$. Here, $\Phi(\epsilon,1,-z)\equiv
_1F_1(\epsilon,1,-z)$ is the confluent hypergeometric function,
whose properties will be reviewed in the Appendix. The
$\epsilon\to 0$ limit in this last expression is finite. After
also combining this with the non-linear contribution, which simply
reads $-(1/2) \Gamma^2(0,z)$, we finally find :
\begin{equation}\label{10phitilde2}
    \tilde{\varphi}_0=
    \frac{\pi^2}{12}
    -\frac{1}{2}
    \left[
    \gamma_E + \ln z + \Gamma(0,z)
    \right]^2
    +\frac{1}{2}\,
    \Phi^{200}(0,1,-z),
\end{equation}
where $\gamma_E = 0.577$ is the Euler--Mascheroni constant, and
$\Phi^{200}(0,1,-z)$ is a particular derivative of the confluent
hypergeometric function, to be defined in the Appendix. It is
illuminating to study the small-- and large--$z$ limits of
$\tilde{\varphi}_0$, which are given by
\begin{equation}\label{10limits}
    \tilde{\varphi}_0=
    \begin{cases}
        \displaystyle{\frac{\pi^2}{12}-\frac{z^2}{4}
        +\cdot \cdot \cdot} &
        \text{ if  $z \ll 1$}
        \\ \\
        \displaystyle{\frac{1}{z}+\frac{1}{2 z^2}
        + \cdot \cdot \cdot} &
        \text{ if  $z\gg 1$}.
    \end{cases}
\end{equation}
When $z$ is small, we simply obtain a constant correction. This is
consistent with the fact that the low--$\kk$ limit of $\varphi$ is
not affected by the evolution --- this is rather fixed by
Eq.~(\ref{10phi}) as $\varphi(k,Y) \to \ln \kk^2$ when $\kk\to 0$
---, but it is only the scale in this logarithm which changes with
$y$ : this is the saturation momentum $Q_s(A,y)$, since this is
the scale at which the dipole starts to be completely absorbed:
$\mathcal{N}(r,Y)\simeq 1$ for $r\simge 1/Q_s(A,y)$. This
discussion suggests that the term $(\pi^2/12)\, \Delta Y$ coming
from the one--step evolution could be interpreted as the first
contribution to the evolution of the saturation scale.

The crucial point about the correction induced by evolution, as
displayed in Eqs.~(\ref{10phitilde2})--(\ref{10limits}), is that
this contains {\sf power--law} tails at large $z$, despite the
fact that it has been generated by evolving the {\sf compact}
(saturating) part of the initial distribution alone. This is due
to the fact that the (perturbative) gluon interactions --- as
encoded in the BFKL kernel
--- are long--ranged, in both the momentum and the coordinate space
(see Eq.~(\ref{Kdef})). A similar phenomenon is the generation of
power--law tails in impact parameter space when solving the
Kovchegov equation (\ref{10BK}), as found in Ref. \cite{GBS03}.

Thus, in addition to the original ``twist'' contributions encoded
in the function $\varphi_1$,  which are parametrically suppressed
at large $A$, the evolved gluon distribution in Eq.~(\ref{10phi2})
contains other power--law contributions
{\sf which are not parametrically suppressed}, because they are
generated by the evolution of the gluons which were originally at
saturation: ${\varphi}_0$ and $\tilde{\varphi}_0$ are
parametrically of the same order, and the induced piece $\Delta
\varphi =\Delta Y \tilde{\varphi}_0$ is small only as long as
$\Delta Y$ itself is small. But although we cannot rely on the
previous approximations to go up to larger values of
$Y=\bar\alpha_s y$, it is clear that, for $\bar\alpha_s y \sim 1$
and momenta $\kk \sim Q_s(A,y)$, the twist terms induced by the
evolution become as large as the saturating distribution.
(Incidentally, this justifies the fact that, for $\bar\alpha_s y
\simge 1$, we can rely on the linear, BFKL, evolution to approach
the saturation region from the above; see also the discussion
after Eq.~(\ref{BFKLp0}))

We are now in a position to explain the rapid flattening of the
Cronin peak, as seen in the numerical simulations in Ref.
\cite{Nestor03}. We have already noticed, in the analysis of the
Cronin peak in the MV model in Sect. \ref{CroninMV}, that the
presence of long--range tails has the effect to flatten the peak,
and to push it towards larger values of $\kk$. At $y=0$ and for a
large value of $A$, a well pronounced peak exists just because the
compact contribution to the gluon distribution due to the
saturated gluons (i.e., ${\varphi}_0\equiv \varphi^{\rm sat}$) is
the dominant contribution at  $\kk\simle Q_s(A)$ (see Figs.
\ref{phi} and Figs. \ref{RpA}). However, with increasing $y$, the
power--law tails are enhanced, and the effect of this enhancement
is particularly pronounced in the early stages of the evolution
($\bar\alpha_s y \ll 1$), when the modes at high $\kk$ receive
large contributions from the evolution of the gluons which were
originally at saturation. In this way, the exponential tail which
was present in the saturating distribution at $y=0$ is rapidly
washed out, and replaced by a power law tail which starts already
at $\kk \sim Q_s(A,y)$. When this happens, the peak in the ratio
${\mathcal R}_{pA}$ flattens out completely and disappears. In
view of the previous estimates, we expect a complete flattening
for $\bar\alpha_s y\sim 1$. But the effect shows up already for
smaller rapidities, where our one--step evolution is a good
approximation.

\begin{figure}[]
\begin{center}
\includegraphics[scale=1.33]{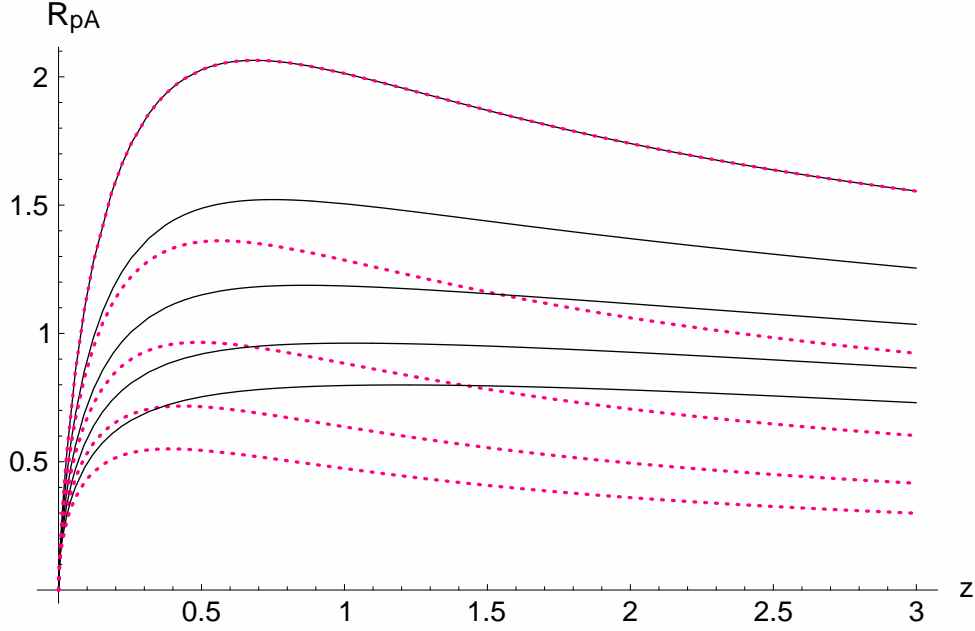}
    \caption{\label{tilt} {\sl The Cronin ratio
    $\mathcal{R}_{pA}(z)$ below and near the saturation scale for
    $\rho_A=6$.} {\small The black (solid) lines correspond to an
    evolved nuclear wavefunction by $\Delta Y \ll 1$. The red
    (dotted) lines correspond to an unevolved one (MV). The proton
    wavefunction is always given by the full DLA solution. The
    curves, from top to bottom, correspond to $\Delta Y=n/(2
    \rho_A)$ with $n=0,1,...,4$.}}
\end{center}
\end{figure}

To illustrate this, we have represented in Fig. \ref{tilt} the
ratio $\mathcal{R}_{pA}$ (for $\rho_A=6$) obtained by using the
result in Eq.~(\ref{10phi2}) for the nuclear gluon distribution
together with the DLA approximation for the proton
distribution\footnote{To be able to perform this calculation also
for very small values of $Y$, where the saddle point approximation
is not appropriate anymore, we have used the exact solution to the
DLA equation, which is well known to read (compare to
Eq.~(\ref{DLAp0})) : $\varphi_p(\kk,y) =\frac{1}{\alpha_s
N_c}\,\frac{Q_p^2}{k_{\perp}^2}\,I_0\big(\sqrt{4\bar\alpha_s y
\rho}\big)$, with $\rho=\ln(k_{\perp}^2/Q_p^2)$ and $I_0(z)$ the
modified Bessel function of the first kind.}. For comparison, we
have also represented on the same plot the ratio which is obtained
when the non--evolved, MV model, distribution is used for the
nucleus. The rapid suppression of the peak, due to the fast rise
in the proton distribution, is clearly seen in both cases. But in
the absence of nuclear evolution the peak is always there; just
its amplitude gets smaller and smaller. By contrast, when using
the evolved nucleus distribution from Eq.~(\ref{10phi2}) the
flattening of the peak is manifest, and in fact the maximum has
almost disappeared after an evolution $\Delta Y=2/\rho_A \approx
0.3$.

It is also visible on Fig. \ref{tilt} that with increasing $y$,
the position of the peak moves towards larger momenta --- in
agreement with our expectation that the peak should follow the
nuclear saturation momentum ---, but it does that only {\sf very
slowly}. This is so slow because, first, the nucleus evolves only
little for such a small rapidity increment, and, second, there is
an opposite effect due to the DLA evolution of the proton, which
for the small values of $Y$ of interest here is almost
compensating the evolution towards larger $\kk$ due to the
nucleus. Using Eq.~(\ref{10phi2}) for the nuclear distribution,
together with the DLA approximation (the full Bessel function) for
the proton, we have been able to compute  analytically the
position of the peak for very small values of $y$. The result
reads (compare to Eq.~(\ref{RmaxMVf})) : \be\label{Rmaxy}
z_0(Y)\,\approx\, 0.435\,+\,\frac{0.882}{\rho_A}\,+\,0.862 Y
\,-\,0.769 \sqrt{\frac{Y}{\rho_A}}\,,\ee where the neglected terms
are such that, when $Y\sim 1/\rho_A$, they are all of
$\mathcal{O}(\rho_A^{-2})$.

To summarize the previous considerations, the ratio (\ref{Rdef})
 can be suggestively rewritten as follows:
\beq\label{RR} {\mathcal
R}_{pA}(k_\perp,y)\,\equiv\,\frac{\varphi_A(k_\perp,y)}{A^{1/3} \,
\varphi_p(k_\perp,y)}\,=\,\frac{\Phi_A(k_\perp,y)}{\Phi_p(k_\perp,y)}\,
 {\mathcal R}_{pA}(k_\perp,y=0)\,,\nn \Phi_A(k_\perp,y)
\equiv\frac{\varphi_A(k_\perp,y)}{\varphi_A(k_\perp,y=0)}
\,,\qquad \Phi_p(k_\perp,y)
\equiv\frac{\varphi_p(k_\perp,y)}{\varphi_p(k_\perp,y=0)}\,. \eeq
The ratios $\Phi_A$ and $\Phi_p$, which describe the relative
evolutions of the nucleus and of the proton, respectively, play
very different roles for the evolution at small $y$ and momenta
around $Q_s(A,y)$ : $\Phi_p$ rises very fast with $y$, and is
responsible for the rapid suppression of ${\mathcal R}_{pA}$ at
generic momenta. However, this factor varies only slowly with
$\kk$, and thus cannot be responsible for the flattening of the
peak. $\Phi_A$, on the other hand, rises only slowly with $y$, but
its evolution is quite asymmetric around $Q_s(A,y)$ : the
(relative) evolution is substantially larger at momenta above
$Q_s(A,y)$ than below it, and because of this asymmetry, the peak
gets progressively tilted (its `upper' side at larger $\kk$ rises
faster than the lower one) and also flattens, until it eventually
disappears.

 \section{High--$k_\perp$ suppression in the nuclear gluon distribution
at small $x$} \setcounter{equation}{0} \label{HIGHPT}

In this section we shall extend our analysis of the evolution of
the ratio ${\mathcal R}_{pA}(\kk,y)$ with increasing $y$ to
arbitrary values of the transverse momentum $\kk$. The various
regimes of evolution in the $y-\ln\kk^2$ plane are illustrated in
Figs. \ref{EVOL-MAP} and \ref{EVOL-MAP-run} for fixed and running
coupling, respectively. In Sect. \ref{CRevolve}, we have followed
the evolution along the nuclear saturation line $\kk=Q_s(A,y)$.
Below, we shall consider all the physical regimes exhibited in
Figs. \ref{EVOL-MAP} and \ref{EVOL-MAP-run}, but our main focus
will be on the high momentum region at $\kk\gg Q_s(A,y)$, which is
the most interesting region for the phenomenon of ``high--$\kk$
suppression". The natural way to follow this evolution is to
increase simultaneously $y$ and $\kk$, in such a way that the
ratio $z\equiv \kk^2/Q_s^2(A,y)$ remains constant. This ensures
that, with increasing $y$, the nucleus remains in the same
physical regime, which could be either the saturation regime (for
$z < 1$), or the (nuclear) BFKL regime (for $z>1$).

In performing this analysis,  it is useful to keep in mind the
limitations of the analytic approximations that we shall use for
the gluon occupation factor: The transition regions from
saturation to BFKL, and also from BFKL to DLA, are not
analytically under control, nor are the very early stages of the
evolution. For the BFKL regime, as defined by Eq.~(\ref{swindow}),
the saddle point approximation is justified provided $\bar\alpha_s
y\ge 1$, while for the DLA regime at $\rho \gg \bar\alpha_s y$,
the corresponding condition reads $\bar\alpha_s y \rho\gg 1$. We
recall that $\rho\equiv \rho(A,\kk) = \ln\kk^2/Q_s^2(A)$ is much
larger for the proton than for the nucleus (for the same value of
$\kk$). Because of that, we shall be able to study the early
($\bar\alpha_s y\ll 1$) evolution of the proton (as described by
DLA), but not also that of the nucleus (unless $\kk$ is so large
that the nucleus itself enters the DLA regime).

The analysis below will not only confirm the general arguments in
Sect.~\ref{general} about the variation of the ratio ${\mathcal
R}_{pA}$ with $y$, $\kk$ and $A$, but will also allow us to
explicitly compute this ratio and, in particular, its limiting
values at large $y$ and/or large $\kk$. In the fixed coupling
case, and for the linear regime, our results below will confirm
and extend some of the conclusions originally obtained in Refs.
\cite{KLM02,KKT}, with which the present analysis has some
overlap. Also, still for fixed coupling, our results will be seen
to be consistent (within their limited range of validity and to
the present accuracy) with the numerical analysis in Ref.
\cite{Nestor03} based on the Kovchegov equation.

\subsection{Fixed coupling}

In this section we shall explore the various regimes for the fixed
coupling evolution exhibited in Fig. \ref{EVOL-MAP}, and study the
variation of the ratio ${\mathcal R}_{pA}(\kk,y)$ with $y$ and
$\kk$, as well as its parametric dependence upon $A$.

\subsubsection{$\kk < Q_s(A,y)$ : The nuclear saturation region}

The nuclear distribution at saturation is not explicitly known
(except for very small momenta $ \kk \ll Q_s(A,y)$, where
Eq.~(\ref{phiAsat}) applies), but this is some slowly varying
function of $z\equiv \kk^2/Q_s^2(A,y)$ (because of geometric
scaling), which moreover is simply a constant when the evolution
is performed along a line parallel to the saturation line. As for
the proton, it can be either at saturation, or in the linear
regime above saturation, depending upon the value of $\kk$.

{\sf (I) $\kk < Q_s(p,y)$ : Proton at saturation}

In this regime, the gluon distributions in  both the proton and
the nucleus are  universal functions of $\kk^2/Q_s^2(A,y)$ (with
$Q_s(A,y) \rightarrow Q_s(p,y)$ in the case of the proton), which
are only slowly varying. Thus, clearly, the ratio (\ref{Rdef}) is
of order $1/A^{1/3}$, with only weak, logarithmic, dependencies
upon $A$ and $\kk$. This becomes  explicit for sufficiently low
momenta, $\kk \ll Q_s(p,y)$, where one can use Eq.~(\ref{phiAsat})
for both the proton and the nucleus, and thus find that ${\mathcal
R}_{pA}(\kk,y)$ is monotonously increasing with $\kk$, and slowly,
but monotonously, decreasing with $y$ at fixed $\kk$. In
particular, along the proton saturation line $\kk = Q_s(p,y)$,
which defines the upper boundary of this region, one finds: \be
{\mathcal R}_{pA}(\kk=
Q_s(p,y),y)\,\sim\,\frac{\rho_A}{A^{1/3}}\,\qquad({\rm
independent\,\,of\,\,}y)\,.\ee

{\sf (II) $Q_s(p,y) < \kk < Q_s(A,y)$ : Proton in the linear
regime}

The corresponding analysis is very similar to that of the
evolution along the nuclear saturation line, as discussed in Sect.
\ref{CRevolve}. For $y < y_c$, cf. Eq.~(\ref{ycfixed}), the proton
can be either in the DLA, or in the BFKL, regime (depending upon
the value of $\kk$), while for larger $y$, it is always in the
BFKL regime (see Fig. \ref{EVOL-MAP}).

A straightforward analysis shows that the ratio is rapidly
decreasing with $y$ at any fixed $\kk$, due to the fast evolution
of the proton. It is also interesting to consider the large $y$
behavior at fixed $z$. Then, the proton is in the BFKL regime, so
using Eq.~(\ref{BFKLp0}) (with $Q_s(A,y)\, \to\, Q_s(p,y)$) and
writing $\varphi_A(\kk,y) =(1/\alpha_s N_c)\bar\varphi_0(z)$, one
obtains (with $\gamma\equiv\gamma_s\simeq 0.63$ from now on) :
\be\label{RABFKLfix} {\mathcal R}_{pA}(\kk,y) &\sim &
\frac{\rho_A^{\gamma}}{A^{\frac{1-\gamma}{3}}}\,
\frac{z^{\gamma}\,\bar\varphi_0(z)}{\rho_A - \ln 1/z + \Delta}
\,\exp\left\{\frac{(\rho_A + \ln z)^2}{2\beta \bar\alpha_s y}
\right\}\,\nn &\rightarrow
&\frac{\rho_A^{\gamma}}{A^{\frac{1-\gamma}{3}}}\,
\frac{z^{\gamma}\,\bar\varphi_0(z)}{\rho_A - \ln 1/z}\,,\qquad
{\rm for}\quad 2\beta \bar\alpha_s y\gg \rho_A^2,\ee (in this
regime $\rho_A > \ln 1/z$), which is indeed consistent with the
previous result (\ref{RmaxBFKLyy}) on the nuclear saturation line
$z=1$.

\subsubsection{$Q_s(A,y) <\kk < Q_g(A,y)$ : The nuclear BFKL region}

For $y < y_c$, the proton is in the DLA regime for any $\kk$,
while for $y > y_c$, it can be either in the BFKL, or in the DLA
regime, according to the value of $\kk$.

{\sf (I) $y < y_c$ : Proton in the DLA regime}

\begin{figure}[bht]
\begin{center}
\includegraphics[scale=1.5]{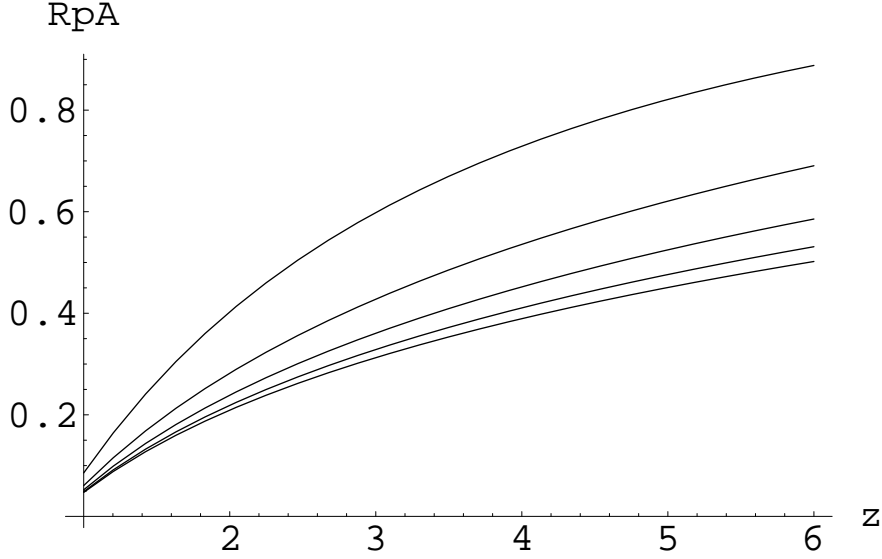}
\caption{{\sl The ratio ${\mathcal R}_{pA}(z,Y)$ in
Eq.~(\ref{BFKLDLAfix}) as a function of $z=\kk^2/Q_s^2(A,y)$ for
$\rho_A=6$ (i.e., $Y_c\approx 0.75$) at different rapidities.}
{\small The lines are for $Y=0.1+0.15\, n,$ with $n=0,1,\dots,4 $,
from top to bottom.  The bottom line corresponds to $Y=0.7 \simeq
Y_c$. The ratio ${\mathcal R}_{pA}$ is measured in arbitrary
units, since the normalization of Eq.~(\ref{BFKLDLAfix}) is not
under control. Note the rapid decrease with increasing $Y$ in the
early stages of the evolution.}} \label{BDF}
\end{center}
\end{figure}

In this region, the ratio ${\mathcal R}_{pA}$ is formed by using
Eq.~(\ref{BFKLp0}) for the nucleus and Eq.~(\ref{DLAp0}) for the
proton. After straightforward manipulations, similar to those
leading to Eq.~(\ref{RmaxDLAfix}), the result can be cast into the
form: \BQ\label{BFKLDLAfix} {\mathcal R}_{pA}(z,Y)\,\sim\, \rho_A
\, z^{1-\gamma}\,\left(\ln z +{\Delta}\right) \exp\left\{cY
-\sqrt{4Y(\ln z + cY +\rho_A)}-\frac{\ln^2 z}{2\beta Y} \right\},
\EQ where $z\equiv \kk^2/Q_s^2(A,y)$ and $Y\equiv\bar\alpha_s y$.
Note that the overall normalization is not under control.
Eq.~(\ref{BFKLDLAfix}) applies for $\kk$ in the nuclear scaling
window (\ref{swindow}), that is: \beq\label{swindow4}
 0<\,\ln z <\, \ln \big[Q_g^2(A,y)/Q_s^2(A,y)\big]=\,\nu cY,\qquad
{\rm and}\qquad 1\simle Y<Y_c \equiv \rho_A/\nu c.\eeq It can be
checked that, within this physical range\footnote{The function in
Eq.~(\ref{BFKLDLAfix}) develops a maximum at very large $z$, well
outside the window (\ref{swindow4}). This is, of course,
unphysical.} the function ${\mathcal R}_{pA}(z,Y)$ is monotonously
increasing with $z$, and also rapidly decreasing with $Y$ at any
fixed $z$. This behavior is in agreement with the general results
in Sect. \ref{general}, and is illustrated in Fig. \ref{BDF}.

{\sf (II) $y > y_c$ \& $Q_s(A,y) <\kk < Q_g(p,y)$  :
Double--scaling regime}

This region is interesting in that the proton and the nucleus are
both in the same physical regime, namely in the window
(\ref{swindow}) for (approximate) geometric scaling, so they are
described by the same approximation --- the BFKL expression
(\ref{BFKLp0}) ---, and the normalization ambiguities cancel in
the ratio ${\mathcal R}_{pA}$. Thus, in this regime, one can also
predict the {\sf amplitude} of this ratio, and not only its
functional dependencies. Besides, this is also the regime for
which the quantum evolution above the saturation line has been
originally invoked, by Kharzeev, Levin and McLerran \cite{KLM02},
as a possible mechanism to explain the high--$\kk$ suppression
observed in the RHIC data.

Using Eq.~(\ref{BFKLp0}) for both the nucleus and the proton, one
finds : \be\label{DSCALING} {\mathcal R}_{pA}(z,y)&\simeq&
\frac{\rho_A^\gamma}{A^{\frac{1-\gamma}{3}}} \,\frac{\ln z
+{\Delta}}{\ln z +\rho_A + {\Delta}}  \,\exp\left[
\frac{\rho_A}{2\beta Y}(2\ln z +\rho_A) \right], \ee
 where we have used $\ln[\kk^2/Q_s^2(p,y)] \simeq \ln z +\rho_A$.
Eq.~(\ref{DSCALING}) is valid within the range:
 \BQ\label{rangeDS}
0<\ln z < \nu cY -\rho_A\qquad {\rm and}\qquad Y > Y_c\equiv
\rho_A/\nu c. \EQ

\begin{figure}[bht]
\begin{center}
\includegraphics[scale=1.5]{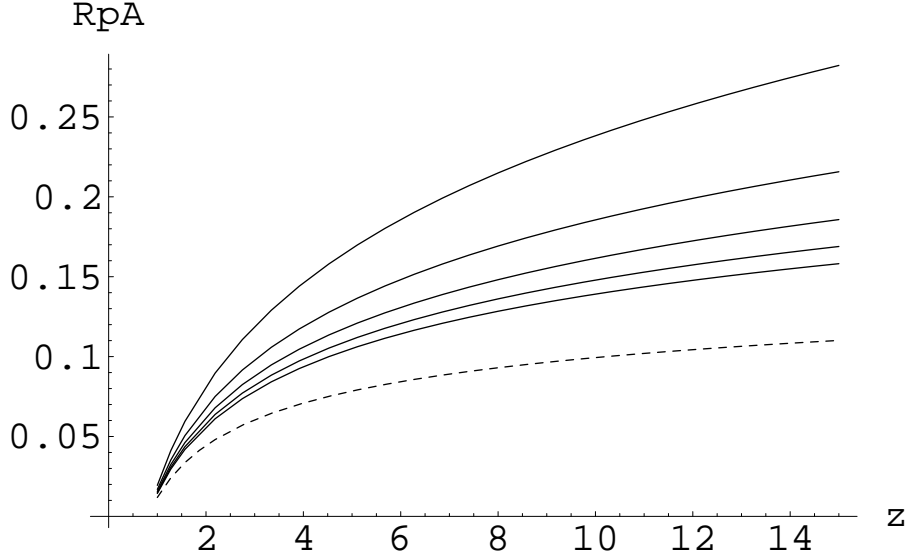}
\caption{{\sl The ratio ${\mathcal R}_{pA}(z,Y)$
 in the double scaling window for $\rho_A=6$ as a function of $z$
at different $Y \simge Y_c\approx 0.75$.} {\small The solid lines
correspond to $ Y=0.75+0.3\, n,$ with $n=0,\dots,4$ from top to
bottom. The dashed line is the asymptotic ($Y\to\infty$) profile
in Eq.~(\ref{ASsat}). The upper limit of the double scaling window
$z_{\rm max}=\exp\{\nu c (Y-Y_c)\}$ rapidly increases with
increasing $Y$. For instance, $z_{\rm max}\approx 11$ for
$Y=1.05\, (n=1)$, and $z_{\rm max}\approx 127$ for $Y=1.35\,
(n=2)$.}} \label{BBF}
\end{center}
\end{figure}

It is easily checked that the ratio (\ref{DSCALING}) is an {\sf
increasing} function of $z$ for arbitrary $Y$, and a {\sf
decreasing} function of $Y$ for arbitrary $z>1$. This  behavior is
illustrated in Fig. \ref{BBF}.

A noticeable difference with respect to the previous case, cf.
Eq.~(\ref{BFKLDLAfix}), is that, in this double-scaling regime,
the dominant $y$--dependencies of the nuclear and proton
distributions, which are exponential, have cancelled in the ratio
${\mathcal R}_{pA}$. Indeed, according to Eq.~(\ref{BFKLp0}),
these dependencies are encoded in the respective saturation
momenta, which however evolve in the same with $y$, cf.
Eq.~(\ref{Qsfix0}), and thus their ratio $Q_s^2(A,y)/Q_s^2(p,y) =
Q_s^2(A)/Q_p^2 \simeq A^{1/3} \rho_A$ is independent of $y$.

The residual dependence on $y$ in Eq.~(\ref{DSCALING}) comes from
the diffusion term in Eq.~(\ref{BFKLp0}), and is rather weak.
(This is also in agreement with the discussion in Sect.
\ref{general}; see especially Eq.~(\ref{dRdyBFKL}).) Thus, for $Y
> Y_c$, the suppression slows down, and for sufficiently large $Y$
(such that the exponent in the second line of Eq.~(\ref{DSCALING})
become negligible), ${\mathcal R}_{pA}$ stabilizes at a small
value which depends slowly on $z$: \BQ\label{ASsat} {\mathcal
R}_{pA}(z,Y\to\infty)\,\simeq\,
\frac{\rho_A}{(A^{1/3}\rho_A)^{1-\gamma}} \,\frac{\ln z
+{\Delta}}{\ln z +\rho_A + {\Delta}} \,.\EQ It can be verified
that most of this suppression has been achieved already in the
early stages of the evolution, while the proton was still in the
DLA regime.

When approaching the nuclear saturation line from the above,
Eq.~(\ref{DSCALING}) yields: \BQ {\mathcal R}_{pA}(z\to
1,Y)\,\simeq\, \frac{1}{(A^{1/3}\rho_A)^{1-\gamma}} \, \, {\rm
e}^{\frac{\rho_A^2}{2\beta
Y}}\,\,\stackrel{Y\to\infty}{\longrightarrow} \,
\frac{1}{(A^{1/3}\rho_A)^{1-\gamma}}\,. \label{RpA_DSW_yc} \EQ
 As expected, this is the same function of $A$ and $y$ as obtained
when approaching the saturation line from the below, cf.
Eq.~(\ref{RmaxBFKLfix}), with the noticeable difference that, now,
also the {\sf normalization} of this result is under control.

It is finally interesting to evaluate the ratio (\ref{DSCALING})
on the proton geometric line (i.e., $\kk = Q_g(p,y)$, or $\ln z =
\nu cY -\rho_A$), which is the upper boundary for the
double--scaling regime considered here. In the large $Y$ limit,
one obtains:
 \BQ \lim_{y\to \infty}{\mathcal R}_{pA}(\kk\sim
Q_g(p,y),y)\,=\,\frac{\rho_A}{(A^{1/3}\rho_A)^\delta}\,,\qquad
\delta=1-\gamma-\frac{\nu c}{\beta}\simeq 0.21\,. \EQ The dominant
power of $1/A^{1/3}$ is lower on this upper boundary than on the
lower one ($\delta \simeq 0.21$ as compared to $1-\gamma \simeq
0.37$), since $\kk$ is increasing faster along the geometric
scaling line, and ${\mathcal R}_{pA}$ is increasing with $\kk$.

{\sf (III) $y > y_c$ \& $Q_g(p,y) <\kk < Q_g(A,y)$  : Proton in
the DLA regime}

In this regime,  the ratio ${\mathcal R}_{pA}(z,Y)$ is given again
by Eq.~(\ref{BFKLDLAfix}), but which now applies to a different
kinematical range : $\nu cY -\rho_A < \ln z < \nu cY$ and $Y>
Y_c$.

\subsubsection{$\kk > Q_g(A,y)$ : The nuclear DLA region}

This is the high momentum limit, in which both the proton and
nucleus are in the DLA (or, more properly, DGLAP) regime, so the
ratio can be again computed without uncertainties related to the
normalization. One finds: \BQA {\mathcal
R}_{pA}(z,Y)\,=\,\exp\left[ \sqrt{4Y(\ln z + cY)} -\sqrt{4Y(\ln z
+ cY +\rho_A)} \right]\,, \EQA valid for $\ln z > \nu cY$. It is
clear that the extra term $\rho_A$ in the exponential comes from
the difference between two saturation scales. It is easy to check
analytically that the ratio is an increasing function of $z$ for
any $Y$, but
 a decreasing function of $Y$ for any fixed $z$.
In particular, at very large $z$ such that $\ln z \gg  cY+\rho_A$,
one obtains \BQ {\mathcal R}_{pA}\,\simeq \,{\rm
e}^{-\rho_A\sqrt{Y/\ln z}}. \EQ Therefore, when $z\to \infty$ the
ratio approaches one {\sf from below}.

\subsection{Running coupling}

\begin{figure}[]
\begin{center}
\includegraphics[scale=0.8]{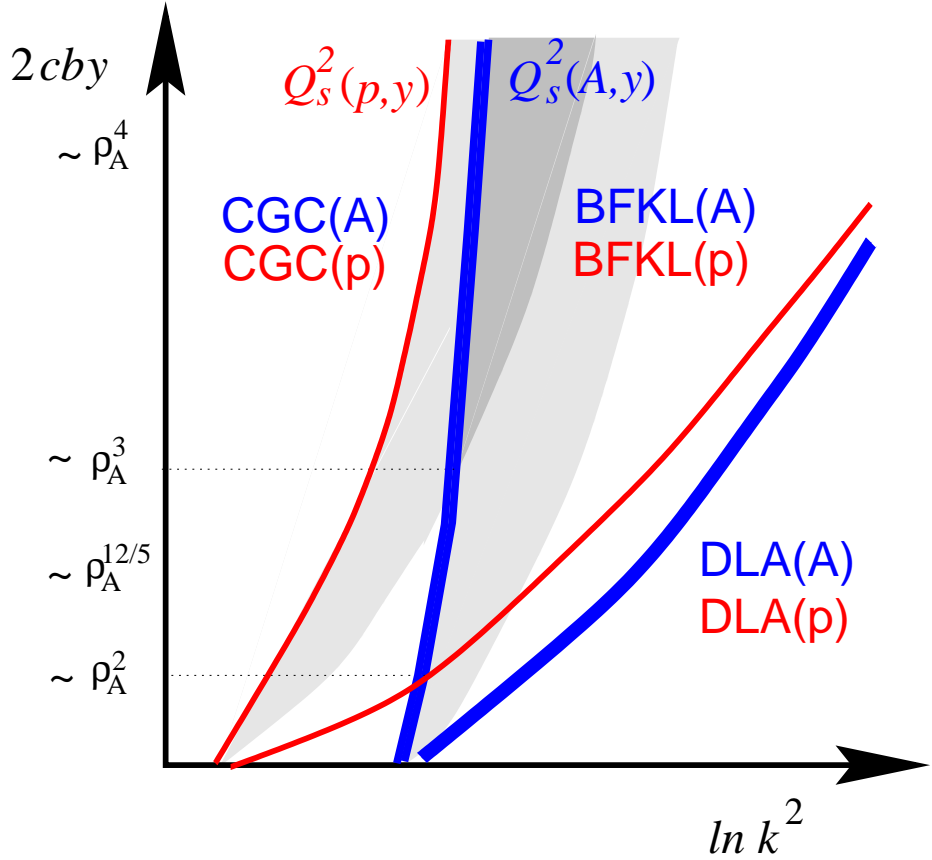}
\caption{{\sl Various regimes of evolution for running coupling.}
The boundaries which separate the three physical regimes (CGC,
BFKL, and DLA) are shown by the red (thin) and blue (thick) curves
for the proton and the nucleus, respectively. For each color
(thickness), the left curve corresponds to the saturation line
($\kk=Q_s(p,y)$ or $\kk=Q_s(A,y)$), while the right curve is the
boundary between BFKL and DLA regimes (which is less under
control). The shaded bands on the right of the saturation lines
are the respective geometric scaling regimes, which overlap with
each other in the "double scaling" regime (the area with a darker
shade).} \label{EVOL-MAP-run}
\end{center}
\end{figure}

The various physical regimes for evolution with running coupling
are exhibited in Fig.~\ref{EVOL-MAP-run}, which should be compared
to the corresponding map for fixed coupling, in
Fig.~\ref{EVOL-MAP}. The differences between the two cases come
from several sources:

First, the two saturation scales evolve differently with $y$ (that
for the proton growing faster) and eventually converge towards a
common limit at very large $y$ (cf.
Eqs.~(\ref{Qsrun})--(\ref{QsrunAp})). A similar behavior holds for
the corresponding lines for geometric scaling, and also for the
boundaries separating BFKL from DLA behavior. This implies that,
for sufficiently large $y$, the gluon distributions in the proton
and in the nucleus become identical for all the interesting
momenta. (Differences between the two distributions persist only
in the (common) DLA regime, which with increasing $y$ is rapidly
pushed towards asymptotic momenta.) This leads to the important
conclusion that the ratio (\ref{Rdef}) approaches the universal
constant $1/A^{1/3}$ at very large $y$, for all non--asymptotic
momenta.

Second, 
in the case of a running coupling, the window for geometric
scaling above saturation becomes narrower (cf.
Eq.~(\ref{20geom})), and, in contrast to the fixed coupling case,
it does not coincide with the BFKL regime anymore : the latter is
now defined by Eq.~(\ref{swindowrun}), and includes a domain where
scaling violations are important. In  Fig.~\ref{EVOL-MAP-run}, the
regions of extended geometric scaling are the shaded bands lying
on the right hand side of the nuclear and proton saturation lines,
respectively.

Third, the precise boundaries of the BFKL and DLA regimes, and
also the transition between these two regimes, are not under
control. Whereas we expect the BFKL approximation (\ref{BFKLrun})
to be valid for $0<\ln z < c_1\tau_A^{2/3}(y)$ (with $z \equiv
\kk^2/Q_s^2(A,y)$), and the DLA approximation (\ref{DLAprun})
should apply at $\ln z > c_2\tau_A(y)$, the constant factors
$c_{1,2}$ in these conditions are not determined, and we do not
dispose of analytic approximations for the intermediate region at
$c_1\tau_A^{2/3}(y) <\ln z < c_2\tau_A(y)$.

To avoid cumbersome discussions and ambiguities associated with
the location of the borderlines for the various regimes, in what
follows we shall focus mainly on the situations in which both the
proton and the nucleus find themselves in the same physical regime
(which could be CGC, BFKL, or DLA). These situations are the most
interesting ones, as they allow for a calculation of the ratio
${\mathcal R}_{pA}$ which is free of renormalization ambiguities.
Besides, they are the only ones to survive at large $y$: The other
regimes visible on Fig.~\ref{EVOL-MAP-run} are squeezed in between
the proton and nucleus saturation lines, or in between the
respective borderlines for BFKL behavior, and therefore shrinks to
zero in the high energy limit.

The analysis to follow is particularly important because, with
running coupling, we have no general arguments to understand the
behavior of the ratio ${\mathcal R}_{pA}$, as we had for a fixed
coupling. As we will see, in spite of interesting differences, the
global behavior of ${\mathcal R}_{pA}$ as a function of $y$, $\kk$
and $A$ is the same as in the fixed coupling case discussed in
Sect. \ref{general}.

\subsubsection{The double CGC region}

On the left of the proton saturation line, both the proton and the
nucleus are in the color glass condensate regime. For sufficiently
low momenta $\kk \ll Q_s(p,y)$, one can compute the ratio
${\mathcal R}_{pA}$ by using Eq.~(\ref{phiAsatrun0}) for both
hadrons. Then one can easily check that ${\mathcal R}_{pA}(\kk,y)$
has the same behavior as in the corresponding regime at fixed
coupling. It is further interesting to follow the proton
saturation line $\kk = Q_s(p,y)$, where Eq.~(\ref{SATcond}) holds
for the proton. One finds (see Eq.~(\ref{Qsrun}) for notations)
\be \label{RpAsatrun} {\mathcal R}_{pA}(\kk=
Q_s(p,y),y)\,\approx\, \frac{1}{A^{1/3}}\,
\frac{\big(\tau_A(y)+\tau_p(y)\big)
\big[a_0\big(\tau_A(y)-\tau_p(y)\big)
+\kappa\big]}{2\tau_p(y)\kappa}, \ee where we have added a
constant $\kappa$ to the nuclear distribution in
Eq.~(\ref{phiAsatrun0}) in such a way to match the boundary
condition (\ref{SATcond}) when $\kk\to Q_s(A,y)$. Of course, this
is only a crude approximation --- the approach towards the
saturation line is not under control ---, but this is interesting
as it allows us to qualitatively study the behavior at large $y$,
where the two saturation momenta converge to each other. For $2cby
\gg \rho_A^2$, one finds: \be {\mathcal
R}_{pA}(\kk=Q_s(p,y),y)\,\approx\, \frac{1}{A^{1/3}} \,
\left(1+\frac{a_0}{\kappa}\, \frac{\rho_A^2}{2\sqrt{2cby}}\right),
\ee which decreases with $y$, and for  $2cby \gg \rho_A^4$
approaches the universal value ${1}/{A^{1/3}}$, as expected.

\subsubsection{The nuclear BFKL region}

Here we consider the region in between the two blue (thick) curves
in Fig.~\ref{EVOL-MAP-run}. For relatively small rapidities $y <
y_c$ with $2cby_c\sim \rho_A^2$, the proton is in the DLA regime
for any $\kk$ within this region. (See the discussion at the
beginning of Sect. \ref{CR_RUN} for more details on the separating
rapidities.) For larger rapidities, the proton BFKL regime starts
to appear, but this is theoretically under control only for $y >
y_c'$, with $2cby_c'\sim \rho_A^{12/5}$. The most interesting
"double scaling" regime appears at the even larger rapidity $2cb
y_c''\sim \rho_A^3$.

{\sf (I) $y < y_c$ : Proton in the DLA regime}

Straightforward manipulations using Eq.~(\ref{BFKLrun}) for the
nucleus and Eq.~(\ref{DLAprun}) for the proton yield (with $z
\equiv \kk^2/Q_s^2(A,y)$, as usual) \BQA\label{BFKLDLArun}
{\mathcal R}_{pA}(z,y)&\sim&\Big(\ln z + \tau_A(y)\Big) \,
\tau_A^{1/3}(y)\, {\rm Ai}\left(\xi_1 + \frac{\ln z +\Delta}{D
\tau_A^{1/3}(y)}\right)\, z^{1-\gamma}\NN &&\qquad
\times\,\exp\left\{\tau_A(y)-\rho_A -\sqrt{4by\ln (\ln z +
\tau_A(y))} -\frac{2\ln^2 z}{3\beta \tau_A(y)} \right\}, \EQA
where the normalization is not under control. This is valid
provided the nucleus is in the BFKL window (\ref{swindowrun}),
which implies ($c_1$ is a constant of ${\mathcal O}(1)$) :
 \beq\label{swindow5} 0<\,\ln z <\, c_1\,\tau_A^{2/3}(y). \eeq It
can be checked that, within this physical range, the ratio
${\mathcal R}_{pA}(z,y)$ is monotonously {\sf
increasing}\footnote{Although the ratio (\ref{BFKLDLArun})
develops a maximum at large $z$, it is well outside the physical
region, similarly to the fixed coupling case. In fact, one can
roughly estimate the position of the maximum to be $\ln z\sim a
\tau_A(y)$ with $a\sim {\mathcal O}(1)$.} with $z$, and also
rapidly {\sf decreasing} with $y$ at fixed $z$.

In particular,  when approaching the nuclear saturation scale from
the above, one recovers the same parametric dependencies upon $A$
and $y$ as in Eq.~(\ref{RmaxDLArun}).

{\sf (II) $y>y_c'$: The double scaling regime}

As before, this regime is interesting since the uncontrollable
normalization factors cancel out in the ratio between the gluon
distributions for the nucleus and the proton. Using
Eq.~(\ref{BFKLrun}), one finds : \be\label{DBFKLrun} {\mathcal
R}_{pA}(z,y)&\,\simeq\,&
\frac{1}{A^{1/3}}\left(\frac{\tau_A}{\tau_p}\right)^{1/3}
\frac{{\rm Ai}\left(\xi_1+\frac{\ln z+\Delta}{D
\tau_A^{1/3}}\right)} {{\rm Ai}\left(\xi_1+\frac{\ln z
+\tau_A-\tau_p + \Delta}{D \tau_p^{1/3}}\right)}\NN && \times
\,\exp\left[ \gamma(\tau_A-\tau_p) +\frac{2(\ln z
+\tau_A-\tau_p)^2}{3\beta \tau_p}- \frac{2\ln^2z}{3\beta \tau_A}
\right], \ee which is free of normalization ambiguities. Since now
$y$ is relatively large, one can replace in this and the
subsequent equations $\tau_A-\tau_p\approx \rho_A^2/2\sqrt{2cby}$
and $\tau_p\approx \sqrt{2cby}$.

\begin{figure}[]
\begin{center}
\includegraphics[scale=1.4]{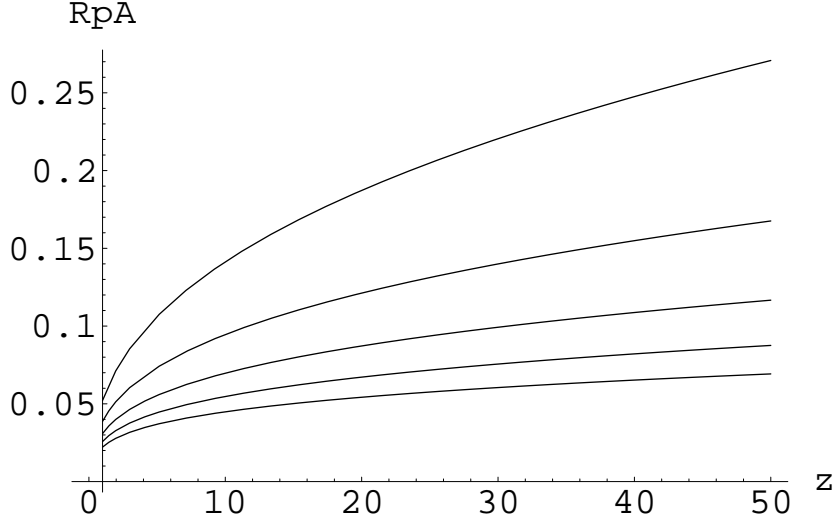}
\caption{{\sl The ratio ${\mathcal R}_{pA}(z,y)$
 in the double BFKL regime for $\rho_A=6$ as a function of $z$ at
different values of $y$.} {\small The lines correspond to
$y=1.1+0.3\, n,$ with $n=0,\dots,4$ from top to bottom. The third
line ($n=2$) corresponds to $y=y_c''=1.7$. The range in $z$ shown
in the figure is in the physical range for all the lines, except
for the top one. Indeed, the upper boundary of the physical regime
is $z_{\rm max}=22.8$ for $y=1.1$, $z_{\rm max}=50$ for $y=1.4$,
$z_{\rm max}=97$ for $y=y_c''$, etc. }} \label{RpA_fig_running}
\end{center}
\end{figure}

The upper limit for the validity range of this expression is fixed
by the borderline for the proton BFKL regime (see
Fig.~\ref{EVOL-MAP-run}). Therefore,  Eq.~(\ref{DBFKLrun}) is
valid for: \BQ\label{rangeDBFKL1} 0<\ln z < c_1\,\tau_p^{2/3}(y) -
(\tau_A(y)-\tau_p(y)). \EQ It can be checked that, the ratio
(\ref{DBFKLrun}) is an {\sf increasing} function of $z$, and a
{\sf decreasing} function of $Y$. This can be explicitly seen on
the numerical plot of the ratio in Fig.~\ref{RpA_fig_running}.
Note the rapid decrease of the ratio with increasing $y$. As
compared to the corresponding plot for fixed coupling, in Fig.
\ref{BBF}, one can see that with a running coupling the decrease
of ${\mathcal R}_{pA}(z,y)$ is pursued up to larger values of $y$,
until the ratio eventually stabilizes at the $z$--independent (and
comparatively smaller) value $1/A^{1/3}$.

This evolution can be more explicitly studied by using the
approximate form of  Eq.~(\ref{DBFKLrun}) valid in the ``double
scaling'' window, which is the regime where both the nucleus and
the proton exhibit geometric scaling. This window opens at a
rapidity $2cb y_c''\sim \rho_A^3$, and extends over the following
interval in $z$ ($c_3$ is a constant of ${\mathcal O}(1)$); see
Eq.~(\ref{20geom})):
 \BQ\label{rangeDS1} 0<\ln z < \ln
\left[Q_g^2(p,y)/Q_s^2(A,y)\right]= c_3\,\tau_p^{1/3}(y) -
(\tau_A(y)-\tau_p(y)), \EQ  which is represented by the dark
shaded area in Fig.~\ref{EVOL-MAP-run}. In this range, the ratio
takes the simpler form (cf. Eq.~(\ref{20phiscale})) :
\be\label{DSCALINGrun} {\mathcal R}_{pA}(z,y)&\simeq&
\frac{1}{A^{1/3}}\, \frac{\ln z+\Delta}{\ln z
+\rho_A^2/2\sqrt{2cby} + \Delta} \ \exp\left\{\frac{\gamma}{2}\,
\frac{\rho_A^2}{\sqrt{2cby}}\right\} , \ee which when approaching
the saturation line from above ($\ln z \to 0$) is consistent with
the previous estimate in Eq.~(\ref{20rsatscale}).

 For large $y$, such that $2cby \gg \rho_A^4$, the ratio
(\ref{DSCALINGrun}) becomes independent of $z$, and approaches the
universal limit $1/A^{1/3}$, as anticipated. This is also the
limit of the full expression (\ref{DBFKLrun}) for $2cby \gg
\rho_A^4$ and any $z$ in the range (\ref{rangeDBFKL1}).

\subsubsection{The nuclear DLA region}

By using Eq.~(\ref{DLAprun}) for both the proton and the nucleus,
 one easily obtains: \BQA \hspace*{-0.7cm}{\mathcal R}_{pA}(z,
y)&=&\exp\left[ \sqrt{4by\{\ln (\ln z + \tau_A(y))-\ln \rho_A\}}
-\sqrt{4by\left\{\ln(\ln z +\tau_A(y))-\ln \rho_p\right\}}
\right]\,,\NN \EQA which is valid, roughly, for $\ln z \simge
\tau_A(y)$. It is easy to check that  the ratio is an increasing
function of $z$ for any $y$. We have numerically verified that,
for $\ln z \simge \tau_A(y)$ at least, the ratio is an decreasing
function of $y$.

At very large $z$ such that $\ln z \gg  \tau_A(y)$, one obtains
\BQ {\mathcal R}_{pA}\,\simeq \,{\rm e}^{- \ln
\frac{\rho_A}{\rho_p}\, \sqrt{by/\ln \ln z}}. \EQ Thus, when $z\to
\infty$ the ratio approaches one {\sf from the below} (but only
very slowly). This concludes our analysis.

\section*{Acknowledgments}

We are particularly grateful to Al Mueller for his careful and
repeated reading of this manuscript, and many useful suggestions.
We would like to thank Stephane Munier for many discussions and
for sharing with us the results of a numerical calculation which
helped us clarifying the argument for the flattening of the Cronin
peak. During the elaboration of this work, we have benefited from
vivid discussions, incisive questions, and insightful remarks from
Jean-Paul Blaizot, Fran\c cois Gelis, Larry McLerran. One of the
authors (K.I.) is thankful to Kirill Tuchin for informative
discussions. This work has been completed while one of the authors
(E.I.) was a visitor at Universidade Federal do Rio de Janeiro. He
wishes to thank Eduardo Fraga for his hospitality and support
during this visit.

\appendix

\section{Appendix}

Here we study in more detail the gluon occupation factor
$\varphi_A$ and the integrated gluon distribution function
$\mathcal{G}_A$ in the McLerran-Venugopalan model, both for fixed
and running coupling. We first deal with the running coupling
case, which turns out to be simpler, and then we continue with the
fixed coupling one. At the end of the appendix we present some
useful properties of the confluent hypergeometric function. Note
that, in order to simplify writing, we shall often use different
normalizations for the various quantities as compared to the main
text. The differences will be indicated at the appropriate places.

\subsection*{The Running Coupling Case}

The twist contribution to the gluon occupation factor, as given in
Eq.~(\ref{phiMVrunexp}), and with the constant factor $1/(b_0
N_c)$ omitted since it can be restored at the end of the
calculation, reads
\begin{equation}\label{phiat}
    \varphi_A^{\rm twist}(z) =
    \int\limits_{0}^{\infty}
    dt \, J_0\left(\sqrt{4 z t}\right)\,
    \frac{1- e^{-t}}{t}\, \ln \frac{1}{t}\,.
\end{equation}
In principle, one should integrate up to $\sim t_{\rm
max}=e^{\rho_A}$, which corresponds to a dipole size $\sim r_{\rm
max}=2/\Lambda$. However, it is harmless to extend the integration
to infinity, so long as we are interested in momenta $k \gg
\Lambda \Leftrightarrow z \gg  e^{-\rho_A}$. We can do the above
integration by replacing $\ln (1/t) \rightarrow t^{-a}$. Then we
obtain an expression involving the confluent hypergeometric
function $_1F_1(-a,1,-z) \equiv \Phi(-a,1,-z)$. Finally we take
the derivative with respect to $a$ at $a=0$ to recover the
quantity in Eq.~(\ref{phiat}), and we obtain
\begin{equation}\label{phiat2}
    \varphi_A^{\rm twist}=
    -\frac{1}{2}\ln^2 z - \gamma \ln z +
    \gamma\, \Gamma(0,z) +
    \frac{\pi^2}{12} - \frac{\gamma^2}{2}
    +\frac{1}{2}\, \Phi^{200}(0,1,-z),
\end{equation}
where $\gamma\equiv \gamma_E = 0.577$ denotes the
Euler--Mascheroni constant throughout this Appendix. In the
small--$z$ limit the hypergeometric term vanishes and one is left
with
\begin{equation}
    \varphi_{A}^{\rm twist} =
    -\frac{1}{2}\ln^2 z - 2 \gamma \ln z +
    \frac{\pi^2}{12} - \frac{3 \gamma^2}{2} +
    \mathcal{O}(z).
\end{equation}
Notice that this is negative, but it never becomes larger in
magnitude than the contribution of the saturation term
$\varphi_{A}^{\rm sat}=\rho_A \Gamma(0,z)$. Even at $k=\Lambda
\Leftrightarrow z=e^{-\rho_A}$, one has
\begin{equation}
    \varphi_{A}^{\rm twist} \simeq
    - \frac{1}{2} \rho_A^2  \simeq
    -\frac{1}{2} \varphi_{A}^{\rm sat}.
\end{equation}
and the gluon occupation factor is positive and well defined.

In the large-$z$ region, we can use Eq.~(\ref{hypasy}), given at
the end of the appendix, to derive the asymptotic expansion of the
index-differentiated hypergeometric function. This will have the
form given in Eq.(\ref{dhypasy}). It is straightforward to show
that
\begin{equation}\label{d2hypasy}
    \Phi^{200}(0,1,-z)=
    \ln^2 z + 2 \gamma \ln z +
    \gamma^2 - \frac{\pi^2}{6} +
    2\sum_{n=1} \frac{\Gamma(n)}{n} \frac{1}{z^n}
    + \mathcal{O}(e^{-z}).
\end{equation}
Then, Eqs.~(\ref{phiat2}) and (\ref{d2hypasy}) imply that all the
logarithmically divergent and constant terms in $\varphi_A$
cancel, and as expected only the inverse powers of $z$ survive.
Therefore one has
\begin{align}
    \varphi_A=\varphi_A^{\rm twist} =
    \sum_{n=1} \frac{\Gamma(n)}{n} \frac{1}{z^n}
    + \mathcal{O}(e^{-z})
    = \frac{Q_s^2(A)}{k^2}+
    \frac{1}{2} \frac{Q_s^4(A)}{k^4}+
    \frac{2}{3} \frac{Q_s^6(A)}{k^6} +
    \cdot \cdot \cdot,
\end{align}
which exhibits the correct bremsstrahlung spectrum, accompanied by
power corrections which are all positive.\\

Now we turn our attention to the integrated gluon distribution
function, which we conveniently normalize as
\begin{equation}
    \mathcal{G}_A=
    \int \limits^{Q}
    \frac{d^2 k}{\pi Q_s^2(A)}\, \varphi_A.
\end{equation}
For the saturation term of the gluon occupation factor the
integration is trivial, and one obtains
    \begin{equation}\label{gsat}
        \mathcal{G}_A^{\rm sat}=
        \left[ 1- e^{-Z} +Z\, \Gamma \left(0,Z\right)
        \right] \rho_A.
    \end{equation}
In the above we have defined the scaled momentum variable
$Z=Q^2/Q_s^2(A)$. For the twist contribution as given in
(\ref{phiat}), one can perform first the integration over $k$,
which simply replaces $J_0 \left(\sqrt{4 z t}\right)$ by
$\sqrt{Z}\, J_1\left(\sqrt{4 z t}\right)/\sqrt{t}$. Then the
integration over $t$ can be done as before and gives
\begin{align}
    \mathcal{G}_A^{\rm twist}=&
    -\frac{1}{2}\, Z\,\ln^2 Z
    +(1-\gamma)\, Z\, \ln Z
    +\left( \frac{\pi^2}{12} -\frac{\gamma^2}{2}
    + \gamma -1 \right) Z
    \\ \nonumber
    & +\gamma \left[ 1- e^{-Z} +Z\, \Gamma \left(0,Z\right)
    \right]
    + \frac{1}{2}\, Z\, \Phi^{200}(0,2,-Z).
\end{align}
When $Q^2 \gg Q_s^2(A) \Leftrightarrow Z \gg 1$, the last two
equations lead to
\begin{equation}
    \mathcal{G}_A =
    \rho_A + \ln Z + 2 \gamma
    -\sum_{n=1} \frac{\Gamma(n)}{n+1} \frac{1}{Z^n}
    +\mathcal{O}(e^{-Z}),
\end{equation}
where the first term comes from $\mathcal{G}_A^{\rm sat}$ and the
remaining from $\mathcal{G}_A^{\rm twist}$. One can rewrite the
above equation, in a more illuminating way, as
\begin{equation}
    \mathcal{G}_A=
    \ln \frac{Q^2}{\Lambda^2} + 2 \gamma
    - \frac{1}{2} \frac{Q_s^2(A)}{Q^2}
    - \frac{1}{3} \frac{Q_s^4(A)}{Q^4}
    +\cdot \cdot \cdot,
\end{equation}
which shows that the higher twist corrections give a small
negative contribution (shadowing) to the integrated gluon
distribution function.

\subsection*{The Fixed Coupling Case}

The twist contribution to the gluon occupation factor as given in
Eq.~(\ref{phiMVexp}) and multiplied by the constant factor $\rho_A
\alpha_s N_c$, reads
\begin{equation}\label{phiatfix}
    \varphi_{A}^{\rm twist}(z) =
    -\int \limits_{0}^{\infty}
    dt \, J_0\left(\sqrt{4 z t}\right)\, e^{-t}
    \sum_{n=1}^{n_{\rm max}}
    \frac{t^{n-1} \ln^n t}{n! \, \rho_A^{n-1}}.
\end{equation}
The integration in the above equation needs to be cut at $\sim
t_{\rm max} = e^{\rho_A}$ which corresponds to a dipole size $\sim
r_{\rm max}=2/\Lambda$. However, and in the ``worst case'' that
$z$ is small, each term gets its most contribution from the region
$t\approx n$, with the ``magnitude'' of the $n$-th term being
$\sim (\ln n/\rho_A)^n$. Therefore the integration in the terms
with $n \lesssim n_{\rm max} = e^{\rho_A}$ can be extended to
infinity, while terms with $n \gtrsim n_{\rm max}$ do not
contribute, since their peak lies outside the region of
integration. Thus, the cutoff in the MV model translates to a
cutoff in the series in Eq.~(\ref{phiatfix}). Furthermore, this
(asymptotic) series is rapidly convergent for $\rho_A \gg 1$,
since $\ln n/\rho_A < 1$ for $n < e^{\rho_A}$. Thus, it is enough
to keep a few terms in (\ref{phiatfix}). They can be calculated by
the same method we used in the running coupling case. The $n$-th
term, apart form the overall suppressing factor $\rho_A^{-n+1}$,
becomes constant when $z \ll 1$, and falls as $z^{-n}$ (plus all
possible higher inverse powers) for $z \gg 1$. The expressions
become more and more complicated as $n$ increases, and therefore
we shall present only the first two terms. We have
\begin{align}
    \varphi_A^{n=1} = &
    \gamma\,e^{-z}
    -\Phi^{100}(1,1,-z),
    \\ \nonumber
    \varphi_A^{n=2} = & -\frac{1}{\rho_A}
    \Bigg[
    \left( \frac{\pi^2}{12} + \frac{\gamma^2}{2} -\gamma  \right)
    (1-z)\, e^{-z}
    \\
    & +(1-\gamma) \,\Phi^{100}(2,1,-z)
    +\frac{1}{2}\, \Phi^{200}(2,1,-z)
    \Bigg].
\end{align}
Since $\Phi^{n00}(a,1,-z)$ vanishes at small-$z$, the limiting
value of $\varphi_A^{\rm twist}$ is finite. One easily finds that
\begin{equation}
    \varphi_A^{\rm twist}=
    \gamma - \frac{1}{\rho_A}
    \left( \frac{\pi^2}{12} + \frac{\gamma^2}{2} -\gamma \right)
    +\mathcal{O}(z).
\end{equation}
In the large-$z$ limit,  we need to expand the hypergeometric
functions and we obtain
\begin{equation}
    \varphi_A=\varphi_A^{\rm twist}= \frac{Q_s^2(A)}{Q^2}+
    \frac{Q_s^4(A)}{Q^4}
    \left[
    1 + \frac{1}{\rho_A}
    \left( \ln \frac{Q^2}{Q_s^2(A)} + 2 \gamma -2 \right)
    \right]
    +\cdot \cdot \cdot,
\end{equation}
where both the $n=1$ and $n=2$ terms contribute to the
$1/z^2=Q_s^4(A)/Q^4$ term. This correction is positive and the
same
will be true for all the $1/z^n$ ones.\\

Now let us calculate the integrated gluon distribution function.
The saturation contribution is the same as the one we found in the
running coupling case and is given in Eq.~(\ref{gsat}). For the
twist contributions we shall again present only the first two
terms, which are
\begin{align}
    \mathcal{G}_A^{n=1}= &
    \gamma -\gamma\, e^{-Z} -Z\,\Phi^{100}(1,2,-Z)
    \\ \nonumber
    \mathcal{G}_A^{n=2} = & -\frac{1}{\rho_A}\,Z\,
    \Bigg[
    \left( \frac{\pi^2}{12} + \frac{\gamma^2}{2} -\gamma  \right)
    \, e^{-Z}
    \\
    &+(1-\gamma) \,\Phi^{100}(2,2,-Z)
    +\frac{1}{2}\, \Phi^{200}(2,2,-Z)
    \Bigg],
\end{align}
and similar, but more complicated expressions can also be obtained
for $n \geq 3$. Each term vanishes as $Z$ becomes small, and all
terms with $n \geq 2$ approach 0 as $Z$ becomes large. Keeping the
$n=1,2$ terms, we can obtain the first power law corrections in
$\mathcal{G}_A$ in the high momentum limit $Q_s^2(A) \gg Q^2$. We
obtain
\begin{equation}
    \mathcal{G}_A=
    \ln \frac{Q^2}{\Lambda^2} + 2 \gamma
    -\frac{Q_s^2(A)}{Q^2}
    \left[
    1 + \frac{1}{\rho_A}
    \left( \ln \frac{Q^2}{Q_s^2(A)} +2 \gamma -1\right)
    \right]\,.
\end{equation}
As in the running coupling case the higher twist terms give a
small negative contribution to the integrated gluon distribution
function.

\subsection*{The Confluent Hypergeometric Function $\Phi(a,b,-z)$}

Here we give some useful properties of the confluent
hypergeometric function $_1F_{1}(a,b,-z) \equiv \Phi(a,b,-z)$.
More precisely we examine its $n$-th derivative with respect to
the first argument, that is $\Phi^{n00}(a,b,-z)$. The range of
parameters that we are interested in is $a \geq 0$, $b \geq 1$ and
$z \geq 0$.

The confluent hypergeometric function is defined as
    \begin{equation}\label{hypdef}
        \Phi(a,b,-z)=1 - \frac{a}{b}\, z +
        \frac{1}{2!}\, \frac{a (a+1)}{b (b+1)}\, z^2 -
        \cdot \cdot \cdot,
    \end{equation}
and its large-$z$ asymptotic expansion reads
    \begin{align}\label{hypasy}
        \Phi(a,b,-z)=&\,
        \frac{\Gamma(b)}{\Gamma(b-a)}\,z^{-a}
        \\ \nonumber
        \times &\left[ 1+
        \frac{a (a-b+1)}{z}+
        \frac{1}{2!}\frac{a (a+1) (a-b+1) (a-b+2)}{z^2}
        +\cdot \cdot \cdot  \right] ,
    \end{align}
where terms that fall as $e^{-z}$ have been neglected.

We can find the behavior of $\Phi^{n00}(a,b,-z)$ in the small-$z$
region from Eq.~(\ref{hypdef}) and in the large-$z$ region from
Eq.~(\ref{hypasy}). We obtain
\begin{equation}\label{dhyplow}
    \Phi^{n00}(a,b,-z)=
    \frac{\Gamma(b)}{\Gamma(b+n)} z^n +
    \mathcal{O}(z^{n+1})
    \qquad {\rm for } \,\, z \ll 1,
\end{equation}
which vanishes when $n \geq 1$, and
\begin{equation}\label{dhypasy}
    \Phi^{n00}(a,b,-z)=z^{-a}
    \left[ \sum_{k \geq 0} \frac{c_k}{z^k}  \right]
    \left[ \sum_{k=0}^{n} d_k \ln^{k} z  \right]
    \qquad {\rm for } \,\, z \gg 1,
\end{equation}
where the coefficients $c_k$ and $d_k$ depend on the set of
parameters $(n,a,b)$ and can be determined from
Eq.~(\ref{hypasy}). The function $\Phi^{n00}(a,b,-z)$ diverges for
large-$z$ as a power of $\ln z$ when $a=0$, while it approaches
zero when $a>0$.


\end{document}